\documentclass[a4paper,11pt]{article}
\pdfoutput=1

\usepackage[a4paper,left=2.73cm,right=2.7cm,top=3cm,bottom=3.5cm]{geometry}

\usepackage{amsmath,amssymb,graphicx,wrapfig,caption,subcaption,enumerate,color,tikz}
\usetikzlibrary{arrows}
\usepackage{colortbl}
\usepackage{booktabs}

\usepackage[mathscr]{euscript}

\usepackage[colorlinks=true,linktocpage=true,linkcolor=blue,citecolor=blue]{hyperref}

\definecolor{gray}{rgb}{.9,.9,.9}
\tikzstyle{block} = [rectangle, text width=5em, text  centered, rounded corners]
\tikzstyle{line} = [draw, very thick, -latex']
\tikzstyle{RGflow}= [->, shorten <=1pt, thick, dashed, color=black!70]

\addtocontents{toc}{\protect\setcounter{tocdepth}{2}}
\numberwithin{equation}{section}

\def\d{\mathrm{d}}
\newcommand{\mt}[1]{\textrm{\tiny #1}}
\newcommand{\sac}{\, , \qquad}
\newcommand{\eqq}[1]{(\ref{#1})}
\newcommand{\fig}[1]{Fig.~\ref{#1}}
\newcommand{\Sec}[1]{Sec.~\ref{#1}}
\newcommand{\eq}[1]{Eq.~(\ref{#1})}
\newcommand{\cM}{{\cal M}}
\newcommand{\rh}{r_\mt{h}}
\newcommand{\orh}{\overline r_\mt{h}}

\newcommand{\lp}{\Lambda_\mt{LP}}

\newcommand{\ls}{\ell_s}
\newcommand{\gym}{g_\mt{YM}}

\newcommand{\Qc}{Q_\mt{c}}
\newcommand{\Qf}{Q_\mt{f}}
\newcommand{\Qst}{Q_\mt{st}}
\newcommand{\nq}{N_\mt{q}}
\renewcommand{\xi}{\overline N_\mt{q}}
\newcommand{\nc}{N_\mt{c}}
\newcommand{\nf}{N_\mt{f}}
\newcommand{\tens}{T_\mt{D7}}

\newcommand{\ratio}{\xi}

\newcommand{\Jkah}{J}
\newcommand{\etakah}{\eta}

\newcommand{\bfunc}{{\cal B}}
\newcommand{\cB}{\mathcal B}
\newcommand{\cfunc}{C}
\newcommand{\cN}{{\cal N}}
\newcommand{\cE}{E}
\newcommand{\cF}{{\cal F}}
\newcommand{\cH}{{\cal H}}
\newcommand{\cG}{\mathsf{G}}
\newcommand{\cC}{\mathsf{C}}
\newcommand{\cA}{\mathcal{A}}
\newcommand{\At}{\cA_{t}}

\newcommand{\ccF}{\mathsf{F}}

\def\sfh{\mathsf{h}}
\def\sff{\mathsf{f}}
\def\sfg{\mathsf{g}}
\def\sfb{\mathsf{b}}
\def\sfc{\mathsf{c}}

\newcommand{\be}{\begin{equation}}
\newcommand{\ee}{\end{equation}}
\newcommand{\bsal}{\begin{align}}
\newcommand{\bal}{\begin{aligned}}
\newcommand{\eal}{\end{aligned}}
\newcommand{\bse}{\begin{subequations}}
\newcommand{\ese}{\end{subequations}}

\newcommand{\jt}[1]{{\bf [JT: #1]}}

\def\d{\mathrm{d}}

\def\temp{\Big|_T}

\def\charge{\Big|_{\Qst}}

\def\energy{\Big|_\cE}

\def\chem{\Big|_{\mu}}

\begin{document}

\begin{titlepage}

\thispagestyle{empty}

\begin{flushright}
\hfill{ICCUB-17-013}
\end{flushright}

\vspace{40pt}  
	 
\begin{center}

{\LARGE \textbf{Towards a Holographic Quark Matter Crystal
}}
	\vspace{30pt}
		
{\large \bf Ant\'on F. Faedo,$^{1}$   David Mateos,$^{1,\,2}$   \\ [2mm]
Christiana Pantelidou$^{1}$  and Javier Tarr\'\i o$^{3}$}

\vspace{25pt}

{\normalsize  $^{1}$ Departament de F\'\i sica Qu\'antica i Astrof\'\i sica and Institut de Ci\`encies del Cosmos (ICC),\\  Universitat de Barcelona, Mart\'\i\  i Franqu\`es 1, ES-08028, Barcelona, Spain.}\\
\vspace{15pt}
{ $^{2}$Instituci\'o Catalana de Recerca i Estudis Avan\c cats (ICREA), \\ Passeig Llu\'\i s Companys 23, ES-08010, Barcelona, Spain.}\\
\vspace{15pt}
{ $^{3}$
Physique Th\'eorique et Math\'ematique, Universit\'e Libre de Bruxelles (ULB) \\
and International Solvay Institutes, Campus de la Plaine CP 231, B-1050, Brussels, Belgium.}

\vspace{15pt}

\texttt{afaedo@ffn.ub.es, dmateos@icrea.cat, cp109@ffn.ub.es, jtarrio@ulb.ac.be}

\vspace{40pt}
				
\abstract{
We construct the gravity dual of $d=4$, $\mathcal{N}=4$, SU($\nc$) super Yang--Mills theory, coupled to $\nf$  flavors of dynamical quarks, 
at non-zero temperature $T$ and non-zero quark density $\nq$.
The supergravity solutions possess a regular horizon if $T>0$ and include the backreaction of $\nc$ color 
D3-branes and $\nf$ flavor D7-branes with $\nq$ units of electric flux on their worldvolume. At zero temperature the solutions interpolate between a Landau pole singularity in the ultraviolet and a Lifshitz geometry in the infrared. At high temperature the thermodynamics is directly sensitive to the Landau pole, whereas at low temperature it is not, as expected from effective field theory. At low temperature and sufficiently high charge density we find thermodynamic and dynamic instabilities  towards the spontaneous breaking of translation invariance. 
}

\end{center}

\end{titlepage}

\tableofcontents

\hrulefill
\vspace{10pt}

\section{Introduction and summary}
\label{intro}
Quantum Chromodynamics (QCD) at non-zero quark density is notoriously difficult to analyze. The only first-principle, non-perturbative tool, namely lattice QCD, is of  limited applicability due to the so-called sign problem \cite{deForcrand:2010ys}. It is therefore useful to construct toy models of QCD in which  interesting questions can be posed and answered. The gauge/string duality, or holography for short \cite{Maldacena:1997re,Gubser:1998bc,Witten:1998qj}, provides a framework in which the construction of  such models is possible. The goal is not to do precision physics but to be able to perform first-principle calculations that may lead to interesting insights applicable to QCD (see e.g.~\cite{Mateos:2011bs} for a discussion of the potential and the limitations of this approach). In the case of QCD at non-zero temperature, the insights obtained through this program  range from static properties to far-from-equilibrium dynamics of strongly coupled plasmas (see e.g.~\cite{CasalderreySolana:2011us} and references therein). 

The simplest four-dimensional example of holographic duality can be obtained from the supergravity solution for a collection of $\nc$ D3-branes. In this setup one finds an equivalence between four-dimensional, 
$\mathcal{N}=4$ super Yang--Mills (SYM) theory with gauge group SU$(\nc)$ living on the stack of D3-branes and string theory on the near-horizon limit of the geometry sourced by the D3-branes \cite{Maldacena:1997re}. Since the matter in these theories is in the adjoint representation of the gauge group, in order to consider a non-zero quark density new degrees of freedom in the fundamental representation must be included. On the gravity side this can be done by adding $\nf$ so-called flavor D7-branes to the D3-brane geometry \cite{Karch:2002sh}. For conciseness  we will refer to these new degrees of freedom as `quarks' despite the fact that they include both bosons and fermions. Placing the theory at a non-zero quark density $\nq$ then corresponds to turning on $\nq$ units of electric flux on the flavor branes \cite{Kobayashi:2006sb}. This flux sources the same supergravity fields as a density $\nq$ of fundamental strings dissolved inside the flavor branes. We will thus refer to $\nq$ as the quark density, as the electric flux on the branes, or as the string density interchangeably.  Note that $\nc$ and $\nf$ are dimensionless integer numbers, whereas $\nq$ is a continuous variable with dimensions of (energy)$^{3}$.

If $\nf \ll \nc$ and $\nq$ is sufficiently small then there exists an energy range in which one can study this system in the so-called `probe approximation' (see \cite{Kruczenski:2003be,Babington:2003vm} for early references and \cite{Erdmenger:2007cm,CasalderreySolana:2011us} for reviews), meaning that the gravitational backreaction of the flavor branes and the strings on the original 
D3-brane geometry can be neglected. However, this approximation inevitably breaks down both at sufficiently high and at sufficiently low energies. 

At high energies the probe approximation breaks down because it ignores the positive \mbox{$\beta$-function} of the gauge theory. 
One way to see this is to note that the $\beta$-function is proportional to $\nf/\nc$ and hence the logarithmic running of the coupling is a small correction to the physics over energy ranges that are not exponentially large in $\nc/\nf$. However, at sufficiently high energies the coupling constant eventually diverges because the theory develops a Landau pole. In this region the backreaction cannot be ignored and the probe approximation ceases to be valid. Nevertheless, this high-energy regime can be correctly described holographically  
\cite{Faedo:2016cih} by including the backreaction of the D7-branes \cite{Benini:2006hh}.

At low energies the probe approximation breaks down because the backreaction of the charge density always dominates the geometry sufficiently deep in the infrared (IR)  no matter how small $\nq$ is. Changing the value of $\nq$ simply shifts the energy scale at which this happens. Therefore the inclusion of  backreaction  is not an option but a necessity in order to identify the correct ground state of the theory. 

In this paper we will find the fully backreacted supergravity solutions for the D3-D7 system at non-zero temperature $T$ and non-zero quark-density 
$\nq$. As we will explain in Sec.~\ref{sec.setup}, we will distribute, or smear, the flavor branes \cite{Benini:2006hh} (see \cite{Nunez:2010sf} for a review) and the strings \cite{Kumar:2012ui} over the compact part of the geometry. We will focus on the case in which this geometry is an S$^5$, but  our results are also valid (with the sole modification of some numerical coefficients) if this is replaced by any other  five-dimensional Sasaki-Einstein (SE) manifold.  On the gauge theory side this  corresponds to replacing the maximally supersymmetric SU$(\nc)$ gauge  theory by an $\mathcal N =1$ quiver theory.

Our results are pictorially summarised in \fig{fig.RGflows}, in which each solution on the gravity side, or equivalently each Renormalization Group (RG) flow on the gauge theory side, is represented by a curve running from top to bottom. Each curve corresponds to a different value of the charge density. A point on a curve corresponds to a given energy scale.  At zero temperature a solution is described by an entire curve. Cutting off a curve at different points would correspond to different solutions with the same charge density and different non-zero values of the temperature. In the limit  $T\gg \nq^{1/3}$ the solutions were constructed perturbatively in 
$\nq^{1/3}/T$ in  \cite{Bigazzi:2011it,Bigazzi:2013jqa}. In this limit, however, all the relevant IR physics  is hidden behind the horizon. 
\begin{figure}[t!!!]
\begin{center}
\begin{tikzpicture}[auto]


\node [block, draw, text width=8em] (YM) { \large Landau pole};
\node [below right of=YM, node distance=7cm] (dummy) {};
\node [block, draw,  right of=dummy, text width=6em, node distance=1.5cm] (CS) {\large log AdS};
\node [block, draw, below left of=dummy, node distance=7cm, text width=7em] (NR) {\large Lifshitz};


\path [line] (YM) -- node (nochargearrow) {} (CS.north west);
\path [line] (YM) -- node (noflavorarrow)  [left]  {} (NR);
\path [line, double, color=black!90] (CS.south west) -- node (conformalarrow) {} (NR);


\fill (noflavorarrow.east) circle [radius=0pt] node (noflavorcircle) { };
\fill (nochargearrow.south west) circle [radius=0pt] node (nochargecircle) { };
\fill (conformalarrow.north west) circle [radius=0pt] node (conformalcircle) { };


\node [right of=nochargecircle, node distance=2cm] 
{\hspace{-2em}  \large no charge};
\node [ left of=noflavorcircle, node distance=2cm] 
{\hspace{2em} \large large charge};
\node [below left of=CS, node distance=3.6cm] 
{ \hspace{2em} \large small charge};


\draw [RGflow] (YM) to[in=85, out=280]  (NR);
\draw [RGflow] (YM) to[in=65, out=300]  (NR);
\draw [RGflow] (YM) to[in=50, out=312.5,looseness=1.2]  (NR);
\draw [RGflow] (YM) to[in=45, out=315,looseness=1.7]  (NR);
\draw [RGflow] (YM) to[in=45, out=315,looseness=2.2]  (NR);
\end{tikzpicture}
\caption{\small Pictorial representation of the family of supergravity solutions and corresponding RG flows  constructed in this paper.}
\label{fig.RGflows}
\end{center}
\end{figure}

The structure  of the solutions can be simply understood as follows. In the ultraviolet (UV) all solutions approach the neutral solution of 
\cite{Benini:2006hh}, since the geometry is dominated by the backreaction of the D3- and the D7-branes. In other words, in this limit the temperature and the quark density only produce subleading corrections. The asymptotic geometry is singular, as it corresponds to the presence of a Landau pole in the gauge theory. This introduces a physical scale  in the theory, $\lp$,  that can be compared, for example, with $T$ or $\nq^{1/3}$. Despite the presence of the LP singularity, holographic renormalisation can be straightforwardly implemented and physical quantities can be computed unambiguously  \cite{Faedo:2016cih}. This means that, just as in Quantum Electrodynamics, the presence of the Landau pole is no impediment for the extraction  of sensible IR physics. 

At zero temperature and non-zero charge the non-compact part of all solutions approaches a five-dimensional Lifshitz geometry  with dynamical exponent $z=7$ in the IR. In this regime the geometry is dominated by the backreaction of  the strings and the D3-branes, with the D7-branes producing only a subleading effect. We emphasize that the precise IR theory to which the theory flows depends on the value of the charge density and other parameters. In other words, although in all cases the dynamical exponent is the same, other features are different. A simple example is the number of active degrees of freedom at low temperature, as measured by the entropy density. The Lifshitz symmetry implies that this must  scale with temperature  as 
\be
\label{prop}
s = f \, T^{3/z} = f \, T^{3/7} \,,
\ee
but the $T$-independent function $f$ is not fixed by the scaling properties of the IR solution. We will see in Sec.~\ref{insta} that this function depends on $\nq$ and other parameters.

When the charge density vanishes the zero-temperature solution reduces to the supersymmetric solution of \cite{Benini:2006hh}. This configuration flows from the Landau pole geometry in the UV to an IR  that differs ``only logarithmically'' from AdS$_5$, meaning that observables exhibit conformal invariance up to a logarithmic dependence on the energy scale. This flow is represented by the upper diagonal straight line in \fig{fig.RGflows}. The solutions with non-zero temperature that can be obtained by adding a horizon at different points along this line were constructed in \cite{Faedo:2016cih}.

The logarithmic dependence of the log-AdS solution is a consequence of the running of the coupling constant caused by the presence of the flavor branes. It is possible to perform a scaling of all the charges in such a way that this effect disappears while the backreaction of the strings is retained. In this limit there exists a flow from an AdS$_5$ geometry in the UV to a Lifshitz geometry in the IR. This flow was found in \cite{Kumar:2012ui} and is represented by the lower diagonal straight line in \fig{fig.RGflows}. 

We will show in Sec.~\ref{sec.scalings} that, up to simple rescalings, the full set of zero-temperature solutions  can be reduced to a set parameterized only by the dimensionless combination 
\be\label{eq.ratiodependenceINTRO}
\ratio \sim  \ell_s^3\, \frac{\nq \, \nc^{1/4}}{\nf^{1/2}} \,,
\ee
where $\ell_s$ is the string length and we have omitted purely numerical factors (see \eqq{eq.ratiodependence} for the exact expression). 
Solutions with non-zero temperature depend on $\xi$ and
\be
\label{Tintro}
\overline T \sim \ell_s \nc^{1/4} \, T \,,
\ee
where we have omitted purely numerical factors (see \eqq{TTrr} for the exact expression). The factors of $\nc, \nf$ and $\ell_s$ in these equations simply reflect a convenient choice of units on the gravity side. In particular, $\ell_s$ can be traded for a function of $\lp, \nq, \nc, \nf$ and the value of the coupling constant at some reference scale. This  shows that $\xi$ and $\overline T$ are truly gauge-theoretic parameters that can be thought  of as a dimensionless charge density and a dimensionless temperature, respectively. Moreover, the $\ell_s$ factors will always cancel in dimensionless gauge theory quantities and, as we will see, the scaling with $\nc$ of physical quantities suffers from an ambiguity. We will come back to this point in Sec.~\ref{disc}. 

We can now understand the general structure of the results. For large values of $\xi$, which can be thought of  as a large-charge density limit, the supergravity solutions transition directly from the LP geometry to the Lifshitz geometry. Instead, for small values of $\xi$ the solutions exhibit an intermediate region in which they display the physics of the log-AdS region. Adding  temperature simply cuts off the flows at different scales, possibly hiding the Lifshitz and/or the log-AdS regions behind the horizon. 

\begin{figure}[h!!!]
\begin{center}
\includegraphics[width=.89\textwidth]{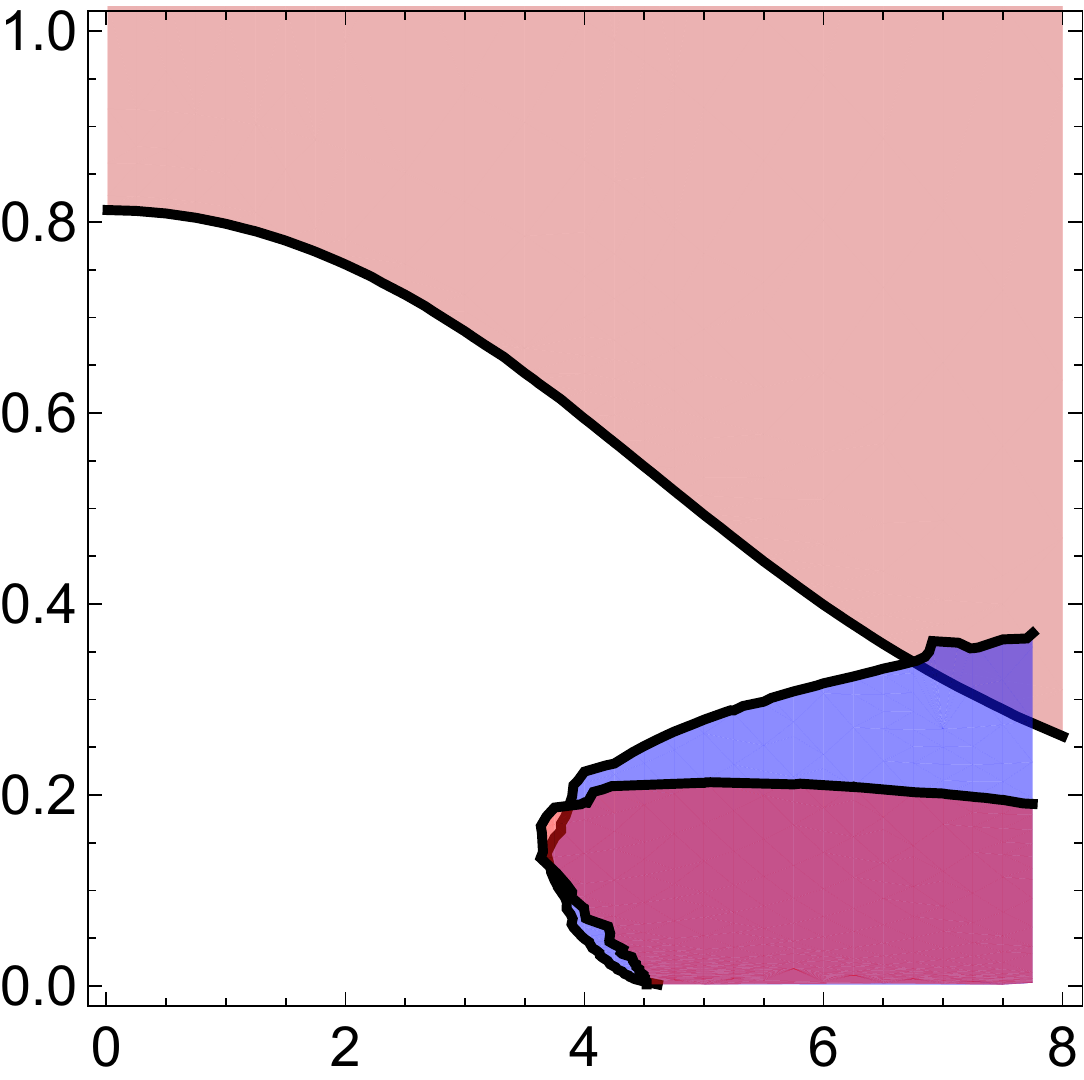} 
\put(-354,323){\large $C_Q=c_s^2=0$}
\put(-60,166){\large $\chi^{-1}=0$}
\put(-56,111){\large $c_s^2=0$}
\put(-140,326){\large  \boxed{\mbox{I}}}
\put(-140,160){\large  \boxed{\mbox{II}}}
\put(-140,116){\large  \boxed{\mbox{III}}}
\put(-140,66){\large  \boxed{\mbox{IV}}}
\put(-231,90){\large \boxed{\mbox{V}} $\longrightarrow$}
\put(-39,144){\large \boxed{\mbox{VI}}}
\put(-430,200){\huge $\overline T$}
\put(-210,-30){\huge $\sqrt{2} \, \xi$}
\caption{\small Phase diagram of the system as a function of temperature and charge density. The dimensionless quantities $\xi$ and $\overline T$ are defined in \eqq{eq.ratiodependence} and \eqq{TTrr}, respectively. The different  curves correspond to the locus where the indicated quantity crosses zero and thus changes sign (see \mbox{Table \ref{signs}}) and divide the diagram into six regions. Note that Region III includes a small area below and to the left of region IV.  
}
\label{phases}
\end{center}
\end{figure}

The main result  of our analysis is the phase diagram of the system. This is schematically summarised in \fig{phases}. We will refer to this figure as a ``phase diagram'' despite the fact that it is a slight abuse of terminology,  since some of the solutions are unstable and we have not identified the putative stable solution that would be thermodynamically preferred.  
The signs of different physical quantities in each region of \fig{phases} are summarised in Table \ref{signs}, where $C_Q$ is the specific heat at constant charge \eqq{cqdef}, $\chi$ is the charge susceptibility \eqq{stab1}, $c_s^2$ is the speed of sound \eqq{cscs}, and $D$ is the charge diffusion constant \eqq{DDD}. Note that the latter equation implies that the last column is the product of the first three. 
\begin{table}[t!]
\centering
 \begin{tabular}{|c || c | c | c | c |} 
 \hline
  & $C_Q$ & $c_s^2$ & $\chi$ & $D$ \\ [0.5ex] 
 \hline\hline
  I & $-$ & $-$ & $+$ & + \\ 
 \hline
 II & + & + & +  & +\\ 
 \hline
 III & + & + & $-$ & $ -$ \\
 \hline
 IV & + & $-$  & $-$ & + \\
 \hline
 V & + & $-$ & + & $-$ \\ 
 \hline
  VI & $-$ & $-$ & $-$ & $-$ \\ 
 \hline
 \end{tabular}
 \caption{Signs of several physical quantities in the different regions shown in \fig{phases}.}
 \label{signs}
\end{table}


Region I describes the high-temperature behavior of the system. In this region  the solutions are locally thermodynamically unstable because the specific heat at constant charge,  $C_Q$, is negative. The system is  also afflicted by a  dynamical instability associated to a negative speed of sound squared, $c_s^2<0$.  The behavior of $C_Q$ and $c_s^2$  in  Region I  
 is qualitatively analogous to the behavior  in the high-temperature limit of the neutral solution of  \cite{Faedo:2016cih}. In the neutral case  $C_Q$ and $c_s^2$ become negative at exactly the same point because they are related through $c_s^2=s/C_Q$, with $s$ the entropy density, which is of course positive. It is remarkable that this feature extends to  charged solutions, at least within our numerical precision. In other words, the top curve in \fig{phases} indicates the locus where both $C_Q$ and $c_s^2$ change sign simultaneously. The behavior of $C_Q$ and $c_s^2$ in region VI is the same as in Region I. In addition, in this region the charge susceptibility $\chi$ is negative. This indicates an additional thermodynamic instability towards charge clustering, and also results in an additional dynamical instability since the charge diffusion constant becomes negative.  As explained in 
\cite{Faedo:2016cih}, the high-temperature behavior follows directly from the properties of the LP. Since our main interest is in IR  features that are independent of the UV completion of the theory, we will not discuss Regions I and VI  further. 

Region III is both thermodynamically and dynamically unstable, since both $\chi$ and $D$ are negative. In contrast, Region V is only dynamically unstable since $c_s^2$ and $D$ are negative. However, while the existence of the other regions is robust, that of Region V may be a numerical artifact due to our finite resolution in parameter space. 

Region II is the only locally stable region in the phase diagram. Local thermodynamic stability is guaranteed by the fact that both 
$C_Q$ and $\chi$ are positive. Our  analysis shows no sign of a dynamical instability either, since $c_s^2$ and $D$ are both positive. 

Region IV is particularly interesting since it corresponds to the low-temperature, high-charge density regime. This region is locally thermodynamically unstable since $\chi<0$ and also dynamically unstable since $c_s^2 <0$. 

The negative values of $c_s^2$ and/or $D$  that we have encountered imply dynamical instabilities in the hydrodynamic sound and charge diffusion channels, respectively, towards the spontaneous breaking of translation invariance. This suggests  that the putative, stable phase in the corresponding  regions may be a crystalline phase. We will come back to this point in Secs.~\ref{insta} and \ref{disc}.

\section{Gravitational description}\label{sec.setup}

We study holographically a family of  theories consisting of SYM with $\nc$ colors  and $\nf$ massless flavors at non-zero temperature $T$ and non-zero  charge density $\nq$. The string dual to this configuration is given by the geometry sourced by a stack of type IIB $\nc$ D3-branes  coupled to a set of $\nf$ D7-branes with $\nq$ units of electric flux on their worldvolume. The temperature is represented by the presence of a black brane  horizon. We work in the approximation in which the system can be described by means of type IIB supergravity with sources. The regime of validity of this approximation will be discussed in Sec.~\ref{regime}. 

Each D7-brane  extends along the directions parallel to the D3-branes and along the holographic radial direction,  and wraps  a three-cycle inside the five-dimensional compact part of the geometry.  We  smear the D7-branes in these compact directions in the most symmetric way compatible with supersymmetry. In this way we recover some of the isometries that would be broken by a single  D7-brane 
(or by a collection of overlapping D7-branes) and reduce the problem to solving ordinary (as opposed to partial) differential equations.

\subsection{Ten-dimensional ansatz}
Four-dimensional, $\mathcal{N}=4$ SYM theory with $\nc$ colors is holographically dual to supergravity solutions sourced by $\nc$ D3-branes. In the supergravity description this is encoded in a flux of the self-dual Ramond-Ramond (RR)  five-form through an appropriate five-dimensional compact manifold $\cM_5$:
\be
\label{satis}
\ccF_5 = \Qc (1+*) \omega_5 \,.
\ee
Here  $\omega_5$ is the volume form of $\cM_5$, whose total dimensionless volume we denote as $V_5$.  Quantization requires that the D3-brane charge is related to the number of colors through
\be
\label{qqcc}
\Qc = \frac{(2\pi\ls)^4}{2\pi V_5} \, \nc \,.
\ee
Note that, unlike in Refs.~\cite{Faedo:2014ana,Faedo:2015ula,Faedo:2015urf,Faedo:2016jbd}, we follow \cite{Faedo:2016cih} and work with a RR charge quantized in units of $\nc$ instead of $g_s \nc$ (for a comparison between both normalizations, see e.g.~Sec.~4.1 of \cite{Mateos:2011tv}). The latter choice is convenient in situations in which there is a natural factorization of the dilaton $\phi$ of the form $e^\phi=g_s e^{\tilde \phi}$. This is the case, for example, if the gauge theory is conformal, since this means that the dilaton is constant and one can simply normalize it so that $\tilde \phi=0$ everywhere. If the gauge theory is not conformal but approaches a fixed point in the IR or in the UV then it is natural to normalize the dilaton so that $\tilde \phi=0$ at the corresponding fixed point. In contrast, in the solutions that we will consider the dilaton  will run from zero to infinity and there will be no natural factorization  into a constant piece and a running piece. We will therefore work with the full dilaton, which is related to the running YM and 't Hooft couplings through 
\be
\label{coup}
\gym^2 = 2\pi e^\phi \sac \lambda =  \gym^2 \nc \,.
\ee

In the simplest setups  the metric supported by the $\ccF_5$  flux is AdS$_5\times\cM_5$, with $\cM_5$ a Sasaki-Einstein (SE) manifold. The radius $L$ of these two spaces is related to the D3-brane charge through
\be
\label{correct}
\Qc = 4 L^4 \,. 
\ee
If \mbox{$\cM_5=S^5$} then the gauge theory is $\cN=4$ SYM; otherwise it is a non-maximally supersymmetric theory. For example, if $\cM_5=T^{1,1}$ then the gauge theory  is the Klebanov-Witten quiver. In  general  the rank of the gauge group, the radius $L$ and the five-dimensional effective gravitational coupling (see \eqq{eq.5dnewtonsconstant} below) are related through 
\be
\label{kkappa2}
\frac{L^3}{\kappa_5^2} = \frac{\pi \nc^2}{4 V_5}  \,.
\ee

In order to add flavor to any of the theories above it is convenient to view the SE manifold as a  U(1) fibration over a four-dimensional  K\"ahler--Einstein (KE) base. This geometric construction is naturally equipped with an SU(2) structure characterized by a real  one-form, $\etakah$, and a real two-form, $\Jkah$, which is  the K\"ahler form of the  KE manifold. These satisfy the relations
\be\label{eq.SU2structure}
\d \etakah = 2 \Jkah \ , \qquad  \d J=0 \ , \qquad  \frac{1}{2}\Jkah\wedge \Jkah \wedge \etakah = \omega_5 \ .
\ee
They also close on each other under Hodge dualization on the SE manifold:
\be\label{eq.SU2closing}
*_5 \etakah = \frac{1}{2} \Jkah \wedge \Jkah \ , \qquad *_5 \Jkah = \Jkah \wedge \etakah \ .
\ee
We normalize the curvature of the SE manifold so that $R_{ab}=4 \, g_{ab}$. Making use of the fibration structure the most general ten-dimensional metric compatible with the symmetries that we wish to preserve takes the form (in string frame) 
\be\label{eq.10dmetricgeneric}
\d s^2  = G_{tt}(r) \, \d t^2 + G_{rr}(r) \, \d r^2 + G_{xx}(r) \, \d  \vec x\,^2   + G_b(r) \, \d s_\mt{KE}^2 + G_{f}(r) \, \etakah^2 \ ,
\ee
where $t$ and $\vec{x}$ are the gauge theory Minkowski directions and $r$ is the holographic radial coordinate. Note that we have allowed for (i) a squashing between the fiber and the base of the SE manifold, i.e.~$G_b\neq G_f$, since this will be produced by the backreaction of the flavor branes, and (ii)  for the possibility that 
$G_{tt}\neq G_{xx}$, since we will consider solutions with non-zero temperature and/or non-zero charge density that will therefore break Lorentz invariance.

As explained in Sec.~\ref{intro}, the addition of flavor and a quark density on the gauge theory corresponds on the gravity side to the addition of D7-branes with an electric Born-Infeld (BI) field on their worldvolume of the form 
\be\label{eq.FformD7}
\cF = B+ 2\pi\ls^2 \, \d \cA \ ,
\ee
with $B$  the Neveu-Schwarz (NS) potential and 
\be\label{eq.AformD7}
\cA = \At(r) \, \d t 
\ee
the BI potential. 
These objects act as sources for both the RR fields and the NS field strength $H=\d B$, and hence they modify their equations of motion. For the RR fields, through the Hodge-duality relations that these fields obey, this also leads to a modification of their Bianchi identities, and therefore to a modification of their very definition in terms of gauge potentials. The full action in the presence of  the most general set of D7-brane sources is discussed in Appendix \ref{app.action}, to which we refer the reader for additional details.  In this work the general equations simplify because $\cF \wedge \cF =0$. Under these circumstances the Bianchi identities take the form 
\bse\bsal
\label{eqn.F2BIviolation}
\d \ccF_1 & = 2 \, \Qf \, J \,, \\
\d \ccF_3 & = H\wedge \ccF_1 + 2 \, \Qf \, \cF \wedge J \,, 
\label{f3} \\
\d \ccF_5 & = 0 \,.
\label{F5}
\end{align}
\ese
In these equations $\Qf$ is the D7-brane charge, related to the number of D7-branes through
\be
\Qf = \frac{V_3 }{8\pi V_5}  \, \nf \,,
\label{qqff}
\ee
with $V_3 = \int \Jkah \wedge \etakah$ the dimensionless volume of the three-dimensional submanifold \mbox{$\cM_3\subset \cM_5$} wrapped by any of the D7-branes. 
 \eq{F5} is satisfied by \eqq{satis}. \eq{eqn.F2BIviolation} is the ``violation'' of the usual Bianchi identity for $\ccF_1$ that expresses the  fact that D7-branes are magnetic sources for $\ccF_1$. This violation is already present in the case of flavor without strings discussed in \cite{Benini:2006hh} (see \cite{Nunez:2010sf} for a review) and it is solved by taking
\be
\label{already}
\ccF_1 = \Qf\, \etakah \,.
\ee
For later reference, we note that in our conventions the tension of a D7-brane is 
\be
\tens= \frac{1}{2\kappa^2}  = \frac{2\pi}{(2\pi\ell_s)^8} \,,
\ee
with $\kappa^2$ the ten-dimensional gravitational coupling. The second term on the right-hand side of \eqq{f3} is a violation of the usual  Bianchi identity for $F_3$ which was not present for the solutions with strings but without D7-branes  studied in \cite{Faedo:2014ana}. Indeed, this term is only present when the system contains both strings and D7-branes \cite{Bigazzi:2011it}, since it comes from the magnetic components of $\cC_6$ sourced by the term 
\be
\int \cC_6 \wedge \mathcal F
\label{rein}
\ee
contained in the Wess-Zumino (WZ)  part of the D7-branes' action. Through the duality relation
\be\label{hodge}
\ccF_3 = - * \ccF_7 
\ee
 these magnetic components give rise to the second term on the right-hand side of   $\d \ccF_3$. We thus see that, although it is the string density represented by $\mathcal F$ that sources $\cC_6$, the presence of the D7-branes is necessary since the coupling between $\mathcal F$ and $\cC_6$ is supported on their worldvolume. This coupling implies that $\cC_6$ must contain a term of the form 
\be\label{eqn.C6potential}
\cC_6 \supset \bfunc(r) \, \d x^1 \wedge \d x^2 \wedge \d x^3 \wedge J \wedge \eta \,,
\ee
where $\bfunc(r)$ is a function that will be determined below. 
We thus take the following ansatz for the dual three-form:
\be\label{eq.F3form}
\ccF_3  =  \Qst \,\d x^1\wedge \d x^2 \wedge \d x^3  + *\, \d \, 
\Big[ \bfunc(r)\, \d x^1 \wedge \d x^2  \wedge \d x^3 \wedge \Jkah \wedge \etakah  \Big] \,.
\ee
The second term in this expression is the one implied by \eqq{eqn.C6potential}, whereas the first one, as we will see, is related to the string density
\cite{Chen:2009kx,Kumar:2012ui}. This first term can also be interpreted as describing a density of baryon-vertex-like D5-branes \cite{Witten:1998xy} wrapping the compact manifold $\cM_5$. In writing this ansatz we have already made use of the fact that we will seek solutions with $H=0$, i.e.~we have discarded the first term on the right-hand side of \eqq{f3}. 

To see the relation between the first term in \eqq{eq.F3form} and the string density we turn to the equation of motion for the $B$-field:
\be
\label{lhslhs}
\d \, \frac{\delta S}{\delta \d B} = \frac{\delta S}{\delta B} \,, 
\ee
with $S$ the total supergravity-plus-sources action.  
This may be written as
\be
\label{eqn.eomAt}
\d \left( e^{-2\phi} * H \right) = \ccF_1 \wedge *\ccF_3 + \ccF_3 \wedge \ccF_5  + 
\frac{\delta S_{\mt{DBI}}}{\delta B} 
\,,
\ee
with 
\bsal
\frac{\delta S_{\mt{DBI}}}{\delta B} &  \,= \, 
-  \frac{e^{-\phi} \, 2\pi \ls^2 \At'\,\sqrt{G_{xx}^3 \, G_b^2 \, G_f} }{\sqrt{- G_{tt}\, G_{rr} - (2\pi \ls^2 \At')^2 }} \,\, \d x^1 \wedge \d x^2 \wedge \d x^3 \wedge \Jkah \wedge \etakah \wedge \d \ccF_1 
\end{align}
the contribution of the Dirac-Born-Infeld (DBI) part of the D7-brane action. Two observations are important. First, there is no explicit contribution from a variation of the WZ term of the D7-branes because this has combined on the right-hand side of \eqq{eqn.eomAt} with other terms into the gauge-invariant, modified field strengths $\ccF_n$, whose  definition  is given in \eqq{eq.Fdefinitions}. Second, the left-hand side of \eqq{lhslhs} is not the same as the left-hand side of \eqq{eqn.eomAt}, since we have moved terms between both sides of the equation  when going from \eqq{lhslhs} to \eqq{eqn.eomAt}. 

As anticipated above, we will solve \eq{eqn.eomAt} with $B=H=0$, which means that  its right-hand side  must vanish. At first sight it may seem surprising that the presence of strings does not automatically lead to a non-zero 
$H$, but the non-linearities of supergravity imply that the $H$ sourced by the strings can be exactly cancelled by the $H$  sourced by the  products of RR forms in \eqq{eqn.eomAt}. Requiring this cancellation fixes the BI field on the D7-branes to 
\be\label{eq.Atprimesolution}
2\pi\ls^2\,\At' = e^{\phi} \sqrt{-G_{tt} G_{rr} } \frac{\Qc\, \Qst + 4 \Qf \, \bfunc}{ \sqrt{16 \, \Qf^2 \, G_{xx}^3 \, G_b^2 \,G_f + e^{2\phi} \left( \Qc\, \Qst + 4 \Qf \, \bfunc \right)^2  } } \ .
\ee
To close the circle we note that  this is automatically a solution of the equation of motion for the BI  field obtained by varying the full action with respect to $\cA$. The reason is that $\d \cA$ always appears together with $B$ in the gauge-invariant combination \eqq{eq.FformD7}. This means that the so-called electric displacement, i.e.~the momentum conjugate to $\cA$, is given by 
\be 
\frac{\delta S}{\delta \d \cA} = 2\pi\ell_s^2 \, \frac{\delta S}{\delta B}  
\ee
and therefore that  the equation of motion for $\cA$, 
\be 
\d \, \frac{\delta S}{\delta \d \cA} = 0 \,,
\ee
is automatically implied by the exterior derivative of \eqq{lhslhs}. A straightforward calculation shows that the  electric displacement is given by
\be\label{eqn.electricdisplacement} 
\frac{\delta S}{\delta \d \cA}   = 2\pi \ls^2 \, \frac{\Qc\, \Qst}{2\kappa^2} \, \d x^1 \wedge \d x^2 \wedge \d x^3 \wedge \omega_5 \,.
\ee
The string density in the $123$-directions is obtained by integrating this expression over $\cM_5$, 
\be
\nq \, \d x^1 \wedge \d x^2 \wedge \d x^3= 
\int_{\cM_5} \, \frac{\delta S }{\delta \d \cA}   \,,
\ee
which finally yields the relation between the string density and $\Qst$:
\be\label{eq.qstvalue}
 \Qst = (2\pi)^3\ell_s^2 \, \frac{\nq}{\nc} \ .
\ee

\subsection{Five-dimensional effective theory}\label{sec.5dsetup}
In the previous section we have written a ten-dimensional ansatz that includes all the solutions of interest. The RR forms are given by \eqq{satis}, \eqq{already} and \eqq{eq.F3form}. The NS $B$-field vanishes. The BI field is given by \eq{eq.Atprimesolution}. The ten-dimensional metric takes the form 
\eqq{eq.10dmetricgeneric}. The functions that we must solve for are the metric components, 
$\cB$, and the dilaton $\phi$, all of which depend only on the radial coordinate $r$.
For the analysis of the solution it is convenient to reduce the ten-dimensional system to an effective five-dimensional action. This exercise was done in a general setup in Ref.~\cite{Cotrone:2012um},  from where we extract the final result, truncated to the fields of interest for us. 

We begin by parameterizing the ten-dimensional string-frame metric as 
\be
\label{param1}
\d s_{10}^2 = e^{\phi/2} \left( e^{-\frac{10}{3}\sigma} \d s_5^2  + L^2 e^{2\sigma- 2w}\d s_\mt{KE}^2 + L^2 e^{2\sigma+8w} \etakah^2 \right) \,,
\ee
and the effective five-dimensional metric as
\be
\label{5Dmetric}
\d s_5^2 = g_{tt}(r)\, \d t^2 + g_{rr}(r)\, \d r^2 + g_{xx}(r)\, \d x_3^2 \,.
\ee
The RR fluxes $\ccF_1$ and $\ccF_5$ thread only internal directions so, upon dimensional reduction, they simply produce terms in the scalar potential. In contrast, in order to dimensionally reduce the RR three-form we must expand it in terms of five-dimensional forms that we denote $G_n$ (not to be confused with the ten-dimensional RR field strengths $\cG_n$ defined in Appendix \ref{app.action}). Computing the Hodge dual in \eqq{eq.F3form} we see that $\ccF_3$ has the following components:
\be
\ccF_3 =  G_3 + L \, G_2 \wedge \eta + \frac{ L^2 }{ \sqrt{2} } \, G_1 \wedge J \,.
\ee
Imposing now the Bianchi identity \eqq{f3} we see that $G_2$ and $G_1$ can be written in terms of gauge potentials $\cfunc_0$, $\cfunc_1$ (not to be confused with the ten-dimensional RR potentials $\cC_n$ defined in Appendix \ref{app.action}) as 
\be
{G}_2 = \d \cfunc_1 + \frac{\Qf }{L}  \, \cF \ , \qquad G_1 = \d \cfunc_0 - \frac{2\sqrt{2}}{L} \cfunc_1 \,.
\ee
We will see below that, in the gauge $\cfunc_0=0$, the vector $\cfunc_1$ is directly related to the function $\cB$ in \eq{f3}. 
We find it convenient to dualize the three-form $G_3$ to a vector field in five-dimensions.  Since $G_3$ appears in the five-dimensional  action  coupled topologically to the NS potential $B$ \cite{Cotrone:2012um}, 
\be
S_{5}\supset \frac{1}{2\kappa_5^2} \int \left[- \frac{1}{2}e^{\phi+\frac{20}{3}\sigma} G_3 \wedge * G_3 - \frac{2}{L} B \wedge G_3 \right] \,,
\ee
its equation of motion reads 
\be
\d \Big[ e^{\phi + \frac{20}{3}\sigma} * G_3\Big]  =- \frac{4}{L} \, H \,.
\ee
Therefore we define the dual vector $A_1$ (not to be confused with the BI potential $\cA$) and its field strength $F_2$ (not to be confused with the ten-dimensional RR field strengths $\ccF_n$ defined in Appendix \ref{app.action})
through the relation
\be\label{eq.G3dual}
 e^{\phi + \frac{20}{3}\sigma} * G_3 = F_2 = \d A_1 - \frac{4}{L} B \,.
\ee
In summary, the dimensional reduction of $\ccF_3$ gives three gauge potentials 
$\cfunc_0$, $\cfunc_1$, $A_1$ 
with field strengths $G_1$, $G_2$, $F_2$. In terms of these the 
 final result for the five-dimensional effective action is  \cite{Cotrone:2012um}
 \be
S_\mt{5} = S_\mt{grav} + S_\mt{DBI} \,,
 \label{eq.5daction}
\ee
where
\be
\bal
S_\mt{grav} & = \frac{1}{2\kappa_5^2} \int \Big[ (R-V)*1 - \frac{1}{2} \d \phi \wedge * \d \phi - \frac{40}{3} \d \sigma\wedge * \d \sigma- 20\, \d w \wedge * \d w \\[2mm]
& \qquad\qquad\quad - \frac{1}{2} e^{\phi+\frac{4}{3}\sigma-8w} {G}_2 \wedge * {G}_2 - \frac{1}{2} e^{\phi-4\sigma+4w} G_1 \wedge * G_1 
   \\[2mm]
& \qquad \qquad \quad - \frac{1}{2}  e^{-\phi+\frac{20}{3}\sigma} H \wedge * H
- \frac{1}{2} e^{-\phi - \frac{20}{3} \sigma} F_2 \wedge * F_2  \Big] 
\eal\ee
and 
\be
S_\mt{DBI}  = - \frac{2}{\kappa_5^2} \frac{\Qf}{ L^2}  \int \d^5 x\, e^{\phi -\frac{16}{3}\sigma+2w} \left[ \sqrt{-\det\left( g+e^{-\frac{\phi}{2}+\frac{10}{3}\sigma}\cF \right)} -\sqrt{-\det [g]}\right]\,. 
\ee
We emphasize that this five-dimensional action is a useful way to encode the equations of motion of the theory within the ansatz that is of interest to us, but it is not a consistent truncation. The reason is that only solutions of the five-dimensional action for which  
\be
\cF \wedge \cF=0 \sac H=0 
\ee
 can be lifted to a solution of  the original ten-dimensional equations. We have kept $H$ in the five-dimensional action despite the condition that it must vanish for the uplift to exist because its equation of motion \eqq{eqH} is non-trivial even after setting $H=0$. 

The  five-dimensional  Newton's constant in \eqq{eq.5daction} is related to the ten-dimensional  one through
\be\label{eq.5dnewtonsconstant}
\frac{1}{2\kappa_5^2} = \frac{V_5\, L^5}{2\kappa^2} \,.
\ee
We have written the DBI part of the action in a form that vanishes identically if $\cF=0$, so that the contribution of this part of the D7-branes' action in the neutral case is contained entirely in the term linear in $\Qf$ of the scalar potential
\be\label{eq.potential}
\vspace{1mm}
V = \frac{1}{L^2} \left[ 8\, e^{-\frac{40}{3}\sigma} + 4 \, e^{-\frac{16}{3}\sigma+ 2w} \left(\Qf \, e^\phi + e^{10w}-6 \right) + \frac{\Qf^2}{2} e^{2\phi-\frac{16}{3}\sigma-8w} \right] \,.
\vspace{1mm}
\ee
This potential can be derived from the superpotential 
\be\label{eq.superpotential}
{\cal W} =  \frac{e^{-\frac{8}{3}\sigma}}{L}\left[ 4\, e^{-4\sigma}- e^{-4w} \left(6 + 4 \, e^{10w}-\Qf \, e^{\phi} \right) \right]
\ee
via the usual relation 
\be
V = \frac{1}{2} \left[ \left( \frac{\partial {\cal W}}{\partial \phi} \right)^2 + \frac{3}{80}  \left( \frac{\partial {\cal W}}{\partial \sigma} \right)^2 + \frac{1}{40}  \left( \frac{\partial {\cal W}}{\partial w} \right)^2  \right] - \frac{1}{3} {\cal W}^2 \ .
\ee

\noindent
The equations of motion for the  differential forms are
\bse\bsal
\label{eq.5deoms}
\d \left[ e^{-\phi - \frac{20}{3} \sigma} * F_2 \right] & = 0 \ , \\[2mm]
\label{eqG2}
\d \left[ e^{\phi+\frac{4}{3}\sigma-8w} * G_2 \right] & = \frac{2\sqrt{2}}{L}e^{\phi-4\sigma+4w}*G_1 \ ,\\[3mm]
\d \left[  e^{-\phi+\frac{20}{3}\sigma} *H \right] & =  \frac{\Qf}{L}\,e^{\phi+\frac{4}{3} \sigma - 8w}*G_2 -\frac{4}{L}\,e^{-\phi - \frac{20}{3}\sigma} *F_2 
\nonumber \\[2mm]
\label{eqH}
& \quad + \frac{4\, \Qf}{L^2} \frac{e^{\frac{\phi}{2}-2\sigma+2w}\, 2\pi\ell_s^2\,  \At' \sqrt{g_{xx}^3}}{\sqrt{-\left[ g_{tt}\, g_{rr}+e^{-\phi+\frac{20}{3}\sigma}(2\pi\ell_s^2\At')^2\right]}} \d x^1 \wedge \d x^2 \wedge \d x^3 \ .
\end{align}
\ese
Note that the equation for $G_1$ follows from the exterior derivative of \eqq{eqG2}.
\eq{eq.5deoms} is simply the Bianchi identity for $G_3$ --- see \eqref{eq.G3dual} --- and  it is  solved by 
\be\label{eq.F2eom}
e^{-\phi - \frac{20}{3} \sigma} * F_2 = G_3 = \Qst \, \d x^1 \wedge \d x^2 \wedge \d x^3 \,.
\ee
In terms of the vector potential this implies 
\be
A_1 = A_t(r) \, \d t 
\ee
with 
\be\label{eq.Cteom}
A_t'(r) = \Qst \, e^{\phi+\frac{20}{3}\sigma} \, \frac{\sqrt{-g_{tt}\, g_{rr}}}{g_{xx}^3} \,,
\ee
where we have used the fact that in our solution $B=0$.

To treat the remaining equations it is convenient to work with the momentum conjugate  to the massive vector $\cfunc_1$ in the gauge in which $\cfunc_0=0$. Taking the ansatz 
\be
\cfunc_1=\cfunc_t(r) \d t
\ee
 one can define the conjugate momentum as
\be\label{eq.conjugatemomentum}
\bfunc \,\, \equiv \,\, - \frac{ 2\kappa_5^2 L^4 }{4}\frac{\delta S_\mt{5}}{\delta {\cfunc_t'} } 
\,\,=\,\, - e^{\phi + \frac{4}{3}\sigma- 8 w } \frac{L^4\, \sqrt{g_{xx}^3}}{4\sqrt{-g_{tt}\, g_{rr}}} \left( \cfunc_t' + \frac{\Qf}{L}  2\pi\ls^2 \At'\right) \ ,
\ee
and similarly
\be\label{eq.conjugatemomentumprime}
\bfunc' \,\,=\,\, - \frac{ 2\kappa_5^2 L^4 }{4}\frac{\delta S_\mt{5}}{\delta {\cfunc_t} } 
\,\,=\,\, -2e^{\phi -4\sigma+ 4 w } L^2\, \frac{\sqrt{g_{xx}^3\, g_{rr}}}{\sqrt{-g_{tt}}}  \cfunc_t \ .
\ee
We have chosen the normalization in such a way that this momentum coincides exactly with the function $\bfunc(r)$ appearing in \eqq{eq.F3form}. The equation of motion \eqref{eqG2} then becomes simply
\be
\partial_r \eqref{eq.conjugatemomentum} = \eqref{eq.conjugatemomentumprime} \,.
\ee

Within our ansatz $B=0$ the equation of motion \eqq{eqH} for the NS form is satisfied by choosing appropriately $\At'$. With the results \eqref{eq.F2eom} and \eqref{eq.conjugatemomentum} this equation becomes
\be
0 = - \frac{4\Qf}{L^5} \bfunc - \frac{4\Qst}{L} + \frac{4\, \Qf}{L^2} \frac{e^{\frac{4}{3}\sigma+2w}\,  2\pi\ell_s^2 \,  \At' \sqrt{g_{xx}^3}}{\sqrt{-\left[g_{tt}\, g_{rr}+e^{-\phi+\frac{20}{3}\sigma}(2\pi\ell_s^2\At')^2\right]}} \ ,
\ee
whose solution is
\be\label{eq.AtprimesolutionBIS}
2\pi \ls^2 \, \At' = \frac{e^{\frac{\phi}{2} - \frac{10}{3} \sigma}\,   \sqrt{-g_{tt}\, g_{rr}} \left( \Qst\, L^4 + \Qf\, \bfunc \right) }{ \sqrt{ e^{\phi-4\sigma+4w} L^6\, \Qf^2 \,g_{xx}^{3} + \left( \Qst\, L^4 + \Qf\, \bfunc \right)^2} } \,.
\ee
This automatically solves the equation of motion for $\At$ that follows from the action \eqq{eq.5daction}:
\be
\left( \frac{\delta S_\mt{5}}{\delta \At'} \right)'= 0 \,,
\ee
where the electric displacement is 
\be
\frac{\delta S_5}{\delta \At'} = \frac{2\pi\ls^2}{2\kappa_5^2} \, \frac{4\,\Qst}{L} \ ,
\ee
which reproduces  the ten-dimensional  result  \eqref{eqn.electricdisplacement} upon using the relations between $\kappa, \kappa_5, L$ and $\Qc$.

\subsection{Scalings}\label{sec.scalings}

The action \eqref{eq.5daction} enjoys several scaling symmetries that will allow us to work with appropriate dimensionless variables.  To see this let us recall the length dimensions of the different dimensionful  variables in our ansatz (in this section we set the temperature to zero):
\be
[L]=\ell \ , \quad [\Qst]=\ell^{-1} \ , \quad [\bfunc]=\ell^3 \ , \quad [t,x^i,r]=\ell \ ,\quad 
[ \At ] = \ell^{-1}
\,.
\ee
In particular, note that $B_{rt}, \cfunc_t, A_t$ are dimensionless.  
In order to work with dimensionless variables we make the following replacements 
\be\label{eq.scalings}
 r \to  L \, \overline r \ , \quad \bfunc \to L^3 \, \overline{ \bfunc } \ , \quad 
\At \to \frac{L}{2\pi \ell_s^2} \, \overline \cA_t
\,,
\ee
where we recall that $L$ is  a length scale in the solution related to the D3-brane charge through \eqq{correct}. Upon this replacement the action scales homogeneously as
\be\label{eq.actionscaling}
S \to L^{-1} \, \overline S \ ,
\ee
with $\overline S$ independent of $L$. The effective Newton's constant (the normalization in front of $\overline S$)
\be
\frac{L^{-1}}{2\kappa_5^2} = \frac{\nc}{4(2\pi\ell_s)^4}
\ee
 has dimensions of (length)$^{-4}$, in accordance with \eqref{eq.scalings}, since we rescaled all coordinates except the four  Minkowski ones. We will use these redefined variables in our numerical code and, from this point onward, we will drop the over-bars in most places for notational simplicity.

The action \eqref{eq.5daction} and the length scale $L$ are also invariant under the following rescaling of the five-dimensional fields:
\be
\label{imple}
e^\phi \to \Qf^{-1} e^{\phi}\ , \quad \cfunc_t  \to \Qf^{1/2}\, \cfunc_t \ , \quad
\left\{ \bfunc, {A_t }, {B_{rt}} ,  {\At} \right\} \to \Qf^{-1/2} 
\left\{ \bfunc, {A_t} , {B_{rt} },  {\At} \right\}  \,.
\ee
Upon implementation of \eqq{eq.scalings} and \eqq{imple} the rescaled five-dimensional action becomes independent of $L$ and $\Qf$ whereas, in terms of the rescaled fields, the right-hand side of \eq{eq.F2eom}  becomes proportional to the dimensionless combination 
\be\label{eq.ratiodependence}
\ratio = \frac{L\,\Qst}{ \Qf^{1/2}} = \frac{(8\pi V_5)^{1/4}}{V_3^{1/2}} (2\pi )^4 \ell_s^3\, \frac{\nq \, \nc^{1/4}}{\nf^{1/2}} \,.
\ee
This is a crucial observation because it implies that, up to trivial rescalings, the zero-temperature solutions only depend on $\xi$. In practice this means that we can set $\Qf=1$ and $\Qc=4L^4=1$ in our numerical code, use $\Qst$ as the parameter labelling our solutions, and then replace  $\Qst$ by $\xi$ and undo the rescalings above in the final answer to obtain the full dependence of the physics on $\nf$, $\nc$ and $\nq$.  In Appendix \ref{app.scaledeoms} we provide the equations of motion for the rescaled functions  of our ansatz. See the comments below Eqs.~\eqq{eq.ratiodependenceINTRO} and \eqq{Tintro} concerning the $\ell_s$ and the $\nc$ dependence of $\xi$.

There is a final scaling symmetry that leaves the action, $L$ and $\Qf$ unchanged, and is given by
\be
g_{tt} \to \epsilon^2 g_{tt} \ , \quad g_{xx} \to \epsilon^{-2/3} g_{xx} \ , \quad \left\{ A_t , B_{rt} , \cfunc_t  , \At \right\} \to \epsilon \left\{ A_t , B_{rt} , \cfunc_t  , \At \right\} \,.
\ee
Since this involves only dynamical fields it leads, via Noether's theorem, to a radially conserved quantity
\be\label{eq.heat}
h \equiv \frac{1}{2\kappa_5^2} \left[ \sqrt{\frac{-g_{tt}\, g_{xx}^3}{g_{rr}}} \left( \frac{g_{tt}'}{g_{tt}} -\frac{ g_{xx}'}{g_{xx}} \right) - \Qst \, A_t  - 2\pi\ls^2 \frac{4\,\Qst\, \At}{L}  + 4 \frac{\cfunc_t \, \bfunc}{L^4} \right] \,.
\ee
Notice that this quantity, evaluated at a horizon with the boundary conditions that the vector fields vanish there, equals the product of the Bekenstein-Hawking entropy and the Hawking temperature, therefore corresponding to the heat energy.

\section{Numerical solutions}
To solve the equations of motion given in Appendix \ref{app.scaledeoms} we resort to numerical methods. The approach we follow is to begin with the $\Qst=0$ black brane solutions of \cite{Faedo:2016cih}. These solutions depend on one parameter, the radius of the black hole horizon $\rh$,  which is related to the temperature of the dual field theory. As we  will explain below, adding a non-zero string density has an important effect in the IR part of the geometry if the temperature is  low, but produces only a subleading correction in the UV. Therefore we can use the $\Qst=0$ solutions of \cite{Faedo:2016cih} with large  $\rh$ as seeds for solutions with non-zero but small string density. By taking small increments we can then span the $(\Qst,\,\rh)$ space  and construct in this way solutions for different values of $T$ and $\nq$.  

We begin by fixing the gauge choice for the radial coordinate. We do so by specifying the functional form of the spatial component of the ten-dimensional  Einstein-frame metric. In terms of the string-frame metric \eqref{eq.10dmetricgeneric} we require that 
\be\label{eq.gaugefixing}
G_{xx} \, e^{-\phi/2} = \frac{r^2}{L^2} \ .
\ee
In Eqs.~\eqq{param1}-\eqq{5Dmetric} we introduced one parameterization of the ten-dimensional string-frame metric that is particularly convenient for the purpose of reducing the system to five dimensions. In contrast, in order to perform the numerical integration of the equations of motion it is more convenient to use the  parameterization
\be\label{eq.10dansatzfornumerics}
\d s^2= e^{\phi/2} \left[ \sfh^{-1/2} \left( -\sfb\, \d t^2 + \d  x_3^2 \right) + \sfh^{1/2} e^{2 \sff} \left( \frac{\d r^2}{ \sfc} + 2\,L^2 \,e^{2\sfg-2\sff} \d s_\mt{KE}^2  + 2\,L^2 \,\etakah^2 \right) \right] \ ,
\ee
in terms of which the condition \eqref{eq.gaugefixing} becomes 
\be
\sfh = \frac{L^4}{r^4} \ .
\ee
This radial gauge fixing and parameterization are exactly those used in \cite{Faedo:2016cih} for the neutral case. As we will review below, as $r\to\infty$ the geometry approaches a hyperscaling-violating (HV) geometry with $\theta=7/2$ that encodes the physics of the Landau pole. Consequently, the dilaton diverges at $r\to\infty$.

Using the equation of motion for $\sfh$ and the radial Einstein equation one can solve algebraically for $\sfc$ and $\sfc'$. Plugging this back into the remaining equations of motion we are left with five second-order differential equations for the functions $\sfb$, $\sff$, $\sfg$, $\phi$ and $\bfunc$ which we do not show explicitly since they  are not particularly illuminating. The bottom line is that  we have reduced the problem to solving for five different functions subjected to ten boundary conditions. This is consistent with the fact that the boundary conditions in the UV and in the IR each depend on five parameters, as we will show now.

\subsection{Boundary conditions in the UV}
\label{sec.susysolution}
In the neutral case it was shown in Ref.~\cite{Faedo:2016cih} that the asymptotic UV solution is given by an HV metric with parameter $\theta=7/2$, supported by running scalar fields. Setting $\Qst=0$ in our five-dimensional action 
 \eqref{eq.5daction}  this asymptotic solution takes the form 
\be\label{eq.LandaupoleasHV}\bal
\d s_5^2 & = 3\cdot 6^{1/3} \, \left(\frac{R}{L}\right)^{-7/3} 
\left[ \frac{1}{3 } \left( \frac{R}{L}\right)^2 \, \eta_{\mu\nu}\d x^\mu \d x^\nu +  L^2 \frac{\d R^2}{R^2} \right]  \ , 
\\[2mm]
e^\phi & = \frac{1}{\Qf} \left( \frac{R}{L} \right)^{1/2} \ , \qquad e^\sigma = \sqrt{2}\,6^{1/10} \left( \frac{R}{L} \right)^{-7/40} \ , \qquad e^w = 6^{1/10} \left( \frac{R}{L} \right)^{-1/20} \ ,
\eal\ee
where subleading corrections are of order $1/R^{1/2}$. In this expression we have changed the radial coordinate  according to 
\be
\left( \frac{r}{L} \right)^4=\left( \frac{R}{L} \right)^{1/2} 
\ee
in order to make the HV-form of the metric explicit.

The inclusion of a non-zero charge density does not modify this asymptotic solution at leading order but produces only subleading corrections (except for the function $\bfunc$, see below). Returning to the parameterization of \eq{eq.10dansatzfornumerics} the corrected solution takes the form
\be\label{eq.UVnumerics}
\bal
e^\sff & = \sqrt{6}\, \frac{L^2}{r^{2}} \left[ 1 + \frac{\kappa_\sff}{r^4} +  \frac{\kappa_{\sff2}}{r^8} + {\cal O}(r^{-12}) \right] \ , \\[2mm]
e^\sfg & = 1 - \frac{\kappa_\phi}{4r^4} + {\cal O}(r^{-8}) \ , \\[2mm]
e^\phi & = \frac{r^4}{L^4}  \frac{1}{\Qf} \left[ 1 + \frac{\kappa_\phi}{r^4} + {\cal O}(r^{-8}) \right] \ , \\[2mm]
\sfb & =  1 + \frac{\kappa_{\sfb}}{r^4} + {\cal O}(r^{-8})  \ , \\[2mm]
\bfunc & = -\frac{L^4 \Qst}{\Qf}  \left[ 1 + \frac{\kappa_{\bfunc}}{r^4} + {\cal O}(r^{-8}) \right] \,.
\eal
\ee

\noindent
In our numerical code we have constructed the expansion in $r^{-4}$ up to order $r^{-54}$. All the non-leading coefficients in this expansion are determined in terms of the five unspecified constants $\kappa_\sfb$, $\kappa_\sff$, $\kappa_{\sff2}$, $\kappa_\phi$, and $\kappa_\bfunc$ that will depend on the temperature and the charge density.

We note that the only difference at leading order between the neutral and the charged configurations  is in the asymptotic value of the function $\bfunc$: In the neutral case this tends to zero, whereas in the charged one it approaches a non-zero constant. Looking at \eqref{eq.Atprimesolution} we see that the specific value of this constant  ensures the physical requirement that  $\At'$ vanishes at, and therefore that there is no electric flux through, the Landau pole.  This should  be contrasted with the situation in the lower-dimensional setup of Ref.~\cite{Faedo:2015urf}. In that configuration  there is no Landau pole (since the field theory is asymptotically free) and both $\bfunc$ and $\At'$ vanish asymptotically.

\subsection{Boundary conditions at the horizon}
\label{bchor}

The presence of a finite temperature in the field theory is captured by the existence of a black brane in the gravitational solution. The boundary conditions at the horizon are the usual requirement that the blackening factor in \eqref{eq.10dmetricgeneric} has a simple zero at $r=\rh$ and the rest of the fields attain a finite value:
\be\label{eq.IRparameters}\bal
e^\sff & = e^{\sff_\mt{h}}+ {\cal O}(r-\rh) \ , \quad e^\sfg = e^{\sfg_\mt{h}} + {\cal O}(r-\rh)  \ , \quad \sfb = \sfb_\mt{h} (r-\rh) + {\cal O}(r-\rh)^2 \ , 
\\[2mm]
e^\phi & = e^{\phi_\mt{h}} + {\cal O}(r-\rh) \ , \quad 
\bfunc = \bfunc_\mt{h} + {\cal O}(r-\rh) \,.
\eal\ee

\noindent
In our numerical code we have gone to seventh non-trivial order in the distance to the horizon in these expansions. All the non-leading coefficients are determined in terms of the five  unspecified parameters $\sff_\mt{h}, \sfg_\mt{h},  \sfb_\mt{h}, \phi_\mt{h}$ and $\bfunc_\mt{h}$. 

\begin{figure}[t]
\begin{center}
\includegraphics[width=.7\textwidth]{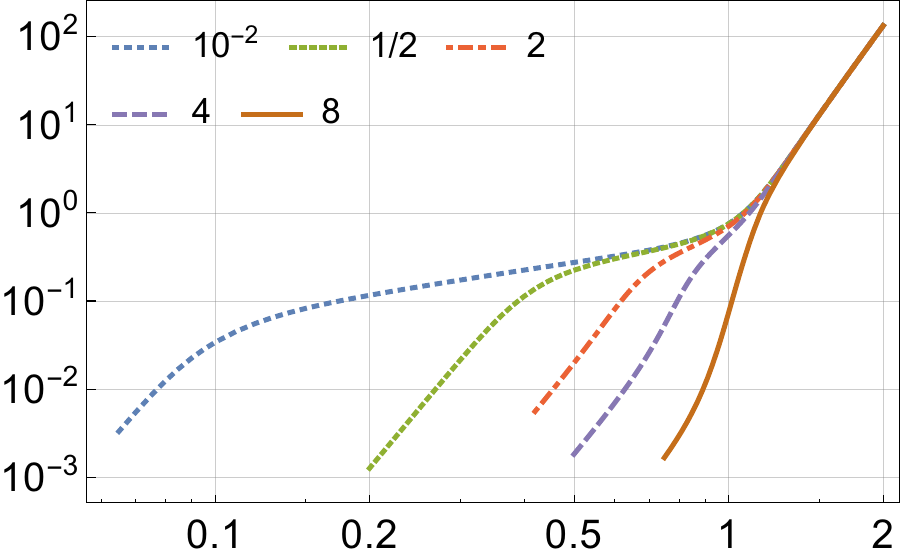} 
\put(-340,100){\Large $\overline T$}
\put(-150,-15){{\Large $\overline \rh$}$ / \sqrt{2}$}
\caption{\small \small Relation between temperature and horizon position at fixed values of the charge density $\sqrt{2}\, \xi$. }\label{Tpdf}
\end{center}
\end{figure}

In the presence of a non-zero charge density the temperature of the solution depends not only on $\rh$ but also on $\Qst$. Nevertheless, 
$\rh$ can still be thought  of as a proxy for $T$ since the latter is still a monotonically increasing function of $\rh$ for fixed charge, as illustrated by \fig{Tpdf}. From this point onwards we will typically work with the dimensionless temperature
\be
\label{TTrr}
\overline T = \Qc^{1/4}\, T  \,,
\ee
as well as with the dimensionless radial coordinate introduced in \eqq{eq.scalings}:
\be
\overline r = \frac{r}{L}= \sqrt{2} \, \frac{r}{\Qc^{1/4}} \,.
\ee

\subsection{Zero-temperature limit}\label{sec.IRisLifshitz}
Our numerical analysis  below will show that the IR geometry of the zero-temperature solutions  can be understood on general grounds. Indeed, we will see that it is given by a Lifshitz spacetime with dynamical exponent $z=7$ and running dilaton and $\bfunc$ fields. In other words, despite the fact that in order to construct the solutions with non-zero temperature we only impose regularity at the horizon,  this boundary condition automatically approaches a Lifshitz boundary condition as $T\to 0$.  The physical reason for this is that, in the deep IR, the tension of the D7-branes is a subleading effect with respect to the backreaction of the electric field on the D7-branes. The IR-limit of the zero-temperature solution is therefore the same as in the D3-brane\,+\,strings system without D7-branes of \cite{Kumar:2012ui,Faedo:2014ana}, and it takes the form 
\be\label{eq.IRlifshitz}\bal
\d s^2 & =   e^{\phi/2} \left[ - c_\mt{t}\, \frac{r^{14}}{L^{14}}\, b\, \d t^2 +  \frac{r^{2}}{L^{2}} \d x^2 + \frac{10 \cdot 136^{1/4}}{11^{1/2}} \frac{L^2}{r^2} \left( b^{-1} \d r^2 + \frac{r^2}{10} \d \Omega_5^2 \right) \right] \ , \\[2mm]
e^{\phi} & = \frac{72\,\sqrt{22}}{(34)^{5/4}\, \Qst^2\, L^2} \frac{r^6}{L^6} \ ,
\eal\ee
together with a rapidly vanishing $\bfunc\propto r^{10}$.
 With respect to the radial variable $u$ in \cite{Faedo:2014ana} we have performed the rescaling
\be
u = \frac{2^{5/8} \, 17^{5/24}}{11^{5/12}}\,  r \ .
\ee
At zero temperature the  blackening function is simply $b(r)=1$. At low temperatures it is approximately given by
\be\label{eq.blacklifshitz}
b = 1- \left(\frac{\rh}{r} \right)^{10} \ .
\ee
The dynamical exponent $z=7$ can be seen  by inspection of the Einstein-frame metric given within the square brackets of \eqref{eq.IRlifshitz}.\footnote{Since  dimensional  reduction on the compact manifold does not change the dynamical exponent \cite{Perlmutter:2012he}.} 
We have left undetermined the normalization of the time coordinate, encoded in the constant $c_\mt{t}$. This factor contains UV information via the RG flow connecting the UV-normalized time coordinate to the IR one of \eqref{eq.IRlifshitz}. 

The ten-dimensional  solution \eqref{eq.IRlifshitz} can be seen to arise as the uplift of an exact solution of our setup \eqref{eq.5daction} in the following limit.\footnote{This limit is performed explicitly in the probe approximation in Appendix \ref{app.probe}.} First one should perform a Legendre transform in the action to trade the radial derivatives of $\At$ in favor of the charge density parameter $\Qst$, with the help of the relation \eqref{eq.Atprimesolution}. Then one must take $\Qf\to0$ keeping $\Qst$ finite. The result of this procedure is that the DBI action of the D7-branes becomes a Nambu-Goto action for a set of fundamental strings \cite{Kobayashi:2006sb}. In other words, in this limit the tension of the D7-branes is subleading with respect to the effect of the charge density --- see Eqn.~\eqref{eq.NGapproximation}.  Once this limit is taken it is a simple exercise to check that the equations of motion for $\bfunc$ and for the metric are compatible with setting $\bfunc=0$ and $G_f = G_b$, meaning that the full symmetry of the five-dimensional compact manifold is recovered in the deep IR, i.e.~the squashing vanishes in this limit. An exact scaling solution of the resulting equations  was first found in \cite{Azeyanagi:2009pr} and studied in detail in  \cite{Kumar:2012ui,Faedo:2014ana}. 

The argument in the previous paragraph just shows  that \eqref{eq.IRlifshitz} is an asymptotic IR solution in the flavorless limit with an external charge density. However, in Appendix \ref{app.Lifshitzmodes} we provide an analysis of the deformations away from this limit  once the flavor effects are included. It is shown that  the irrelevant deformations of the solution \eqref{eq.IRlifshitz} (those that vanish in the IR) are still determined by five independent  constants of integration, in agreement with the analysis in \Sec{bchor}. In particular, the function $\bfunc$ vanishes for $r\to 0$ as
\be\label{eq.IRbfunc}
\bfunc \sim  - \frac{2^5\, 3^3}{34^{3/2}} \frac{\Qf}{\Qst^3} \frac{r^{12}}{L^{12}} + \beta_1 \, r^{10}  \,,
\ee
with $\beta_1$ one of the integration constants.

\subsection{Numerical integration}\label{sec.numerics}
We have seen above that the UV and IR  limits of the solutions that we are seeking depend on five integration constants in both instances. 
Our goal in this section is to show that there is a choice of these ten parameters that allows us to join smoothly the UV and the IR limits. As explained below \eqq{eq.ratiodependence}  we can set 
$\Qc=\Qf=1$  without loss of generality. In particular this makes the radial coordinate $r$ dimensionless. In most places in this section we will not indicate this explicitly, but in the plots we will do so by writing $\overline r$ instead of $r$.

Since we set $\Qc=\Qf=1$ our equations  depend only on one external parameter $\Qst$, which at the end can be replaced by $\ratio$ through  \eqq{eq.ratiodependence}. The  horizon radius $\rh$ enters  the equations of motion via the IR boundary conditions and the range of integration. Therefore the solutions depend on $\rh$ and $\Qst$, or equivalently on $T$ and $\nq$. 

 Our integration strategy is simple. We use the UV and the near-horizon expansions \eqq{eq.UVnumerics} and \eqq{eq.IRparameters} to shoot from $r=60$ and $r-\rh=10^{-8}$, respectively, at fixed values of $\rh$ and $\Qst$. Imposing continuity of the functions and their derivatives at an intermediate radius gives us ten conditions that allow us to completely fix the five UV and the five IR free parameters for every value of $\rh$ and $\Qst$, by an iterative Newton-Ralphson method. We declare continuity when the difference between the UV- and the IR-integrated functions is no larger than $10^{-50}$. 
As an initial seed we use the neutral solution constructed in Ref.~\cite{Faedo:2016cih}. 
Using these new charged solutions  we then fix the charge density and start to decrease $\rh$ (equivalently,  the temperature) until the Lifshitz scaling  of \eq{eq.IRlifshitz} is observed.  The smallest non-zero value of the charge for which  we construct a solution is $\Qst=1/100$. 

One outcome of the numerical integration is the value of the IR and UV parameters as a function of $\Qst$ and $\rh$. These parameters encode much of the physics of the solutions. For example,  we will see in \Sec{thermoSec} that the thermodynamic properties can be extracted from the UV parameters. Here we will see that they  help elucidate the geometric properties of the solutions. 

\begin{figure}[h]
\begin{center}
\begin{subfigure}{.49\textwidth}
\includegraphics[width=\textwidth]{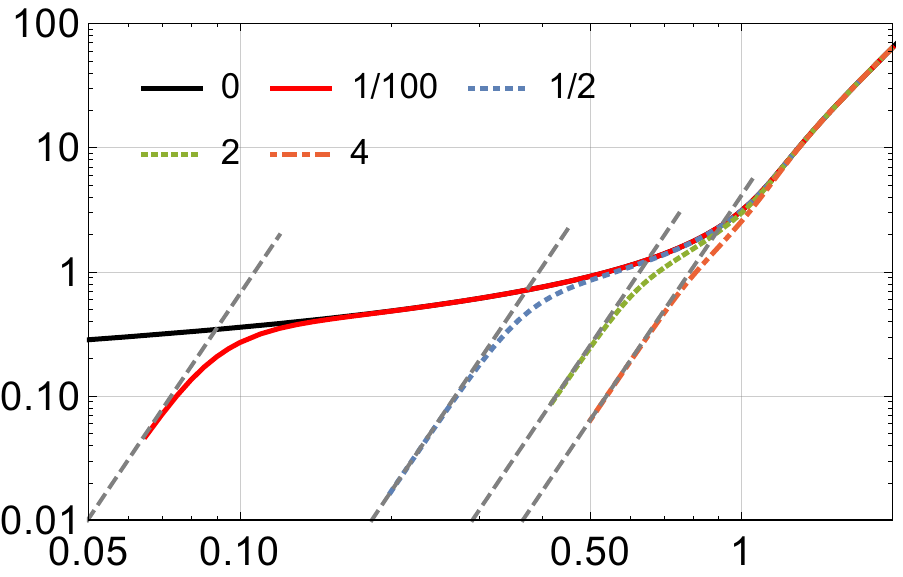} 
\put(-210,145){\large $\Qf \,e^{\phi_\mt{h}}$}
\put(-110,-15){{\Large $\orh$}$ / \sqrt{2}$}
\end{subfigure} 
\begin{subfigure}{.49\textwidth}\vspace{-3pt}
\includegraphics[width=\textwidth]{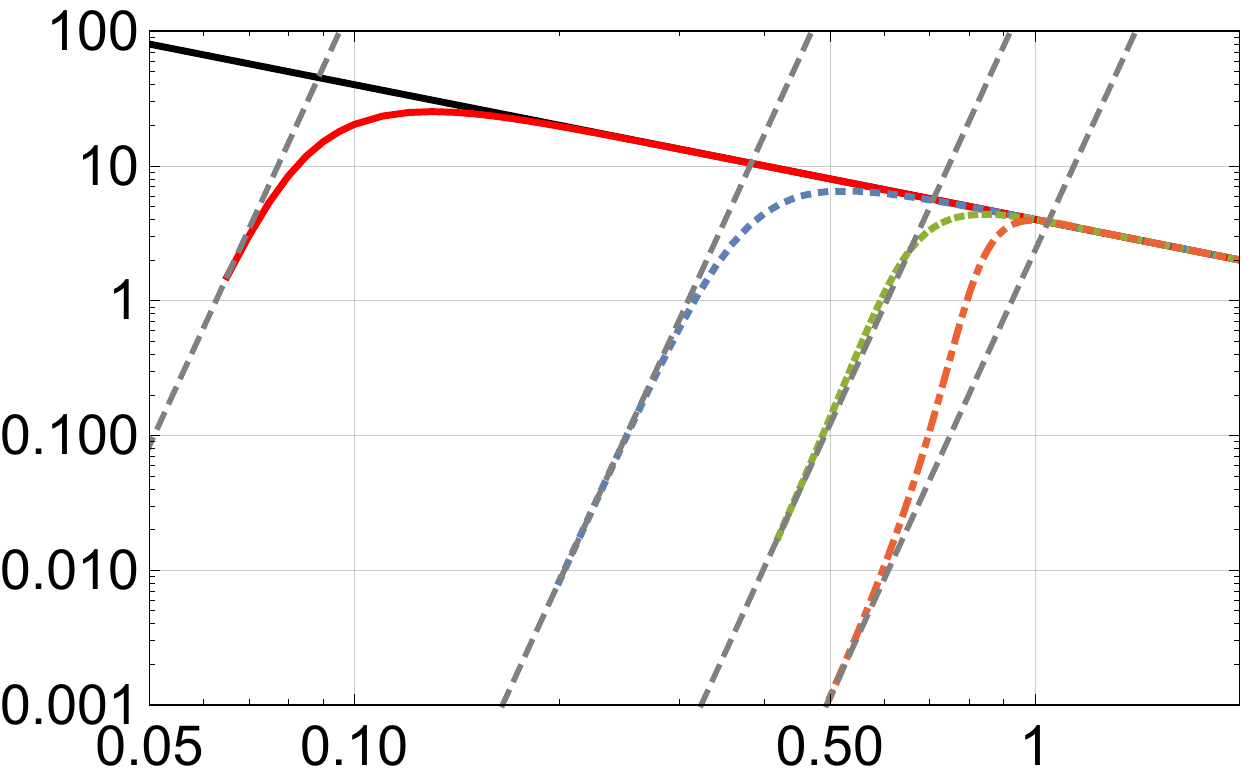} 
\put(-210,145){\large $\Qc^{1/4} \sfb_\mt{h}$}
\put(-130,-15){{\Large $\orh$}$ / \sqrt{2}$}
\end{subfigure}
\caption{\small \small IR coefficients $\phi_\mt{h}$ and $\sfb_\mt{h}$ for different values of the charge density $\sqrt{2}\, \xi$ (given in the legend of the plot on the left-hand side). The dotted gray lines show the behavior determined by the IR Lifshitz solution.}\label{fig.chargedIR1}
\end{center}
\end{figure}

To see the emergence of the Lifshitz solution \eqref{eq.IRlifshitz} for small values of $\rh$ we examine  the behavior of the functions with respect to the horizon radius  at fixed charge. First of all, comparing the metrics \eqq{eq.10dansatzfornumerics} and \eqref{eq.IRlifshitz} and using the form of the blackening factor $b$ in \eqq{eq.blacklifshitz} we expect that, for sufficiently small $\rh$, the constant $\sfb_\mt{h}$ in \eqq{eq.IRparameters} will scale as (recall that  $\Qc=\Qf=1$)
\be\label{eq.sfbhfromLif}
\sfb_\mt{h} \simeq \sfb_\mt{h}^0 \, \rh^{11}  \qquad (\text{small horizon and finite charge}) \,,
\ee 
which is a large deviation with respect to the $\sfb_\mt{h}=4 /\rh$ behavior of the neutral configuration. The \mbox{$\Qst$-dependent} proportionality constant $\sfb_\mt{h}^0 $ in \eqq{eq.sfbhfromLif}  cannot be obtained analytically since it depends on the relative normalization between  the IR and the UV time coordinates, and hence on the entire  numerical solution. For the dilaton we see from \eqref{eq.IRlifshitz}  that its value at the horizon should scale with the horizon radius as  
\be\label{eq.ephihfromLif}
e^{\phi_\mt{h}} \simeq  \frac{72\,\sqrt{22}}{(34)^{5/4}} \frac{16}{\Qst^2} \, \rh ^6 \qquad (\text{small horizon and finite charge})
\ee
where we have substituted $\Qc=4L^4=1$. The numerical results for $\phi_\mt{h}$ and $\sfb_\mt{h}$ for the neutral solution and for solutions with several different values of the charge density are shown in Fig.~\ref{fig.chargedIR1}. 
\begin{figure}[h]
\begin{center}
\includegraphics[width=.5\textwidth]{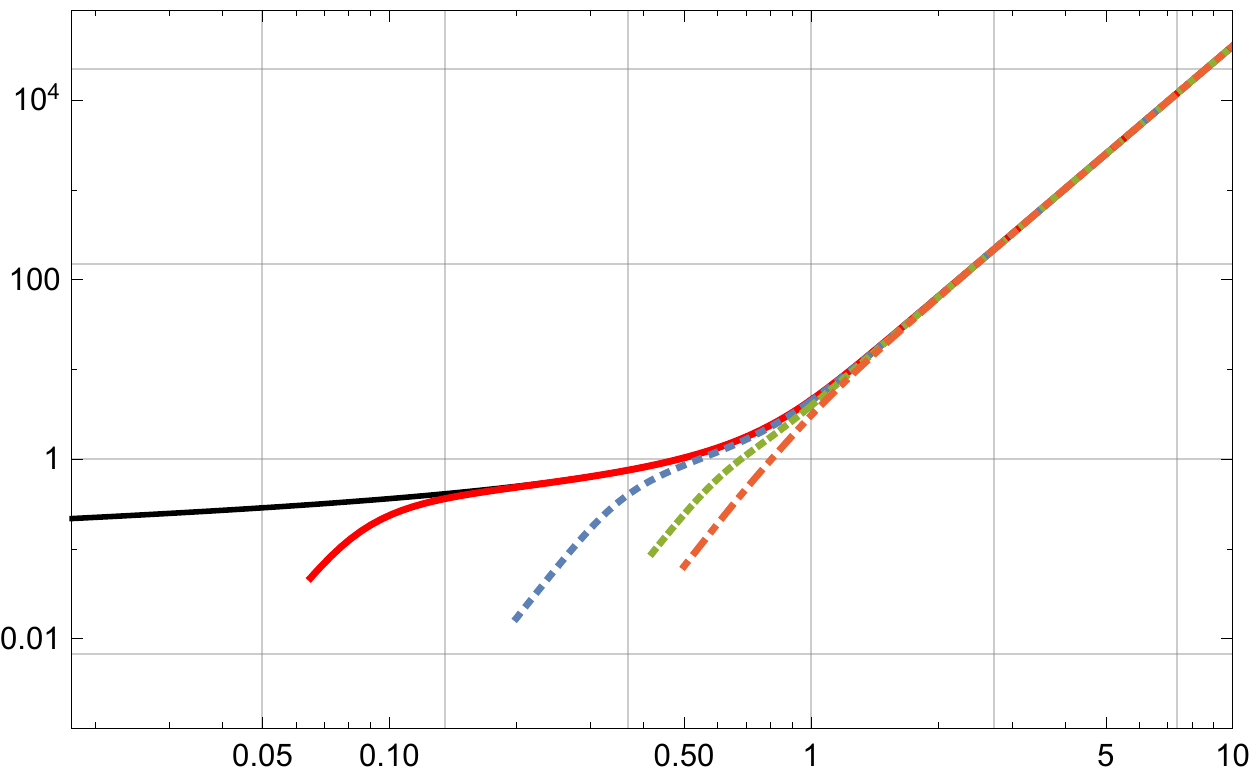} 
\put(-250,80){\large $\Qf \, e^\phi$}
\put(-115,-15){{\Large $\overline r$}$ / \sqrt{2}$}
\caption{\small \small Radial profile of the dilaton for solutions with the smallest value of $\rh$ (where the curves end) for different values of $\sqrt{2} \, \xi$ (see legend in  \fig{fig.chargedIR1}).}
\label{fig.dilatons}
\end{center}
\end{figure}

 We see that for large values of $\rh$ all the curves converge to a single one, reflecting the fact that the leading UV behavior is controlled by the neutral solution. In contrast, for sufficiently small values of $\rh$ the charged curves deviate from the neutral one and approach straight lines with slopes that agree precisely with those predicted by Eqs.~\eqref{eq.sfbhfromLif} and \eqref{eq.ephihfromLif}. For the dilaton the normalization also agrees with that in  \eqref{eq.ephihfromLif}, whereas for the $\sfb_\mt{h}$ parameter we have performed a fit to \eqref{eq.sfbhfromLif}. As expected, the value of $\rh$ at which the IR behavior sets in decreases as the charge density  decreases.  In particular, this means that solutions with small 
$\xi$ develop an intermediate region controlled by the log-AdS geometry of the neutral solution, clearly visible in the left plot of Fig.~\ref{fig.chargedIR1}, whereas solutions with large $\xi$ transition directly from the UV controlled by the neutral solution to the IR controlled by the Lifshitz solution.

At a qualitative level, one would expect the value of the dilaton at the horizon for a solution with non-zero temperature to be approximately the same as the value of the dilaton at the position  $r=\rh$ for a zero-temperature solution. In other words, the scalar fields  in a solution with non-zero temperature should be similar to those in a solution at zero temperature cut off at the corresponding value of the radial coordinate. This is confirmed by the similarity between \fig{fig.chargedIR1}(left) and  \fig{fig.dilatons}, where we plot the radial dependence of the dilaton for the coldest solution that we have constructed for each of the values  $\Qst$. 

\begin{figure}[h]
\begin{center}
\includegraphics[width=.5\textwidth]{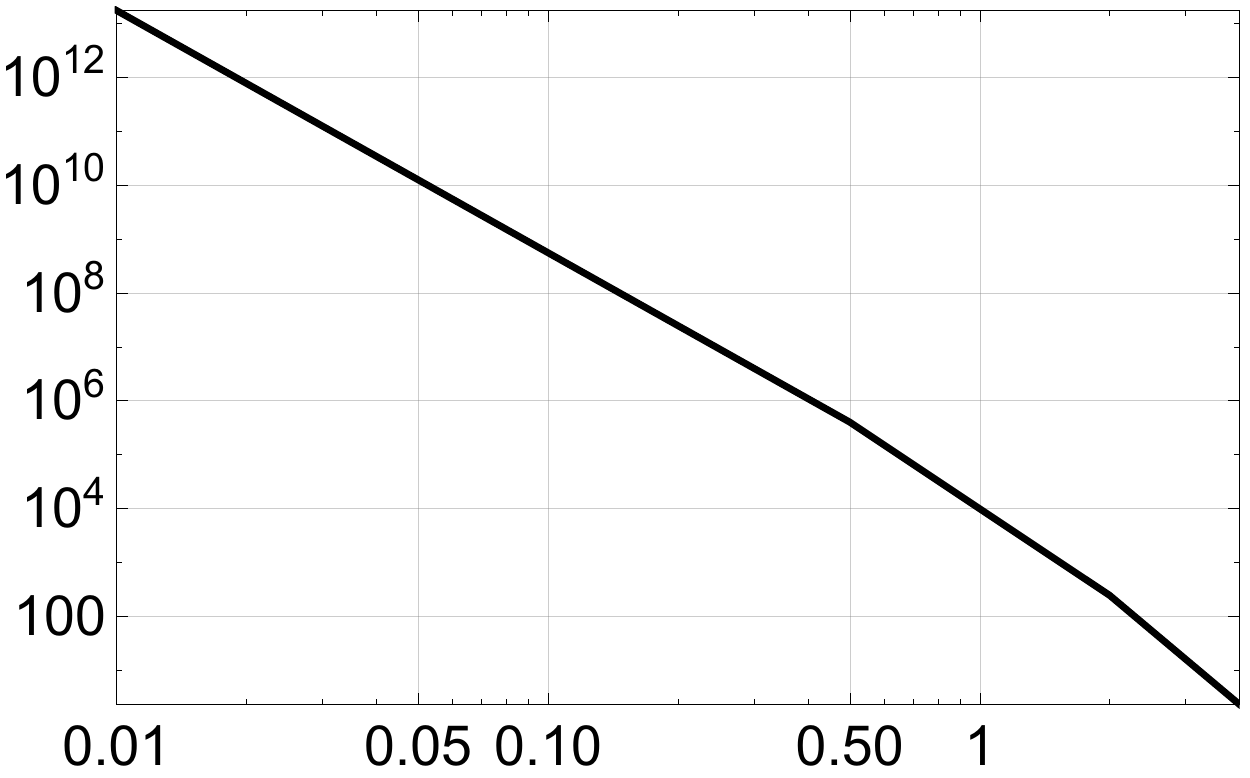} 
\put(-215,150){$\Qc^3\, \sfb_\mt{h}^0$}
\put(-115,-15){$\sqrt{2} \, \xi $}
\caption{\small \small Fit of the function $\sfb_\mt{h}^0$ defined in \eqref{eq.sfbhfromLif}.}\label{fig.b0fit}
\end{center}
\end{figure}

The fit of the quantity $\sfb_\mt{h}^0$ as a function of $\Qst$ is presented in Fig.~\ref{fig.b0fit}. From this figure we observe that $\sfb_\mt{h}^0$ diverges as $\Qst\to0$ approximately as $\Qst^{-9/2}$. 
%
\begin{figure}[h]
\begin{center}
\begin{subfigure}{.49\textwidth}
\includegraphics[width=\textwidth]{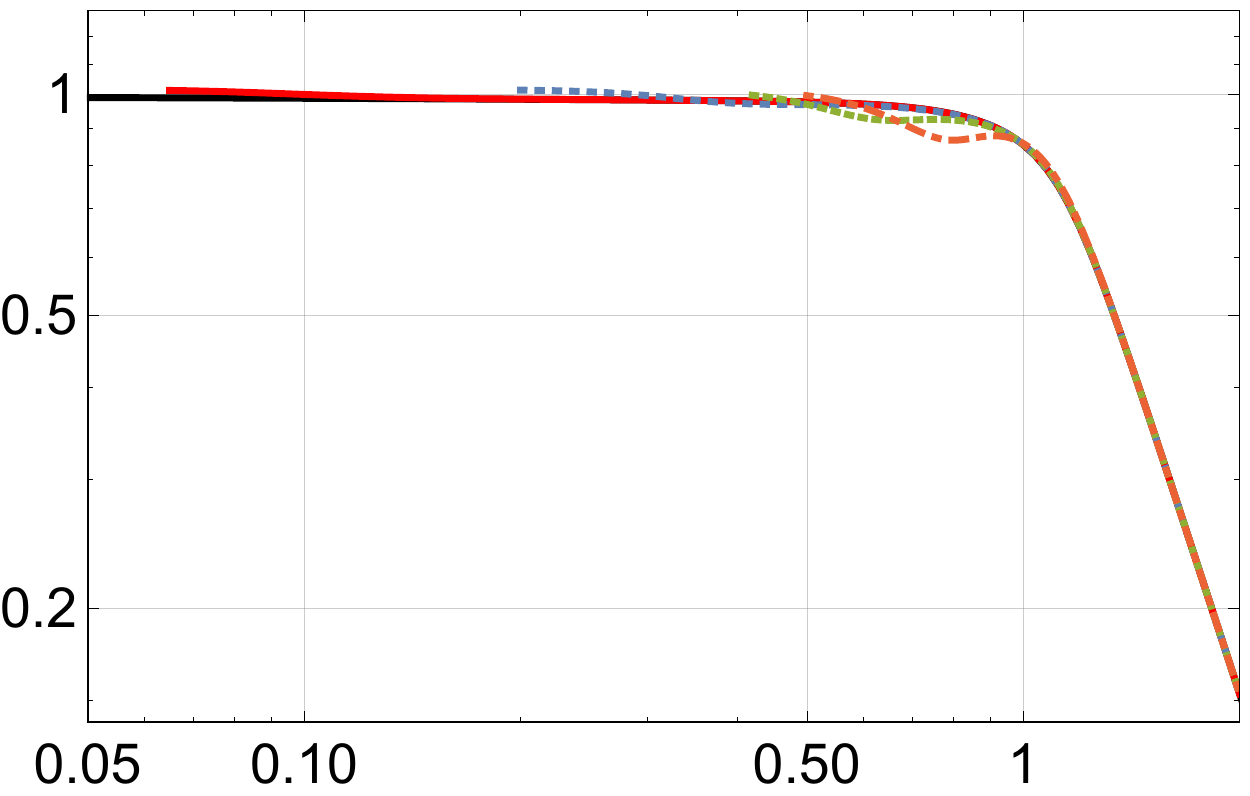} 
\put(-220,145){$\Qc^{1/4}\, e^{\sff_\mt{h}} / \rh$}
\put(-120,-15){{\Large $\orh$}$ / \sqrt{2}$}
\end{subfigure}
\begin{subfigure}{.49\textwidth}
\includegraphics[width=\textwidth]{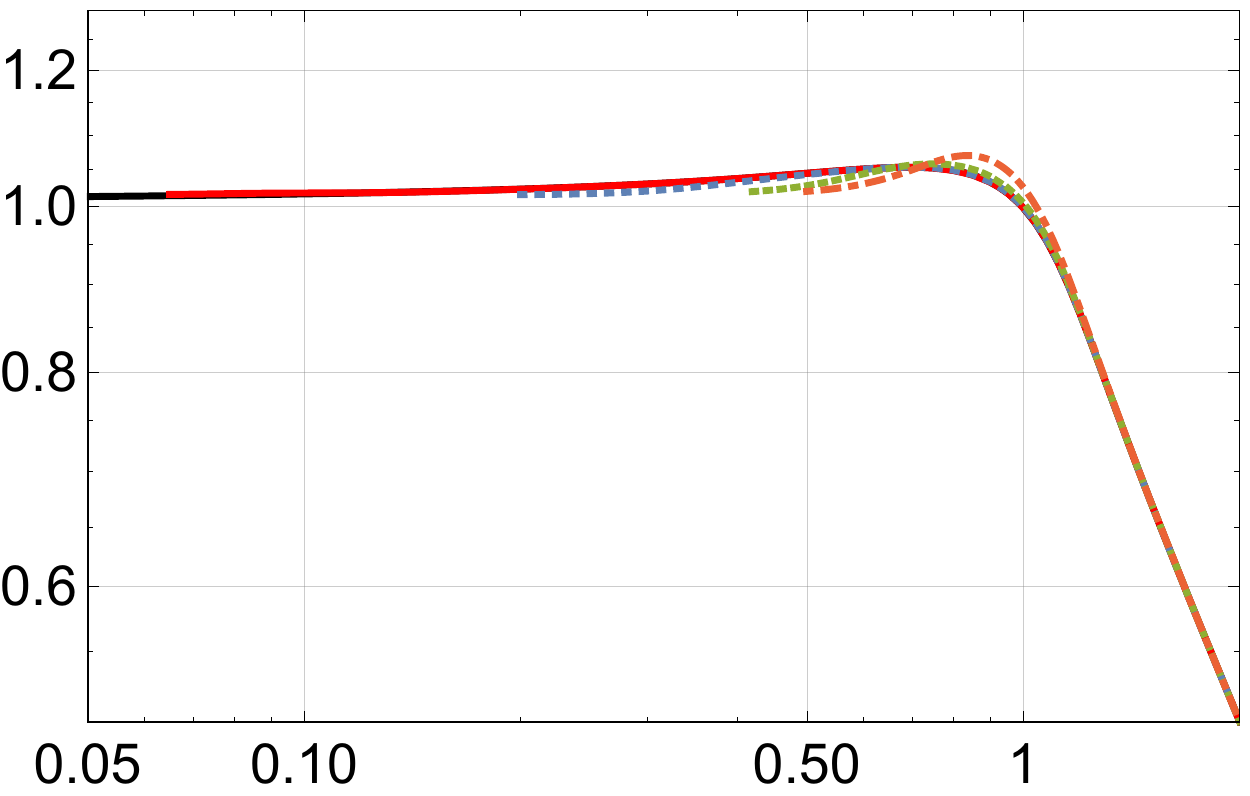} 
\put(-220,145){$\Qc^{1/4}\, e^{\sfg_\mt{h}} / \rh$}
\put(-120,-15){{\Large $\orh$}$ / \sqrt{2}$}
\end{subfigure}
\caption{\small \small IR coefficients $\sff_\mt{h}$, $\sfg_\mt{h}$  for different values of $\sqrt{2} \, \xi$ (see legend in \fig{fig.chargedIR1}).}\label{fig.chargedIR2}
\end{center}
\end{figure}
The two IR parameters of the metric, namely $\sff_\mt{h}$ and $\sfg_\mt{h}$, are shown in Fig.~\ref{fig.chargedIR2}, where we plot the ratio with respect to the horizon radius, such that the curves asymptote a constant at small values of $\rh$. 

 Both in the log-AdS region and in the Lifshitz region the squashing of the S$^5$ disappears and $\sff_\mt{h}$ and $\sfg_\mt{h}$ become equal to one another. For the neutral solution the two parameters approach $1$ logarithmically, whereas in the charged case they approach $136^{1/4}/11^{1/2}\approx 1.029$. Moreover, at large $\rh$ both parameters must approach their value in the neutral solution, where squashing is present. The value of $\rh$ at which the curves transition from the squashed behavior at large-$\rh$ to the non-squashed behavior at small $\rh$ depends on $\Qst$.

\begin{figure}[h!!!]
\begin{center}
\begin{subfigure}{.49\textwidth}
\includegraphics[width=\textwidth]{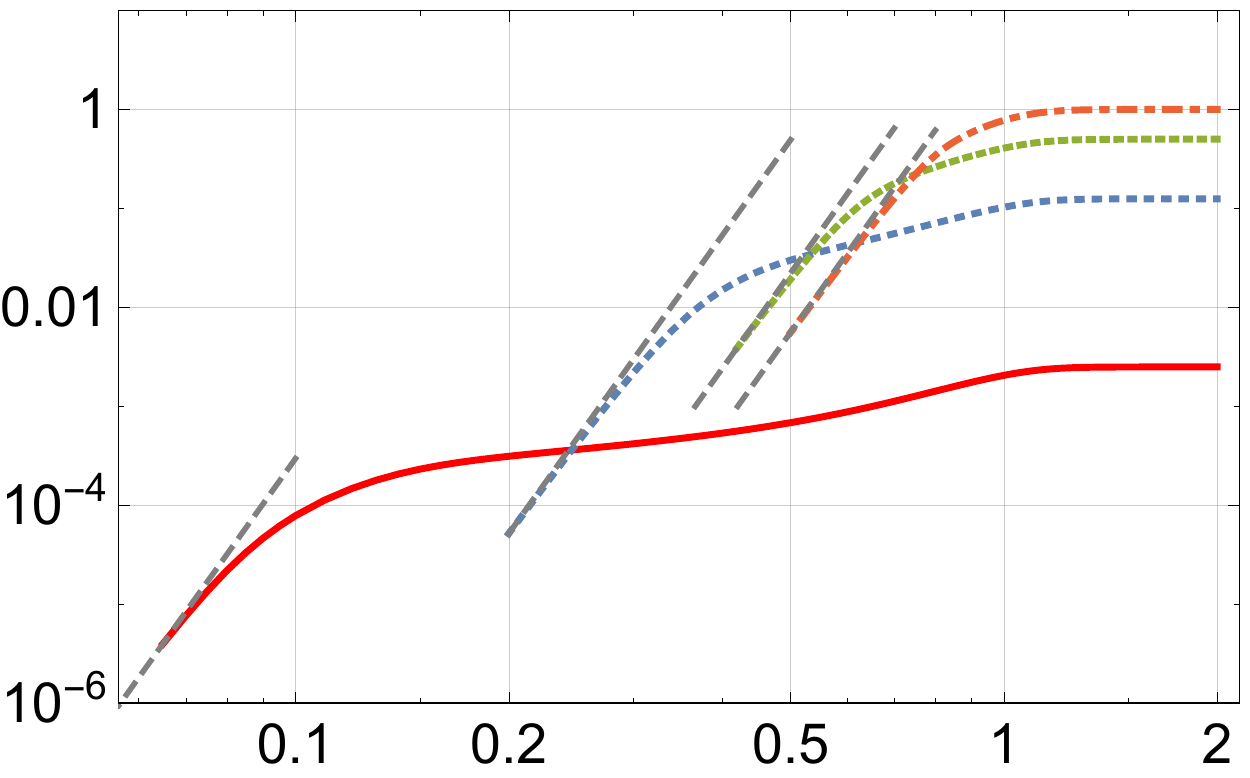} 
\put(-230,145){$-(\Qf^{1/2}/\Qc^{3/4} )\, \bfunc_\mt{h}$}
\put(-110,-15){{\Large $\orh$}$ / \sqrt{2}$}
\end{subfigure}
\begin{subfigure}{.485\textwidth}
\includegraphics[width=\textwidth]{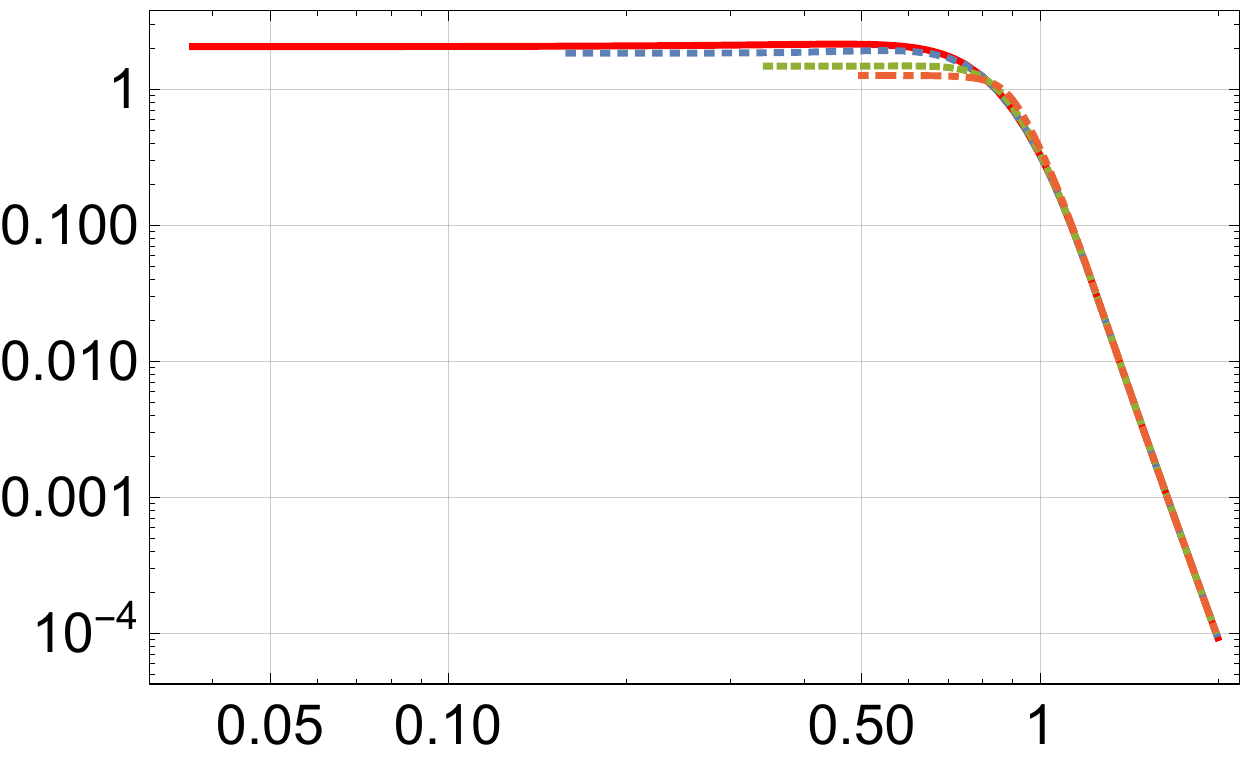} 
\put(-210,145){$-\kappa_{\bfunc}/\Qc$}
\put(-110,-15){{\Large $\orh$}$ / \sqrt{2}$}
\end{subfigure}
\caption{\small \small IR and UV coefficients associated to the function 
$\bfunc$  for different values of $\sqrt{2} \, \xi $ (see legend in \fig{fig.chargedIR1}). The dotted gray lines show the behavior determined by the IR Lifshitz solution.}\label{fig.chargedBfunc}
\end{center}
\end{figure}
\begin{figure}[h]
\begin{center}
\begin{subfigure}{.482\textwidth}
\includegraphics[width=\textwidth]{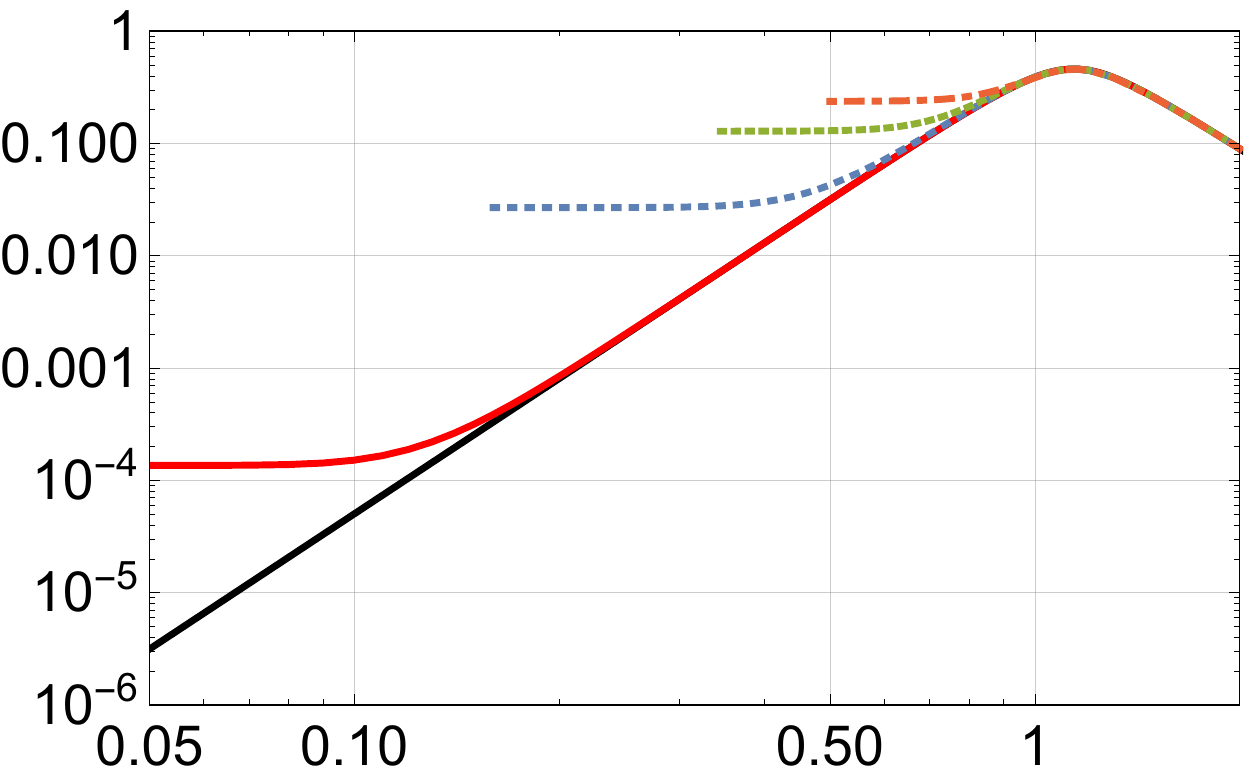} 
\put(-210,145){$-\kappa_\phi/\Qc$}
\put(-110,-15){{\Large $\orh$}$ / \sqrt{2}$}
\end{subfigure}
\begin{subfigure}{.49\textwidth}
\includegraphics[width=\textwidth]{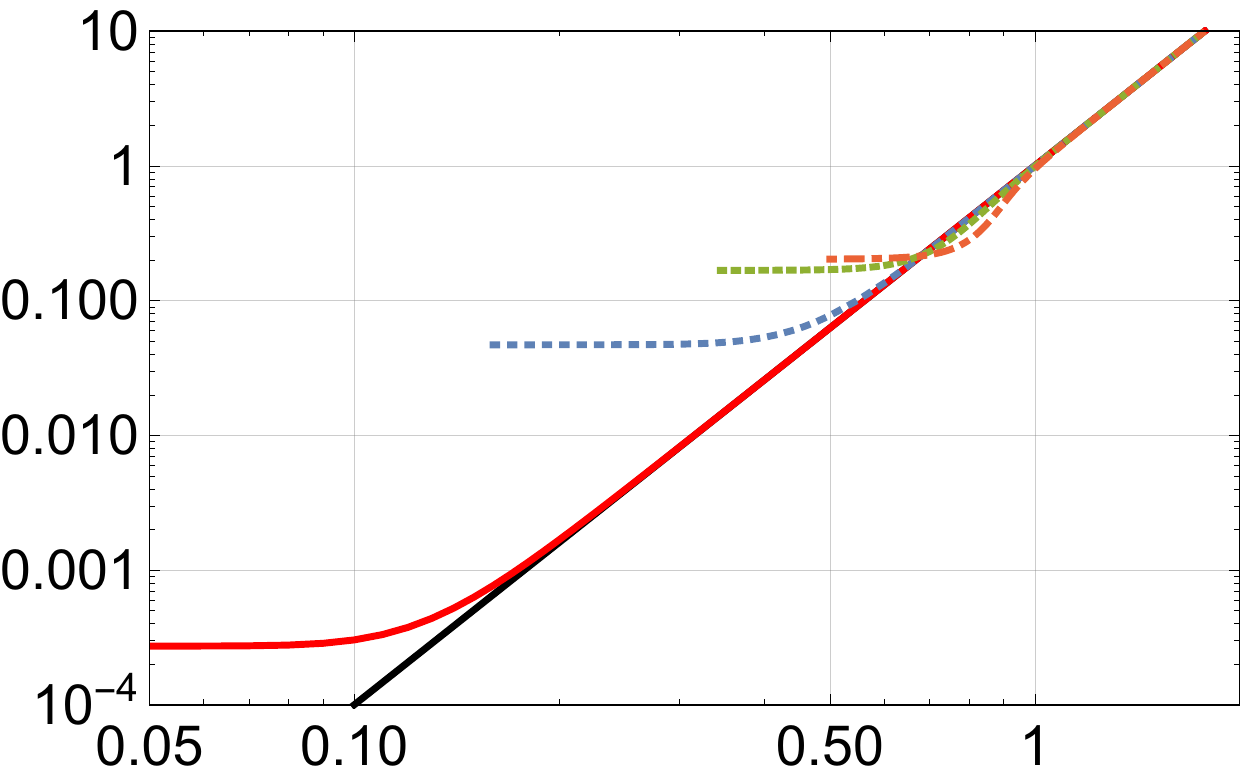} 
\put(-220,145){$-\kappa_\sfb/\Qc$}
\put(-110,-15){{\Large $\orh$}$ / \sqrt{2}$}
\end{subfigure}
\caption{\small \small UV coefficients  $\kappa_\phi$ and $\kappa_\sfb$  for different values of $\sqrt{2} \, \xi$  (see legend in \fig{fig.chargedIR1}).}\label{fig.chargedUV1}
\end{center}
\end{figure}
Finally, \fig{fig.chargedBfunc}(left) shows the horizon value  $\bfunc_\mt{h}$. We see that at small $\rh$ the slope is that determined by the leading term in \eqq{eq.IRbfunc}, i.e.~$\bfunc_\mt{h} \sim \rh^{10}$, whereas at large $\rh$ the horizon value of $\bfunc$ approaches the negative constant corresponding to the leading term in \eqq{eq.UVnumerics}. 

To summarise so far, we have seen that the large- and small-$\rh$ values of the IR parameters of our numerical solutions match precisely those predicted by the  LP  and  Lifshitz geometries, respectively. This strongly supports the fact that our configuration interpolates between the UV LP geometry of the neutral solution and the IR  Lifshitz one.  Moreover, the behavior of the dilaton at intermediate values of $\rh$ for small values of $\Qst$ supports  that in this situation the transition between the LP and the Lifshitz geometries  proceeds through an intermediate log-AdS region.

\begin{figure}[h!!!]
\begin{center}
\begin{subfigure}{.49\textwidth}
\includegraphics[width=\textwidth]{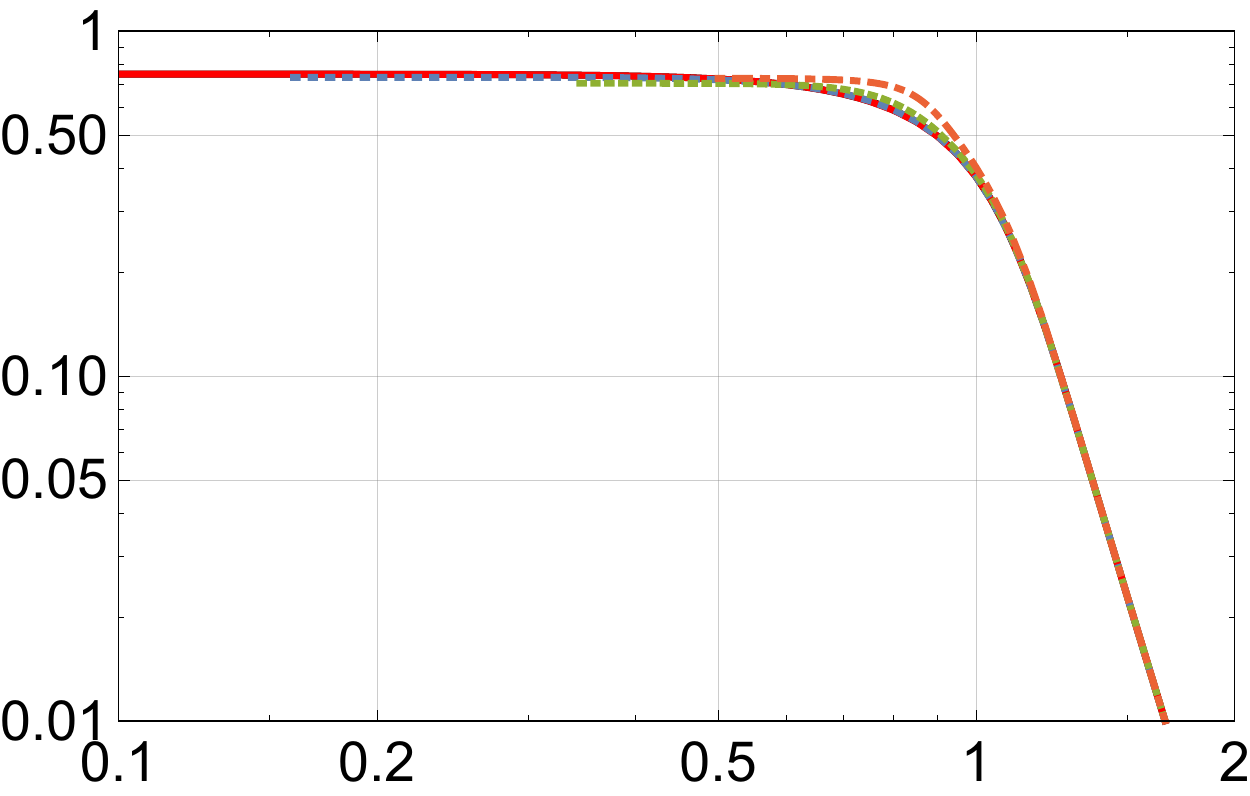} 
\put(-210,145){$-\kappa_\sff/\Qc$}
\put(-110,-15){{\Large $\orh$}$ / \sqrt{2}$}
\end{subfigure}
\begin{subfigure}{.4875\textwidth}\vspace{-1pt}
\includegraphics[width=\textwidth]{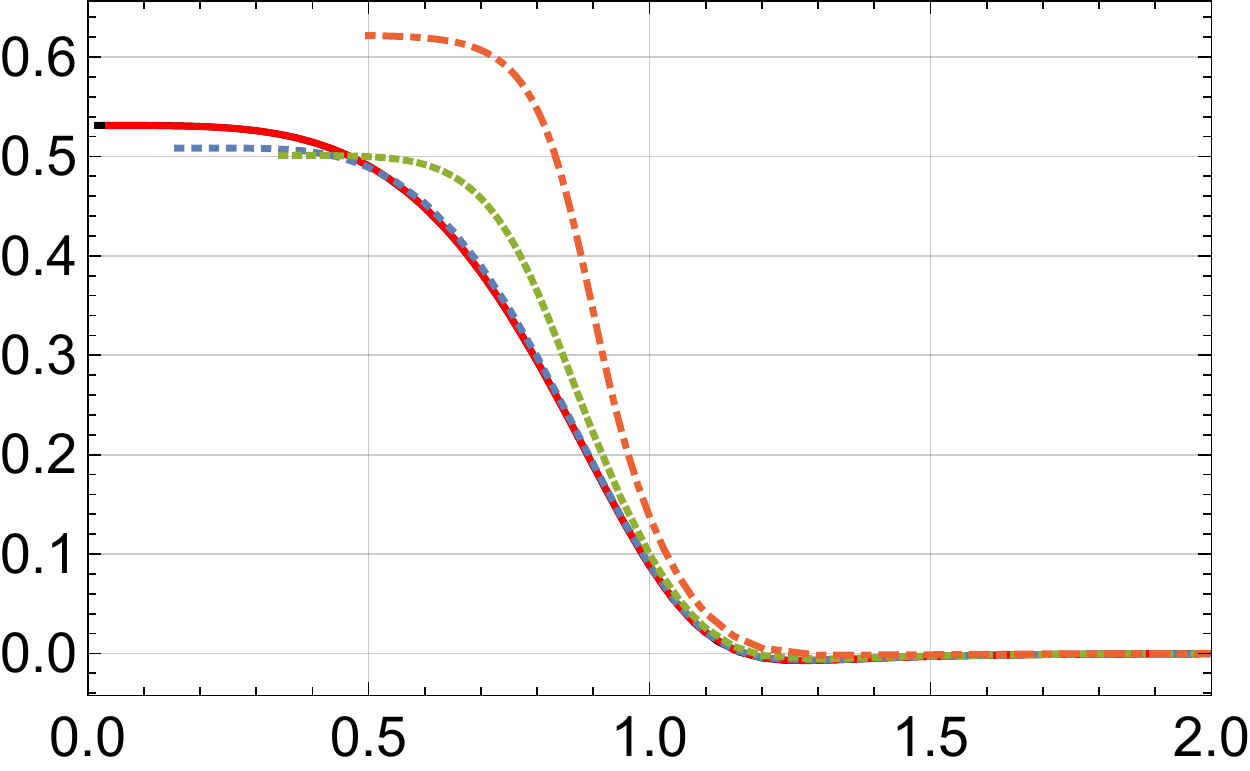} 
\put(-30,145){$\kappa_{\sff2}/\Qc^2$}
\put(-130,-15){{\Large $\orh$}$ / \sqrt{2}$}
\end{subfigure}
\caption{\small \small UV coefficients  $\kappa_\sff$ and $\kappa_{\sff_2}$  for different values of $\sqrt{2} \, \xi$  (see legend in \fig{fig.chargedIR1}).}\label{fig.chargedUV2}
\end{center}
\end{figure}

We now turn to the UV parameters in our numerical solutions. We expect their large-$\rh$ values to match those of the neutral one. This is confirmed by Fig.~\ref{fig.chargedBfunc}(right), Fig.~\ref{fig.chargedUV1} and Fig.~\ref{fig.chargedUV2}. 
In particular, $\kappa_\bfunc$ and $\kappa_{\sff2}$ decrease as $\rh^{-12}$ and $\rh^{-8}$, respectively.
In contrast, the small-$\rh$ values of the UV parameters are not predicted by any general argument. The reason is that, even if the Lifshitz geometry emerges in the IR, the UV is still governed by the LP geometry of \eqref{eq.UVnumerics}, with the free parameters entering in the subleading terms. Therefore the $\Qst$-dependent values of these parameters for small $\rh$ are a prediction of our numerical solutions. One particular value that will play an important role is that of  
$\kappa_\sfb$. This approaches a negative constant at small $\rh$ 
as can be seen in Fig.~\ref{fig.chargedUV1}. 

\begin{figure}[t]
\begin{center}
\includegraphics[width=.6\textwidth]{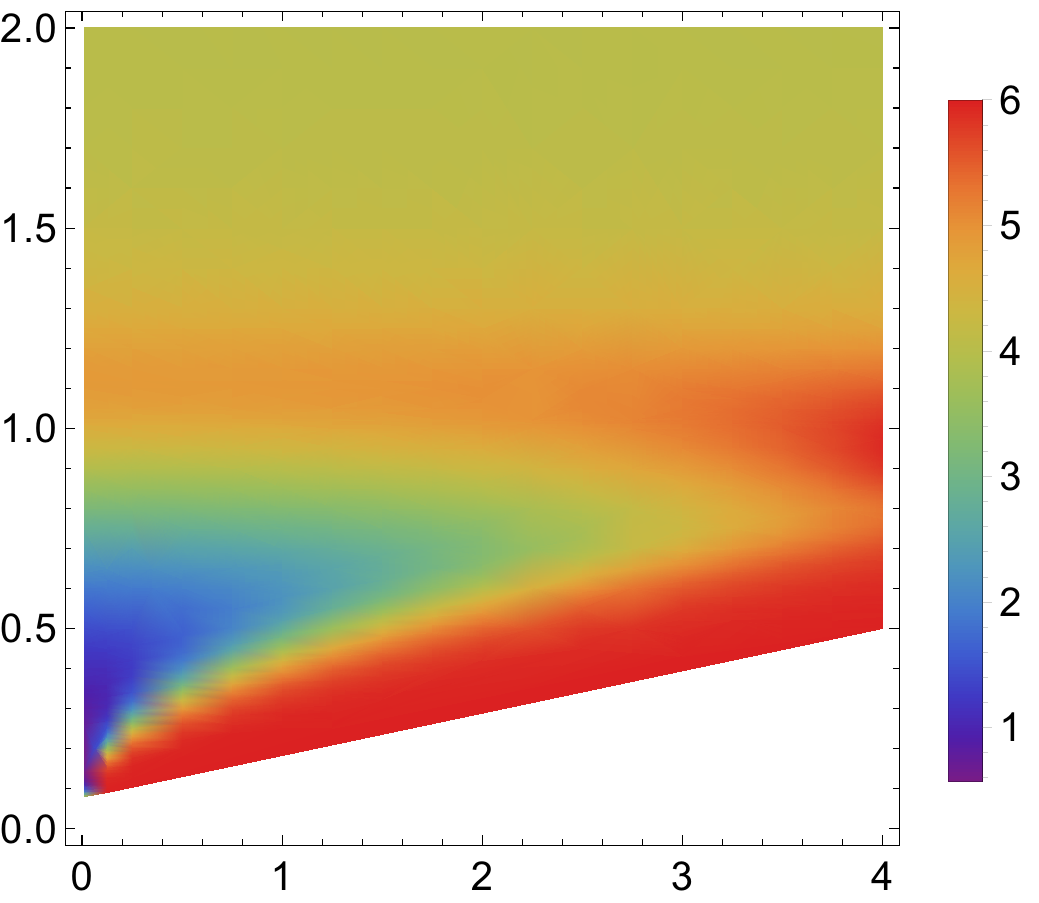} 
\put(-305,115){{\Large $\orh$}$ / \sqrt{2}$}
\put(-160,-15){\large $\sqrt{2} \, \xi $}
\put(-160,235){\Large $\mathcal{R}_\phi$}
\caption{\small \small Dilaton scaling $\mathcal{R}_\phi$  as defined in \eq{dilatonscaling}.}\label{fig.dilscaling}
\end{center}
\end{figure}

The presence of an IR Lifshitz geometry in our solutions can be further verified from the scaling of the different functions.  For any $\psi(r)$ this scaling behavior is extracted as  $ r\, \psi' / \psi$. Since we are interested in tracking the properties of the solutions as the position of the horizon changes we evaluate this expression at the horizon. Thus we define 
\be
\label{dilatonscaling}
\mathcal{R}_\phi \equiv \frac{r \, \partial_r \, e^\phi}{e^\phi} \big|_{r= \rh}=
 r \, \partial_r \phi \big|_{r= \rh} \,,
\ee
and similarly 
\be
\mathcal{R}_\sff = r \, \partial_r \sff \big|_{r= \rh} \sac
\mathcal{R}_\sfg = r \, \partial_r \sfg \big|_{r= \rh} \,.
\ee
From the Lifshitz solution \eqq{eq.IRlifshitz} we expect to find that in the small-$\rh$ limit  
\be\label{eq.fgphiIRscaling}
\lim_{\rh\to 0}\mathcal{R}_\phi = 6 \sac 
\lim_{\rh\to 0}\mathcal{R}_\sff = \lim_{\rh\to 0}\mathcal{R}_\sfg = 
1 \,,
\ee
whereas from \eqq{eq.UVnumerics} we expect that at large $\rh$ 
\be
\lim_{\rh\to \infty}\mathcal{R}_\phi = 4\,.
\ee
The value of $\mathcal{R}_\phi$ extracted from our numerical solutions is shown in \fig{fig.dilscaling}. As expected, we see that $\mathcal{R}_\phi$ approaches 4 and 6 at large and small $\rh$, respectively, for all solutions. Moreover, at small values of $\Qst$ an intermediate region appears where  
$\mathcal{R}_\phi$ becomes very small. This is another manifestation of the intermediate log-AdS region, in which $\mathcal{R}_\phi \sim -1/\log \rh$ for small $\rh$.  

\subsection{Regime of validity}
\label{regime}
We are now ready to determine the regime of validity of the solution. Since we are only interested in parametric dependences, in this analysis we will ignore all purely numerical factors. Moreover, we will focus on the zero-temperature limit, since turning on a non-zero temperature would simply cut-off the zero-temperature solutions. Similar discussions can be found in \cite{Faedo:2014ana,Faedo:2016cih}.

We must require that both supergravity and the smeared DBI action for the D7-branes be valid. We begin with supergravity. The first condition that we must impose is 
\be
e^\phi \ll 1 \,.
\label{firstcondition}
\ee
If this  is not satisfied then string loop corrections become important and degrees of freedom not included in supergravity, such as D-branes, become light. Since \fig{fig.dilatons} shows that the dilaton is a monotonically increasing function of 
$\overline r$,  we expect that the most stringent implication of  \eqq{firstcondition} is obtained from the  UV behavior of the dilaton. Using the asymptotic form of the solutions given in Secs.~\ref{sec.IRisLifshitz} and \ref{sec.susysolution} the dilaton condition becomes
\begin{eqnarray}
\label{dil1}
\mbox{UV:} \qquad e^\phi &\sim& \frac{1}{\nf} \, 
\overline r^ 4 
\,\,\, \ll \,\,\, 1\,.
\\[2mm]
\label{dil2}
\mbox{IR:} \qquad e^\phi &\sim& \frac{1}{\nf} \, \frac{\overline r^6}{\xi^2} 
\,\,\, \ll \,\,\, 1\,.
\end{eqnarray}
As expected, the UV result coincides with that obtained for the neutral solution in \cite{Faedo:2016cih}.

\begin{figure}[t]
\begin{center}
\includegraphics[width=.6\textwidth]{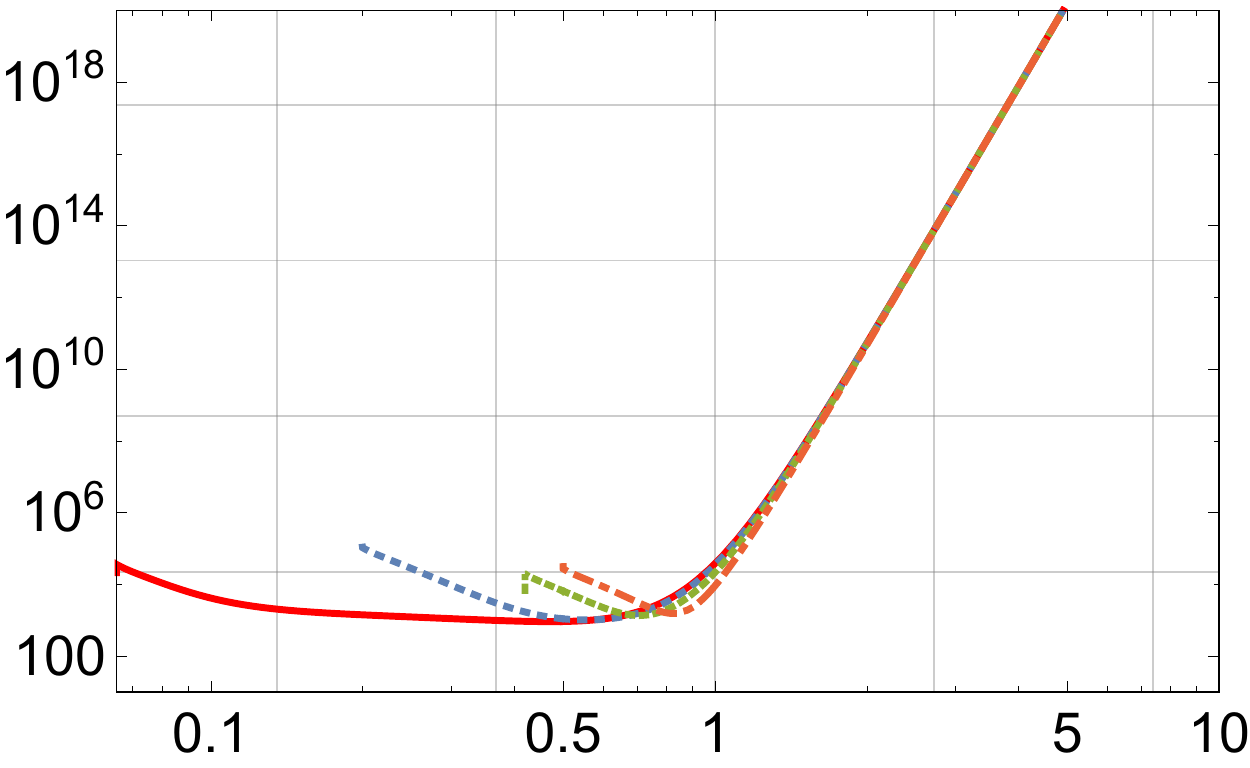} 
\put(-315,90){\large $\ell_s^4 \,  \mbox{Ric}^2_\mt{st}$}
\put(-120,-15){{\Large $\overline r$}$ / \sqrt{2}$}
\caption{\small \small Square of the Ricci tensor, $\mbox{Ric}^2=R^{mn} R_{mn}$,  for several zero-temperature solutions with different charge densities  (see legend in \fig{fig.chargedIR1}).}
\label{Ric}
\end{center}
\end{figure}

The other two conditions that we must require in order for supergravity to be valid are that the curvature of the string-frame metric be small in string units, and that the curvature of the Einstein-frame metric be small in Planck units.\footnote{Note that in our conventions $\ell_p^4 \sim \ell_s^4$ because we are not factoring the dilaton into a constant times a position-dependent part.} The result can be understood simply by considering the two asymptotic forms of the solution in the IR and in the UV. The reason is that, as shown in Fig.~\ref{Ric}, the curvature as measured by the square of the Ricci tensor,  $\mbox{Ric}^2=R^{mn} R_{mn}$, exhibits two simple behaviors separated by a rapid crossover  around $\overline r \sim 1$. 
We will see below that the only exception to this statement occurs when the charge density $\xi$ becomes parametrically small. In this situation the flow follows the neutral flow and penetrates arbitrarily deeply in the log-AdS region, where eventually the supergravity + DBI description  ceases to be valid \cite{Faedo:2016cih}. This situation will not be of interest to us since, as we anticipated in Sec.~\ref{intro}, all the interesting physics takes place at values of $\xi$ of order 1. 

As in the  case of the dilaton,  the two characteristic behaviors of the curvature seen in Fig.~\ref{Ric} at small and large $\overline r$ are controlled by the IR and the UV asymptotic solutions of Secs.~\ref{sec.IRisLifshitz} and \ref{sec.susysolution}, respectively. We have checked that the Ricci scalar, $\mbox{R}=G^{mn} R_{mn}$,  the square of the Ricci tensor, $\mbox{Ric}^2=R^{mn} R_{mn}$ and the square of the Riemann tensor, $R^{mnpq} R_{mnpq}$, behave in the same way parametrically. Since  presumably the same is true for other curvature invariants it suffices to examine $\mbox{Ric}^2$. Through explicit calculation we find that the conditions that the string-frame curvature be small in string units take the form 
\begin{eqnarray}
\label{cond1}
\mbox{UV:} \qquad \ell_s^4 \,  \mbox{Ric}^2_\mt{st} &\sim& \frac{\nf}{\nc} \, 
\overline r^{24} 
\,\,\, \ll \,\,\, 1\,,
\\[2mm]
\label{cond3}
\mbox{IR:} \qquad \ell_s^4 \,  \mbox{Ric}^2_\mt{st} &\sim& 
\frac{\nf}{\nc} \, \frac{\xi^2}{\overline r^6}  \,\,\, \ll \,\,\, 1 \,.
\end{eqnarray}
These $\overline r$-dependences match  the $\overline r \gg 1$ and the $\overline r \ll 1$ behaviors shown in Fig.~\ref{Ric}, respectively.
The curvature of the Einstein-frame metric in Planck units turns out to be proportional to the string-frame curvature in string units,  both in the IR and in the UV:  
\be
\ell_p^4 \, \mbox{Ric}_\mt{Ein}^2 \sim e^{\phi} \, \ell_s^4 \,  \mbox{Ric}_\mt{st}^2  \,.
\ee
It follows that the condition \eqq{firstcondition} and the requirement that the string-frame curvature be small in string units imply that the Einstein-frame curvature are small in Planck units. In fact, the Einstein-frame curvature is non-divergent in the IR, as expected for the Lifshitz times a constant-size sphere solution \eqq{eq.IRlifshitz}.\footnote{Nevertheless the zero-temperature solutions in Einstein frame  still possess a singularity at $\overline r=0$  because tidal forces diverge at this point \cite{Copsey:2010ya}.}
  Therefore we will ignore the Einstein-frame curvature in the following. 

Since Eqs.~\eqq{dil1} and \eqq{cond1} must be valid at the transition point $\overline r \sim  1$, it immediately follows that we must have the hierarchy 
\be
\label{less}
1 \ll \nf \ll \nc \,.
\ee
Under these circumstances  the IR conditions \eqq{dil2}  and \eqq{cond3} are automatically satisfied at the transition point for values of $\xi =\mathcal{O}(1)$, and the supergravity description is valid over the range
\be
\label{range}
\xi^{1/3} \, \left( \frac{\nf}{\nc} \right)^{1/6} \ll \overline r \ll 
\min \left\{  \nf^{1/4} \,, \left( \frac{\nc}{\nf} \right)^{1/24} \right \} \,,
\ee
where the lower bound comes from \eqq{cond3} and the upper bounds come from \eqq{dil1} and \eqq{cond1}.

We now turn to  the constraints imposed by the requirement that the Abelian DBI action for the D7-branes be valid \cite{Bigazzi:2008zt,HoyosBadajoz:2008fw}. The first requirement concerns the characteristic distance between nearby D7-branes, and it consists of two complementary conditions. On the one hand, this distance must be small in macroscopic terms in order for the distribution to be treated as continuous. This simply implies that  $\nf \gg 1$. On the other hand, this distance must be large in string units, since otherwise strings stretching between nearby D7-branes would become light and 
the non-Abelian nature of the DBI action would become important. Since all the D7-branes wrap the $\eta$-fiber in the internal geometry of the metric 
\eqq{eq.10dansatzfornumerics} we must consider their separation in the KE base. The characteristic size of this manifold is 
\be
\ell = \sqrt{G^\mt{st}_\mt{KE}} \,,
\ee
with $G^\mt{st}_\mt{KE}$ the coefficient in front of $\d s_\mt{KE}^2$ in the metric 
\eqq{eq.10dansatzfornumerics}. 
Since inside the four-dimensional KE base the branes are co-dimension two objects, one may effectively think of them as points in a two-dimensional space of volume $\sim \ell^2$. The volume available to each of the branes is therefore $\ell^2/\nf$. As a consequence, the typical inter-brane distance is 
$\sqrt{\ell^2/\nf}$. The requirement that this distance is large in string units is thus
\be
\label{inter}
\frac{\ell^4}{\nf^2\, \ell_s^4} \gg 1\,.
\ee
Using the UV and the IR asymptotics we obtain 
\begin{eqnarray}
\mbox{UV:} \qquad && \frac{\ell^4}{\nf^2\, \ell_s^4} \sim 
\frac{\nc}{\nf^3} \gg 1  \,, \\[2mm]
\mbox{IR:} \qquad && \frac{\ell^4}{\nf^2\, \ell_s^4} \sim  
\frac{\nc}{\nf^3} \, \frac{\overline r^6}{\xi^2} \gg 1 \,.
\end{eqnarray}
The UV condition implies  the hierarchy 
\be
\label{moremore}
1 \ll \nf \ll \nc^{1/3} \,,
\ee
which is more stringent than \eqq{less}. Similarly, the IR condition implies that 
\be
\overline r \gg \xi^{1/3} \, \nf^{1/3}\, \left( \frac{\nf}{\nc} \right)^{1/6} \,,
\ee
which is more stringent than the IR part of \eqq{range}. Note that the hierarchy \eqq{moremore} does not determine which of the two upper bounds on the right-hand side of \eqq{range} is more stringent. 

The second requirement for the DBI action to be valid is that the effective coupling between open strings be small. In the absence of smearing this coupling would be $e^{\phi} \nf$. However, in the presence of smearing not all the $\nf$ branes but only the fraction contained in a volume of string size can participate in a characteristic process involving open strings. As argued above, this fraction is $\nf \, \ell_s^2 / \ell^2$. The requirement that the effective string coupling (squared) be small is therefore 
\be
\frac{e^{2\phi} \, \nf^2  \, \ell_s^4}{\ell^4} \ll 1 \,.
\ee
This follows automatically from \eqq{firstcondition} and \eqq{inter}, so it does not yield any additional independent conditions. 

Putting together the various constraints coming from supergravity and from the DBI action we conclude that, in order for the gravity-plus-branes description to be valid, the number of colors and the number of flavors must obey \eqq{moremore}. The region of validity in terms of the $\overline r$ coordinate is  parametrically large and is given by
\be
\label{low2}
\xi^{1/3} \, \nf^{1/3}\, \left( \frac{\nf}{\nc} \right)^{1/6} \ll \overline r \ll
\min \left\{  \nf^{1/4} \,, \left( \frac{\nc}{\nf} \right)^{1/24} \right \} \,.
\ee

\section{Thermodynamics}
\label{thermoSec}
We will now compute the thermodynamic quantities necessary to study the equilibrium properties of the theory. This goal is greatly simplified by the way in which we wrote the five-dimensional  action \eqref{eq.5daction}. The key point is that, since we traded $G_3=\Qst \d x^{123}$ in favor of $F_2$, none of the fields in our solution has purely spatial components turned on. Combined with the radial dependence of all fields this implies that the on-shell Lagrangian density is given by 
\be
{\cal L} = \frac{1}{2\kappa_5^2} 2 \sqrt{-g} \,{R_x}^x \qquad \text{(no sum)} \ .
\ee
Since this is a total derivative we can write the on-shell Euclidean  action 
$I_\mt{5}$ as
\be
\label{euc}
I_\mt{5} =  \frac{\int \d \tau \d^3 x}{2\kappa_5^2} \sqrt{\frac{  g_{\tau\tau} \,g_{xx}}{g_{rr}}}\, g_{xx}' \Bigg|_{r_\mt{LP}} \ ,
\ee
where   $\tau$ is the Euclidean time and the only contribution comes from the evaluation at the LP.  The relation between the functions in the five-dimensional effective metric \eqq{5Dmetric} and those in the ten-dimensional  metric  \eqref{eq.10dansatzfornumerics} is 
\be\bal
g_{tt} &= - \sfb\, g_{xx} \ ,  \quad g_{xx} = 2\cdot 2^{2/3} \left( \frac{L}{r} \right)^{4/3} e^{\frac{2}{3}(\sff+4\sfg)} \ , \quad g_{rr} = \frac{2\cdot 2^{2/3}}{ \sfc} \left( \frac{L}{r} \right)^{16/3} e^{\frac{8}{3}(\sff+\sfg)}  \ ,\\[2mm]
e^\sigma& = \sqrt{2} \frac{L}{r} e^\frac{\sff+4\sfg}{5} \ , \qquad e^w = e^{\sff-\sfg}\,.
\eal\ee
Using the expansion  \eqref{eq.UVnumerics} in these formulas one easily obtains the asymptotics for the five-dimensional  functions. Substituting the result in \eqq{euc} one finds that the on-shell Euclidean action diverges as
\be
I_\mt{5} =  \frac{\int \d \tau \d^3 x}{2\kappa_5^2} \left[  \frac{8\, r^4}{3L^5} +  \frac{8 \left( \kappa_\sff-\kappa_\phi+\kappa_\sfb \right)}{3L^5}  +  {\cal O} (r^{-4}) \right] \,.
\ee
We must therefore regularize the integral by introducing a cutoff for the radial coordinate, add the necessary boundary terms and take the $r\to \infty$ limit. We first  add  the standard Gibbons-Hawking piece, written in terms of the trace of the extrinsic curvature at a constant-radius slice. Using the UV expansion \eqref{eq.UVnumerics} we get
\be
I_\mt{GH} =  \frac{\int \d \tau \d^3 x}{2\kappa_5^2} \left[ - \frac{32\, r^4}{3L^5} -  \frac{4 \left(8 \kappa_\sff-8\kappa_\phi+11\kappa_\sfb \right)}{3L^5}  +  {\cal O} (r^{-4}) \right] \,.
\ee
As shown in Ref.~\cite{Faedo:2016cih}, counterterms associated to the asymptotically HV metric characterising the LP can be obtained via analytic continuation from the standard counterterms for asymptotically AdS spacetimes, in an analysis similar to the one in \cite{Kanitscheider:2008kd}.
In the present situation the action \eqref{eq.5daction} is more complicated than the one in \cite{Faedo:2016cih} due to the inclusion of massive and massless vector fields, which also  enter via the DBI action, not just via a Maxwell-like action. However, as  stated above, the near-LP behavior of the solutions is dominated by the neutral, supersymmetric sector of Eqn.~\eqref{eq.LandaupoleasHV}. In particular, the divergences are totally determined by the metric and the scalars, so we can simply take the same counterterm as in \cite{Faedo:2016cih}, which is proportional to the superpotential \eqref{eq.superpotential}.
Explicit evaluation of this term gives
\be
I_{\cal W} =  \frac{\int \d \tau \d^3 x}{2\kappa_5^2} \left[  \frac{8\, r^4}{L^5} +  \frac{4 \left( \kappa_\sfb-6L^4 \right)}{L^5}  +  {\cal O} (r^{-4}) \right] \,.
\ee
Adding up the three pieces  we obtain a finite result in the $r\to\infty$ limit:
\be
I_\mt{total} = I_\mt{5} + I_\mt{GH} + I_{\cal W} = 
- \frac{8\int \d \tau \d^3 x}{2\kappa_5^2} \, \frac{  \kappa_\sff-\kappa_\phi+\kappa_\sfb + 3L^4 }{L^5}   \,.
\ee
As usual, we will identify this result with the grand canonical free energy density\footnote{The free energy $G$ should not to be confused with any of the field strengths $G_n$ introduced in Sec.~\ref{sec.5dsetup}, which always carry a subindex.} 
\be
\label{GG}
G =-  \frac{1}{\int \d \tau \d^3 x} I_\mt{total} =  \frac{8}{2\kappa_5^2} \, \frac{  \kappa_\sff-\kappa_\phi+\kappa_\sfb + 3L^4 }{L^5}   \,.
\ee

By varying $I_\mt{GH}$ and $I_{\cal W}$  ($I_5$ gives a vanishing contribution) with respect to the induced metric on the constant-radius slice we obtain two divergent expressions whose sum yields a finite result for the boundary stress tensor:
\be
{T_\mu}^\nu = \text{diag} \left(E , P ,P ,P\right) \,,
\ee
where
\be
\label{EP}
E = \frac{4}{2\kappa_5^2} \, \frac{ 2 \kappa_\sff- 2 \kappa_\phi+\kappa_\sfb + 6L^4 }{L^5} \ , \qquad P = -\frac{8}{2\kappa_5^2} \, \frac{  \kappa_\sff-\kappa_\phi+\kappa_\sfb + 3L^4 }{L^5} = - G \ .
\ee
The fact that the pressure and the free energy \eqq{GG} differ only by a sign confirms that $G$ is the free energy in the grand canonical ensemble. Recalling now that the quantity $h$ defined in \eqref{eq.heat} is equal to $T  s$ and that it is radially conserved we obtain an expression for $T s$ in terms of the UV data:
\be
\label{tsts}
T  s = - \frac{1}{2\kappa_5^2} \left[  \frac{4\,\kappa_\sfb}{L^5} + \Qst\, A_t(r_\mt{LP}) - \frac{4\bfunc(r_\mt{LP})}{L^4} \cfunc_t(r_\mt{LP})+ \frac{4\,\Qst 2\pi\ls^2\At(r_\mt{LP})}{L} \right] \,.
\ee
It is now easy to check that 
\be\label{eq.chemicalpotentialsagogo}
G-E+Ts =  \frac{1}{2\kappa_5^2} \left[ - \Qst\, A_t(r_\mt{LP}) + \frac{4\bfunc(r_\mt{LP})}{L^4} \cfunc_t(r_\mt{LP}) - \frac{4\,\Qst 2\pi\ls^2\At(r_\mt{LP})}{L} \right]\,.
\ee
We will now see that the right-hand side of this expression is precisely of the form $- \mu \, \Qst$, with $\mu$ the quark chemical potential dual to $\Qst$. Indeed, from the action \eqref{eq.5daction} we can compute the momenta conjugate to the three different vectors occurring in \eqq{eq.chemicalpotentialsagogo}. For $A_t$ and $\At$ this leads to two constants
\be
\label{both}
\Pi_{\cA} \equiv \frac{\delta S_5}{\delta \At'} = \frac{2\pi\ell_s^2}{2\kappa_5^2} \frac{4\, \Qst}{L} \ ,\qquad \Pi_{A} \equiv \frac{\delta S_5}{\delta A_t'} = \frac{\Qst}{2\kappa_5^2}  \ ,
\ee
whereas for the massive vector $\cfunc_t$ it leads to a function of $r$:
\be
\Pi_{\cfunc}(r) \equiv \frac{\delta S_5}{\delta \cfunc_t'} =- \frac{4}{L^4} \frac{\bfunc(r)}{2\kappa_5^2} \ .
\ee
In this way, we can rewrite \eqref{eq.chemicalpotentialsagogo} as
\be
\label{this}
G-E+Ts = - \Pi_{A}\, A_t(r_\mt{LP}) - \Pi_{\cfunc}(r_\mt{LP})\, \cfunc_t(r_\mt{LP}) - \Pi_{\cA}\, \At(r_\mt{LP}) \ .
\ee
As usual in holography, the asymptotic values of the conjugate momenta are identified with conserved charges, whereas  the asymptotic values of the vector fields themselves are identified with the dual chemical potentials. Therefore the right-hand side of \eqq{this} has the form 
\be
-\sum_i \mu_i \, Q_i \,. 
\ee
However, due to our boundary conditions there is only one independent charge $\Qst$. This is already apparent in the fact that both conjugate momenta in \eqq{both} are proportional to $\Qst$. Moreover, using the asymptotic expansion \eqq{eq.UVnumerics} we see that it is also true for $\Pi_{\cfunc}$ since 
\be
\Pi_{\cfunc} (r_\mt{LP}) = \frac{4}{2 \kappa_5^2} \frac{\Qst}{\Qf} \ .
\ee
As a consequence, all three charges vary simultaneously when $\Qst$ changes, and the contribution to the change in free energy is of the form $\mu \, \d \Qst$ with an effective chemical potential given by
\be
\label{mutotal}
\mu = \frac{1}{2\kappa_5^2}  \left[ A_t(r_\mt{LP}) + 
\frac{4}{\Qf}\, C_t(r_\mt{LP}) +  
\frac{4}{L} \, 2\pi\ell_s^2 \, \At(r_\mt{LP}) \right] \,.
\ee
In terms of this \eq{this} can be written as 
\be
G=E - Ts - \mu \, \Qst \,,
\ee
which is nothing but the usual definition of the grand canonical free energy. 
Similarly, we can now define the free energy in the canonical ensemble as\footnote{The free energy $F$ should not to be confused with the field strength $F_2$ introduced in Sec.~\ref{sec.5dsetup}, which always carries a subindex.}
\be
F = G + \mu \, \Qst = E- Ts \,,
\ee
which in terms of UV data takes the form
\be\label{eq.freeenergies}
F = \frac{1}{2\kappa_5^2} \left[ \frac{ 8 \kappa_\sff- 8 \kappa_\phi + 8 \kappa_\sfb + 24 L^4  }{L^5} + \Qst\, A_t(r_\mt{LP}) + \frac{ \kappa_\bfunc \, \Qst^2}{6\, L^3\Qf} + \frac{4\,\Qst 2\pi\ell_s^2 \At(r_\mt{LP})}{L} \right] \ .
\ee
Note that the third term inside the brackets comes from the term proportional to 
$C_t(r_\mt{LP})$ in \eqq{tsts} or \eqq{mutotal}. This term has been evaluated explicitly to 
\be
C_t(r_\mt{LP}) = \frac{\kappa_\bfunc \, \Qst}{24 \, L^3} 
\ee
using \eqq{eq.conjugatemomentumprime} to relate 
$C_t(r_\mt{LP})$ to $\bfunc'(r_\mt{LP})$ and \eqq{eq.UVnumerics} to evaluate 
$\bfunc' (r_\mt{LP})$. Note that \eqq{eq.conjugatemomentumprime}, together with the near-horizon expansions, implies that at the horizon
\be
\bfunc' (\rh) = \mbox{const.} \sim \frac{C_t (\rh)}{r-\rh} \,.
\ee
This immediately implies $C_t (\rh)=0$, which is nothing but the regularity condition for a vector field at the horizon. Analogous  conditions must also be imposed on $A_t$ and $\cA_t$ in order to compute $A_t(r_\mt{LP})$ and $\cA_t(r_\mt{LP})$, which are obtained by integrating \eqq{eq.Cteom} and \eqq{eq.AtprimesolutionBIS} from $\rh$ to $r_\mt{LP}$, respectively, with the boundary conditions that $A_t (\rh)=\cA_t(\rh)=0$.

At this point it is useful to comment on the dimensions of several quantities. We have already defined a dimensionless temperature $\overline T$ in \eqq{TTrr}. Similarly, we will  work with a dimensionless entropy density given by 
\be
\label{TTT}
\overline s = \frac{V_5 \, \Qc^{3/4}}{\pi \nc^2} \, s \,,
\ee
where we recall that $V_5$ is the volume of the SE manifold and we have used the expression \eqq{kkappa2} for the five-dimensional Newton's constant. Since the dimensions of $C_Q$ and $s$ are the same we will use the same factor to make $C_Q$ dimensionless:
\be
\label{CQCQCQ}
\overline C_Q = \frac{V_5 \, \Qc^{3/4}}{\pi \nc^2} \, C_Q \,.
\ee
Since  $E$ and $P$ (and $F, G$, etc) have the same dimensions as $Ts$ we  use the product of the two factors above:
\be
\label{EEEPPP}
\overline E =  \frac{V_5 \, \Qc}{\pi \nc^2} \, E \sac 
\overline P =  \frac{V_5 \, \Qc}{\pi \nc^2} \, P \,.
\ee
 We refer to $\Qst$ as the charge density because of the proportionality relation \eqq{eq.qstvalue}, but note that it has dimensions of (length)$^{-1}$. Therefore we use precisely 
\be
\sqrt{2}\,\xi = \frac{\Qc^{1/4}}{\Qf^{1/2}} \Qst 
\ee
as its dimensionless counterpart, as defined in \eqq{eq.ratiodependence}. Since $\mu \,\Qst$ must have the same dimensions as $Ts$ it follows that  the chemical potential has dimensions of (length)$^{-3}$. We emphasise that this is simply an unusual  convention with no physical implications. Our  dimensionless chemical potential is therefore 
\be
\label{MMM}
\overline \mu = \frac{\sqrt{2} \, \Qc^{3/4} V_5 \Qf^{1/2}}{\pi \, \nc^2} \mu \,.
\ee
The necessary coefficient to make the charge susceptibility $\chi$ dimensionless then is just the ratio of that in $\Qst$ over that in $\mu$, i.e.
\be
\overline \chi = \frac{\pi \, \nc^2}{2 V_5 \, \Qc^{1/2}\, \Qf} \, \chi \,.
\ee 
Finally, we anticipate here that the dimensionless versions of the functions $f, g$ to be introduced in Eqs.~\eqq{introf} and \eqq{FF}  are 
\be
\label{fgfgfg}
\overline f = \frac{V_5 \, \Qc^{9/14}}{\pi \nc^2} \, f \qquad \mbox{and} \qquad
\overline g ' = \frac{\sqrt{2} \Qc^{3/4} V_5 \Qf^{1/2}}{\pi \, \nc^2} \, g' \,.
\ee

\begin{figure}[t!!!]
\centering
\hspace{0.02\textwidth}
{\includegraphics[width=0.42\textwidth]{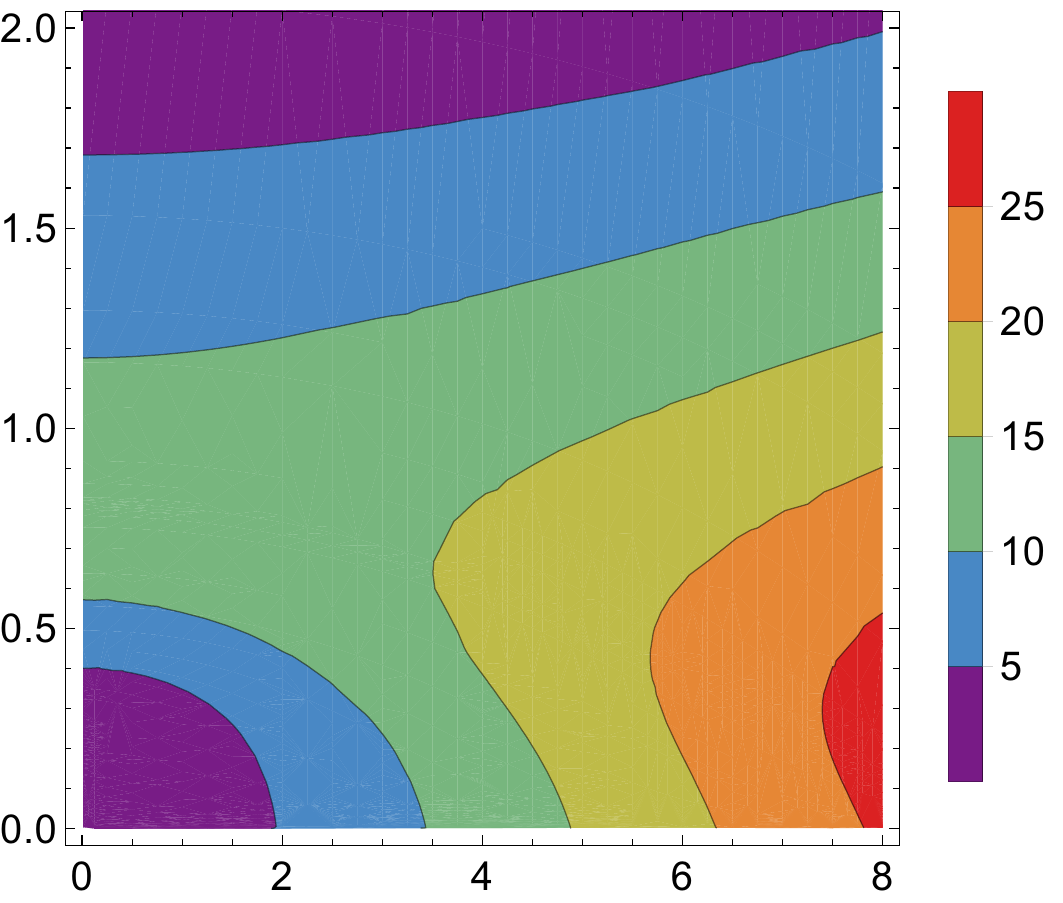}}
\hspace{0.08\textwidth}
{\includegraphics[width=0.42\textwidth]{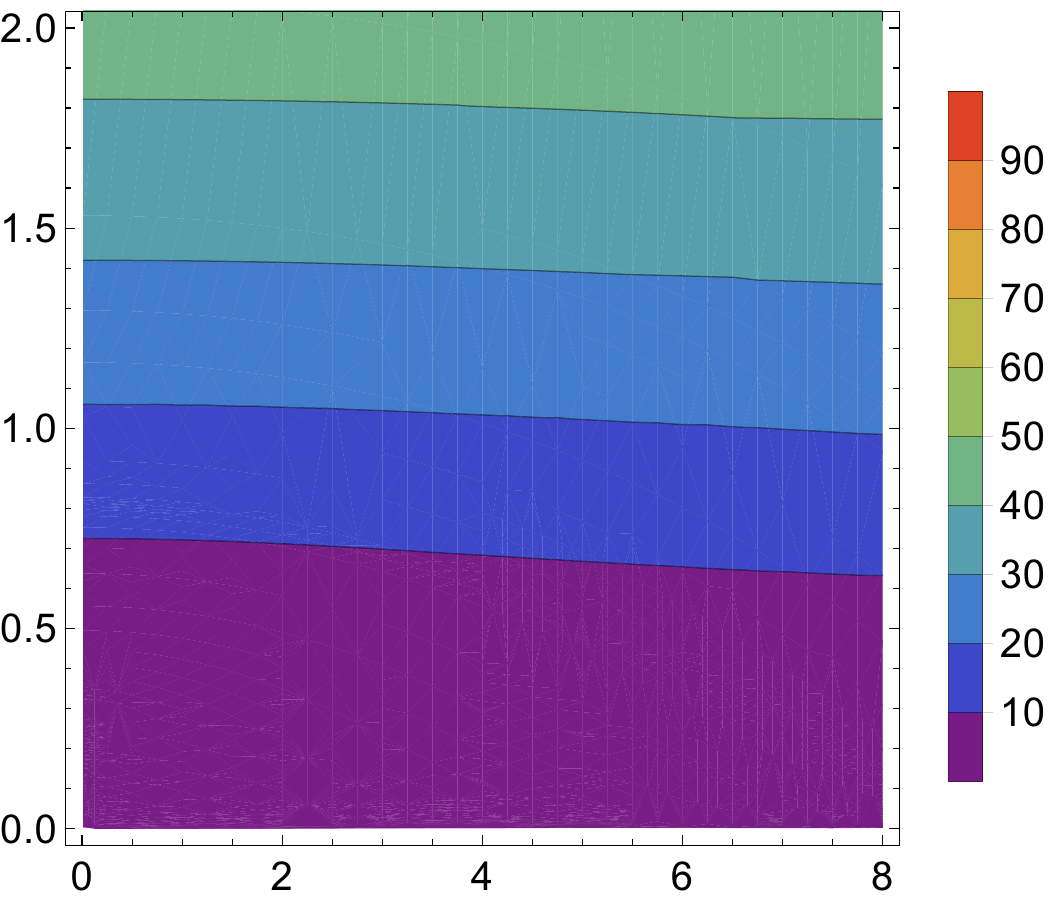}}
\put(-350,-15){ $\sqrt{2} \, \xi$}
\put(-120,-15){ $\sqrt{2} \, \xi$}
\put(-340,160){\Large $\overline E$}
\put(-110,160){\Large $\overline P$}
\put(-432,76){ $\overline T$}
\put(-202,76){ $\overline T$}
 \\[10mm]
 \hspace{0.02\textwidth}
{\includegraphics[width=0.42\textwidth]{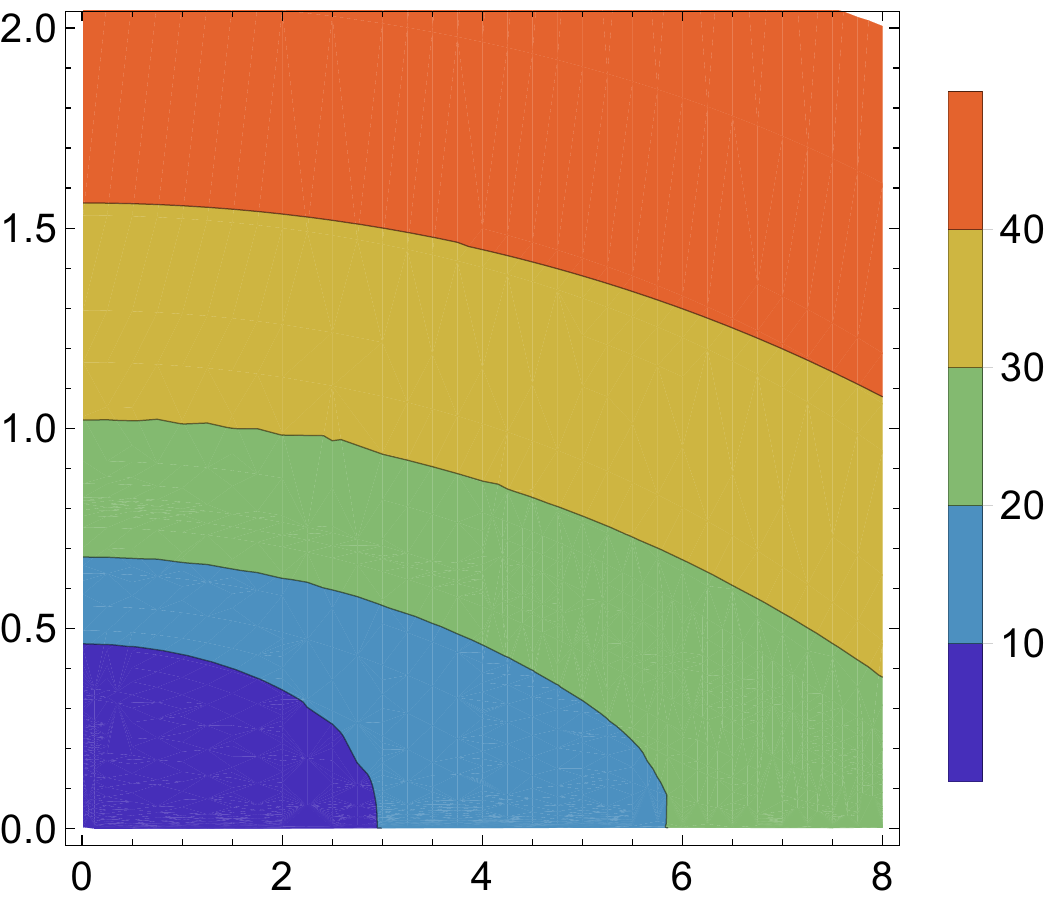}}
\hspace{0.08\textwidth}
{\includegraphics[width=0.42\textwidth]{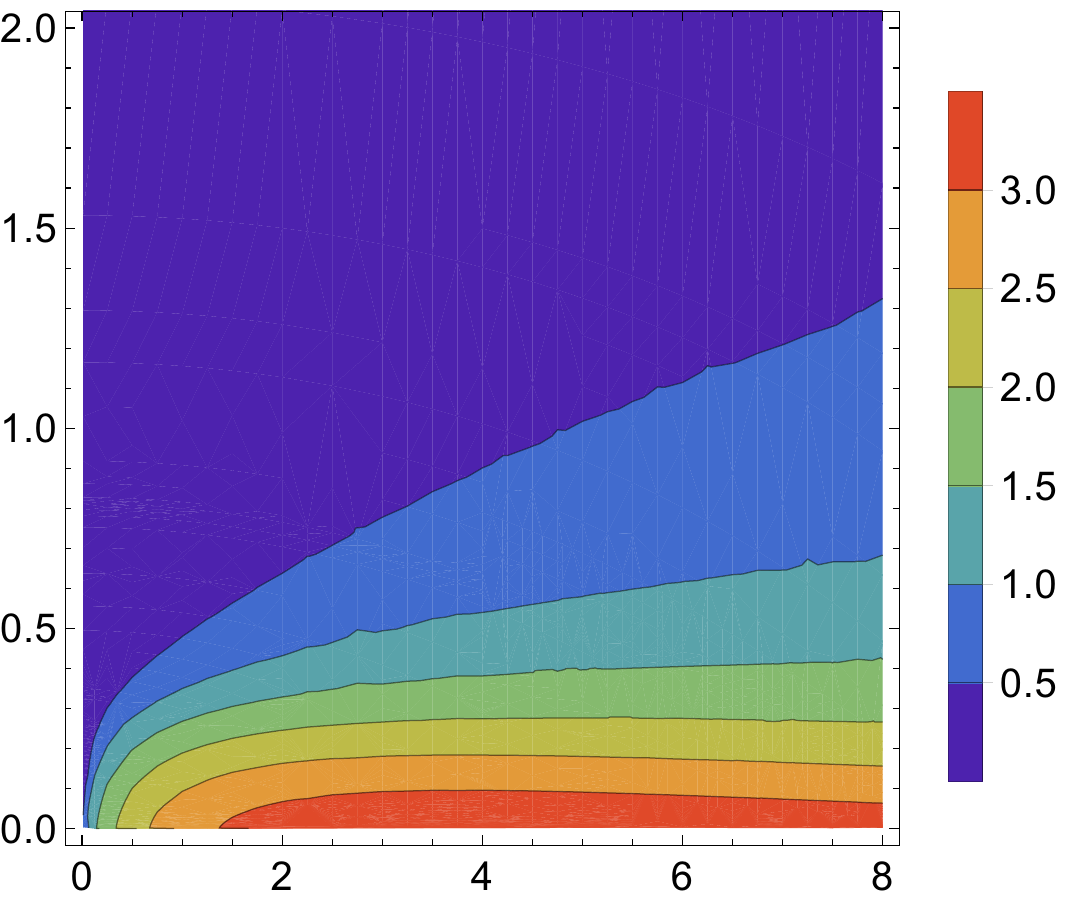}}
\put(-350,-15){ $\sqrt{2} \, \xi$}
\put(-120,-15){ $\sqrt{2} \, \xi$}
\put(-350,160){\Large $\overline E+ \overline P$}
\put(-110,160){\Large $\overline \mu$}
\put(-432,76){ $\overline T$}
\put(-202,76){ $\overline T$}
\caption{\small Dimensionless energy density, pressure, enthalpy and chemical potential, as defined in  \eqq{EEEPPP} and \eqq{MMM}. 
}
\label{EandP}
\end{figure}

We close this section with plots in \fig{EandP} of the energy density, $E$, the pressure,  $P$, the enthalpy, $E+P$, and the chemical potential, $\mu$. 
Note that these quantities are positive for all values of $\overline T$ and $\xi$.
%

\section{Instabilities} 
\label{insta}

We will now examine the local thermodynamic stability of the system. This is equivalent to the positive-definiteness of minus the Hessian  in the grand-canonical ensemble, namely of the susceptibility matrix \be
\label{eq.hessian}
\mbox{$H$} = {\Large
\begin{pmatrix}
 -\frac{\partial^2 G}{\partial T^2} &   
 -\frac{\partial^2 G}{\partial \mu \, \partial T}   \\[2mm]
 -\frac{\partial^2 G}{\partial T\, \partial \mu}   & 
 -\frac{\partial^2 G}{\partial \mu^2}
\end{pmatrix}
=
\begin{pmatrix}
 \frac{\partial s}{\partial T} &   
 \frac{\partial s}{\partial \mu}   \\[2mm]
 \frac{\partial \Qst}{\partial T}   & 
 \frac{\partial \Qst}{\partial \mu}
\end{pmatrix}}
\,,
\ee
where all derivatives with respect to $T$ and $\mu$ are taken with the other one constant. The system is stable if and only if 
\be
\chi = \frac{\partial \Qst}{\partial \mu}\temp > 0 \sac \det H >0 \,.
\label{stab1}
\ee
The first condition is the positivity of the charge susceptibility. This can be conveniently rewritten  in terms of the canonical free energy $F$ as
\be
0 < \chi^{-1} = \left( \frac{\partial \Qst}{\partial \mu} \right)_T^{-1} = 
 \frac{\partial \mu}{\partial\Qst}\temp = \frac{\partial^2 F}{\partial \Qst^2}\temp \,.
\label{stab2}
\ee
The second condition can be related to the specific heat at constant charge: 
\be
\label{cqdef}
C_Q = T \, \frac{\partial s}{\partial T}\charge \,.
\ee
Indeed, using the chain rule and the equality of the crossed derivatives of $F$ we find that 
\be 
C_Q = -T \left[ \frac{\partial^2 G}{\partial T^2} - 
\left( \frac{\partial^2 G}{\partial T \partial \mu} \right)^2 
\left( \frac{\partial^2 G}{\partial \mu^2} \right)_T^{-1} \right] \,.
\ee
It follows that 
\be
\label{eq.detH}
\det H = \frac{\chi \, C_Q}{T} \,.
\ee
Therefore the stability conditions \eqq{stab1} are verified if and only if 
\be
C_Q >0 \sac \chi^{-1} > 0   \,.
\ee

\begin{figure}[t]
\centering
\hspace{0.01\textwidth}
{\includegraphics[width=0.42\textwidth]{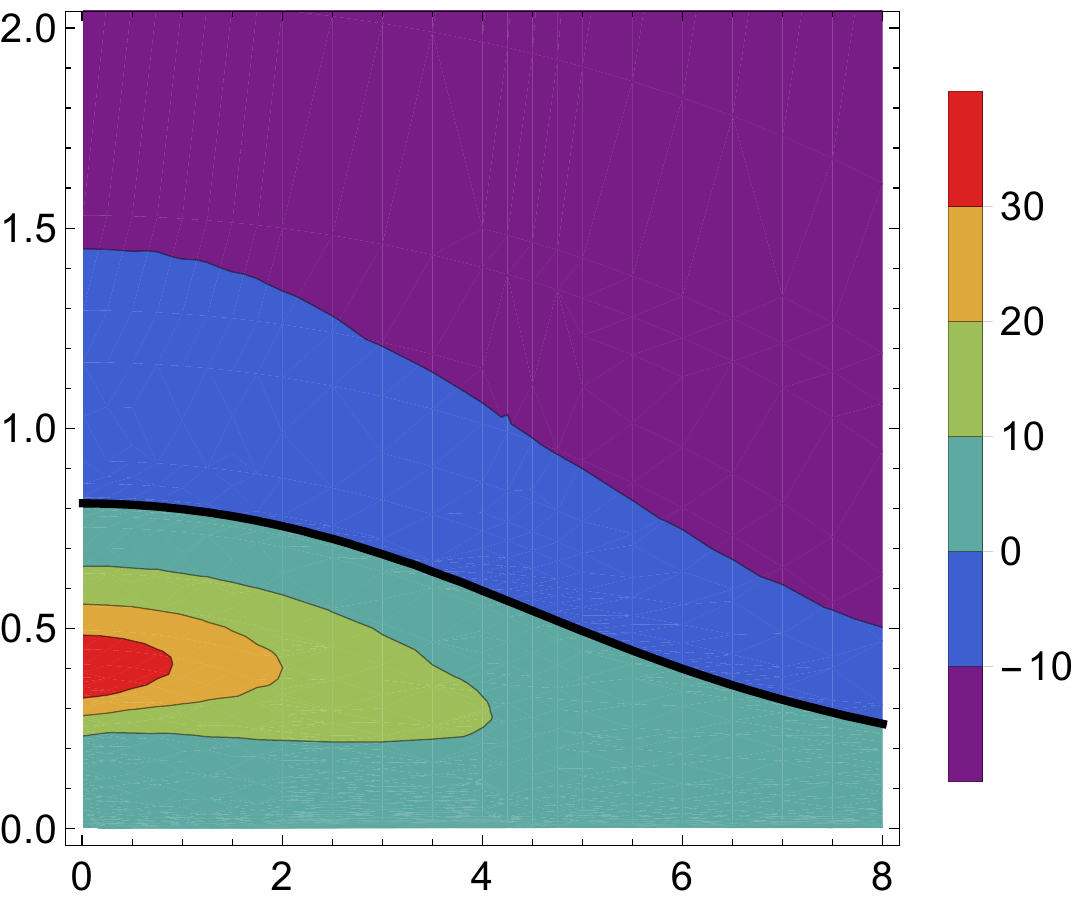}}
\hspace{0.04\textwidth}
\quad{\includegraphics[width=0.48\textwidth]{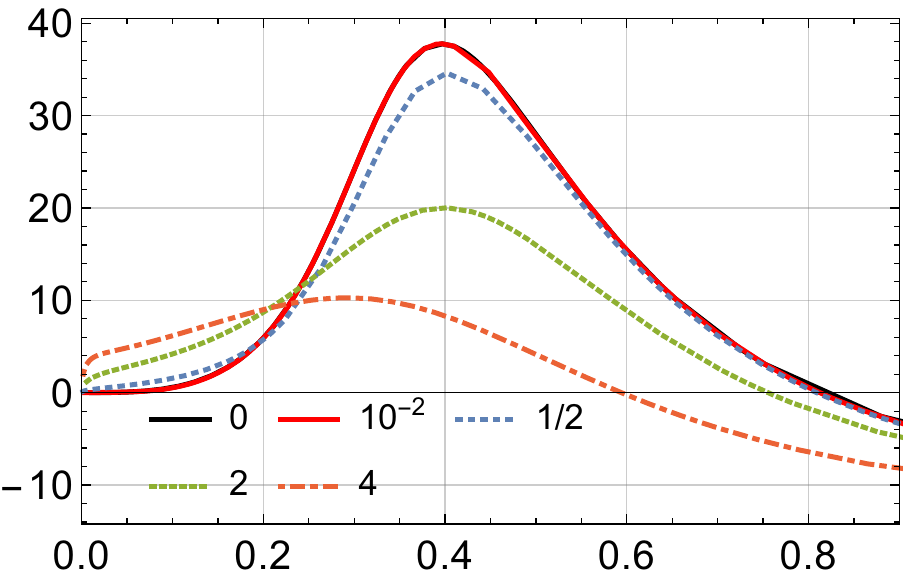}}
\put(-370,-15){ $\sqrt{2} \, \xi$}
\put(-105,-15){ $\overline T$}
\put(-365,163){\Large $\overline C_Q$}
\put(-450,76){ $\overline T$}
\put(-230,70){ $\overline C_Q$}
\caption{\small (Left) Contour plot of the dimensionless specific heat \eqq{CQCQCQ}. The thick
black curve indicates the locus where $C_Q=0$.  (Right) Dimensionless specific heat as a function of temperature for several fixed values of the charge density $\sqrt{2}\,  \xi$. Note that the black curve is hardly visible because it falls almost exactly behind the red curve.
}
\label{fig:SH}
\end{figure}

\begin{figure}[t]
\centering
{\includegraphics[width=0.55\textwidth]{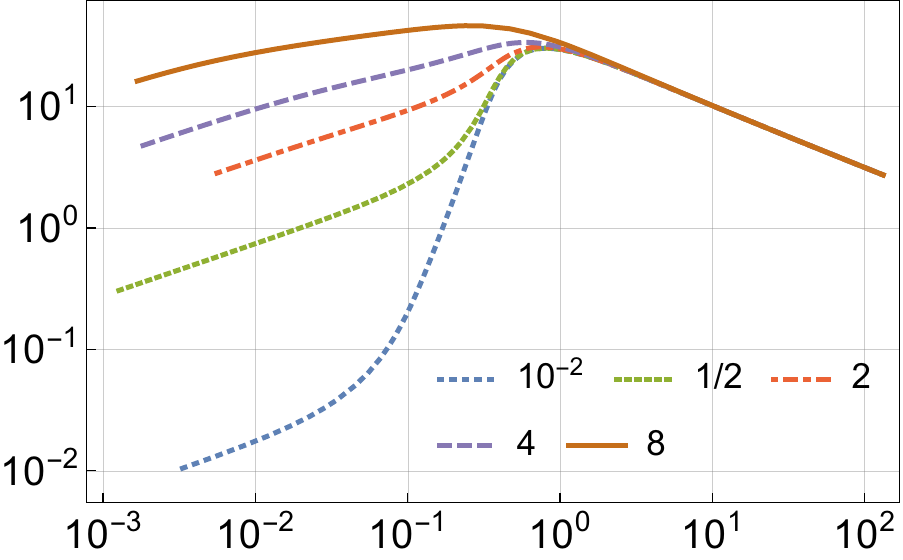}}
\put(-260,75){\Large $\overline s$}
\put(-115,-15){\large $\overline T$}
\caption{\small  Entropy density as a function of temperature for various values of the charge density $\sqrt{2}\,  \xi$.}
\label{fig:Entrop}
\end{figure}

In \fig{fig:SH}(left) we show a contour plot of the specific heat of our solutions as a function of $\overline T$ and $\sqrt{2} \,\xi$, while in \fig{fig:SH}(right) we show slices at several different values of the charge. 
We see that $C_Q$ is positive at low temperature  but becomes negative at high  temperature. This property is also illustrated  by \fig{fig:Entrop}, where the negative slope of the entropy density curves as a function of $\overline T$ is evident at high $\overline T$.  
In Figs.~\ref{fig:SH}(right) and \ref{fig:Entrop} we see that all curves converge to the same one at high  $\overline T$. The common curve is  the same as in the neutral case studied in \cite{Faedo:2016cih}. As explained in that reference, the negativity of $C_Q$ in this region is an UV effect associated to the presence of the LP. Since in this paper we are interested in IR physics that is safe from LP effects, we will not elaborate further on this instability. 

\begin{figure}[t]
\centering
{\includegraphics[width=0.55\textwidth]{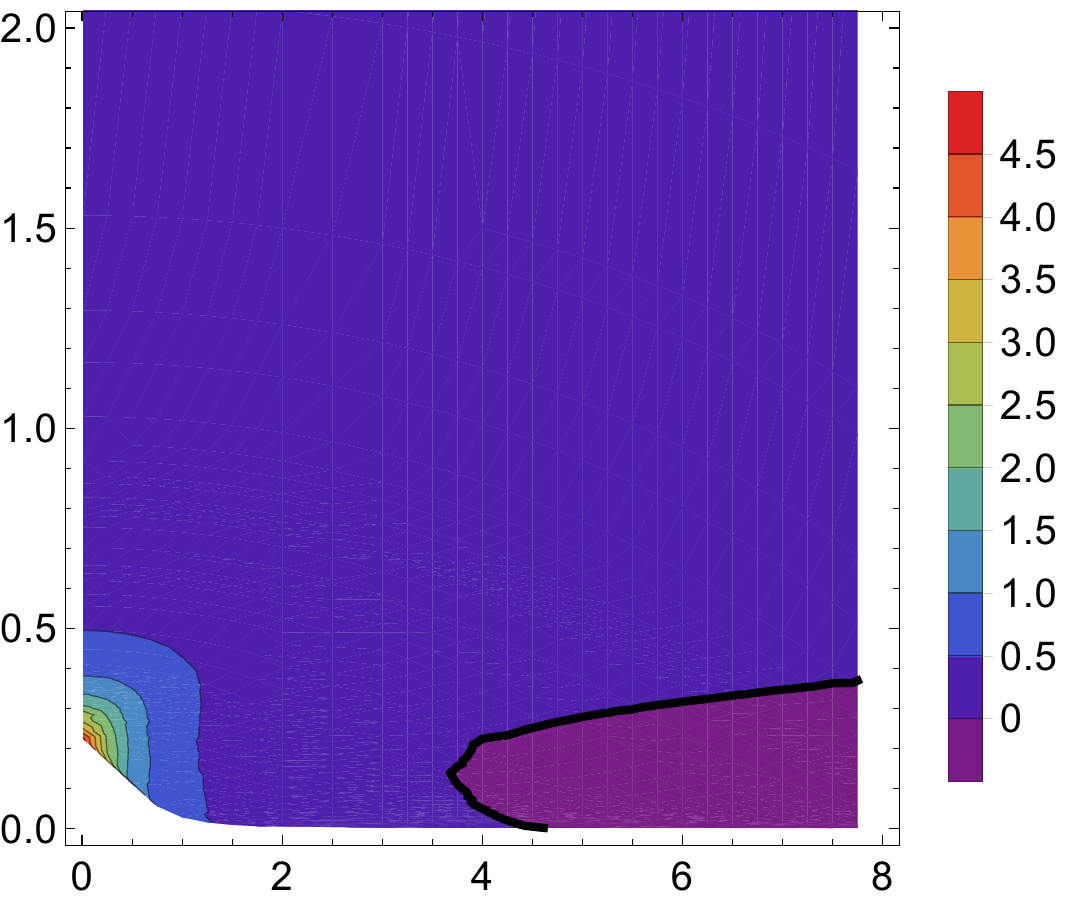}}
\put(-150,210){\Large $\overline \chi^{-1}$}
\put(-260,105){\large $\overline T$}
\put(-155,-15){\large $\sqrt{2} \, \xi$}
\caption{\small  Contour plot of the inverse charge susceptibility. The thick, black, continuous curve indicates the locus where $\overline \chi^{-1}$ changes sign. 
}
\label{fig:H22}
\end{figure}

In \fig{fig:H22} we show a contour plot of the inverse charge susceptibility of our solutions as a function of $\overline T$ and $\sqrt{2}\, \xi$.
\begin{figure}[t]
\centering
\hspace{0.01\textwidth}
{\includegraphics[width=0.42\textwidth]{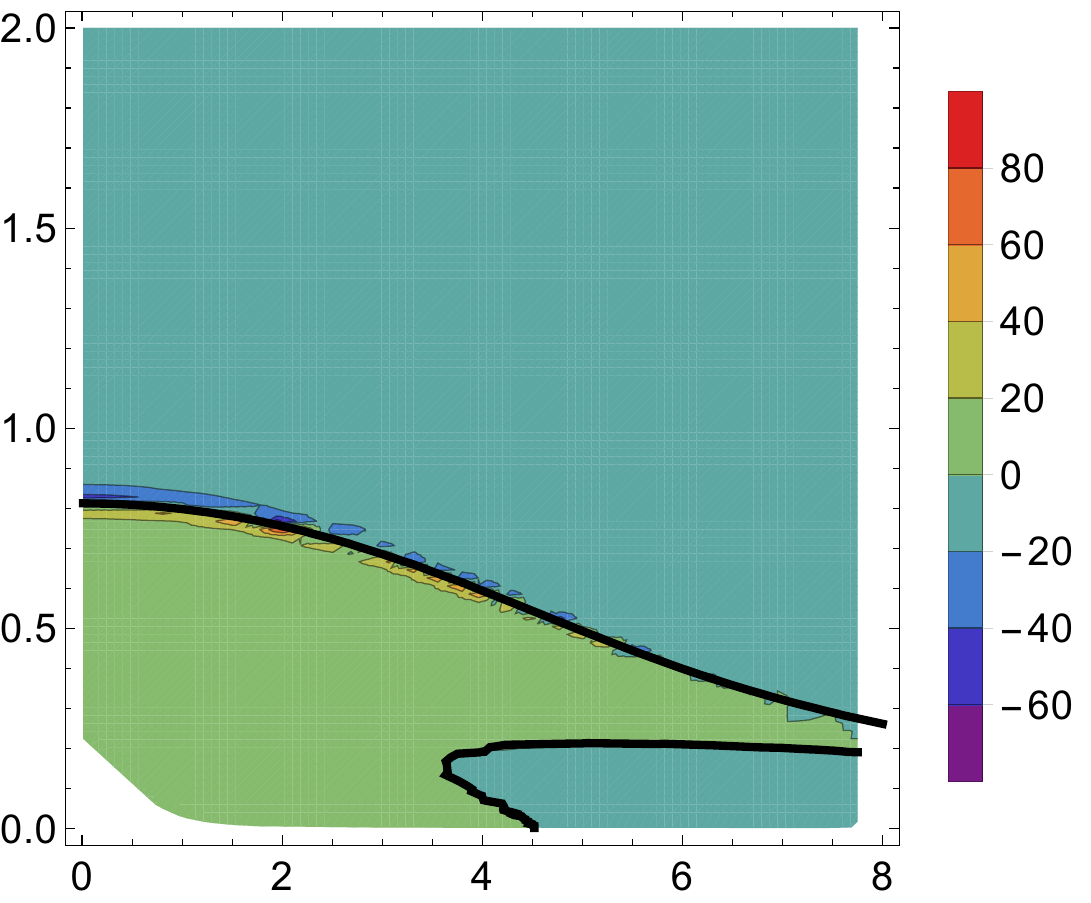}}
\hspace{0.04\textwidth}
\quad{\includegraphics[width=0.48\textwidth]{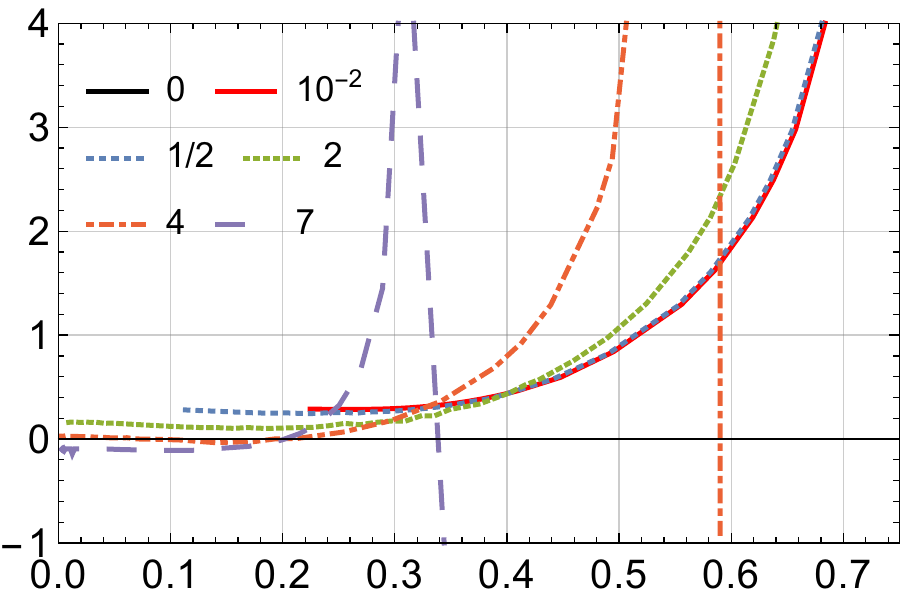}}
\put(-370,-15){ $\sqrt{2} \, \xi$}
\put(-105,-15){ $\overline T$}
\put(-365,160){\Large $ c_s^2$}
\put(-450,76){ $\overline T$}
\put(-230,70){ \Large $c_s^2$}
\caption{\small (Left)  Contour plot of the speed of sound squared. The thick, black, continuous curve indicates the locus where $c_s^2$ changes sign.
(Right) Plots of the speed of sound as a function of temperature for several fixed values of the charge density $\sqrt{2} \, \xi$. The speed of sound diverges  towards positive (negative) infinity as the upper black curve is approached from below (above).  This is why the scale of the contour plot has such a large range.
}
\label{fig:SS}
\end{figure}
In contrast to the case of the specific heat, we see that $\chi^{-1}$ is negative
at arbitrarily low $\orh$ and $\overline T$ for charge densities $\sqrt{2}\, \xi \gtrsim 4.5$. The negativity of $\chi$ signals an instability towards charge clustering, which suggests that the putative, stable phase in this region may break translational invariance spontaneously. We will come back to this point in Sec.~\ref{disc}.

A further instability comes from the speed of sound of the system. For a system with both energy and charge density this is given by (see e.g.~\cite{Kovtun:2012rj})
\be
\label{cscs}
c_s^2 = \frac{\partial P}{\partial \cE}\charge + \frac{\Qst}{\cE + P}\,\frac{\partial P}{\partial \Qst}\energy \,.
\ee
After some manipulations, $c_s^2$ can be expressed in terms of our UV data as 
\be
c_s^2 = -\left[1+\frac{\kappa_b'}{2} \frac{1}{\kappa_f'-\kappa_\phi'+\kappa_b'/2}\right]-\frac{\Qst}{\kappa_b}\left[\kappa_b'\frac{\dot{\kappa_f}-\dot{\kappa_\phi}+\dot{\kappa_b}/2}{\kappa_f'-\kappa_\phi'+\kappa_b'/2}-\dot{\kappa_b}\right]\,,
\ee
where $\dot{}=\d/\d\Qst$ and $'=\d/\d \rh$. In \fig{fig:SS}(left) we show a contour plot of the result, and in \fig{fig:SS}(right) several slices at constant charge.  As for $C_Q$, the region where $c_s^2$ is negative includes a UV region sensitive to LP effects that is of no interest to us. However, in contrast to $C_Q$ and similarly to $\chi$,  $c_s^2$ is also negative in the IR region of arbitrarily low temperature and  $\sqrt{2}\, \xi \gtrsim 4.5$.

\begin{figure}[t]
\centering
{\includegraphics[width=0.45\textwidth]{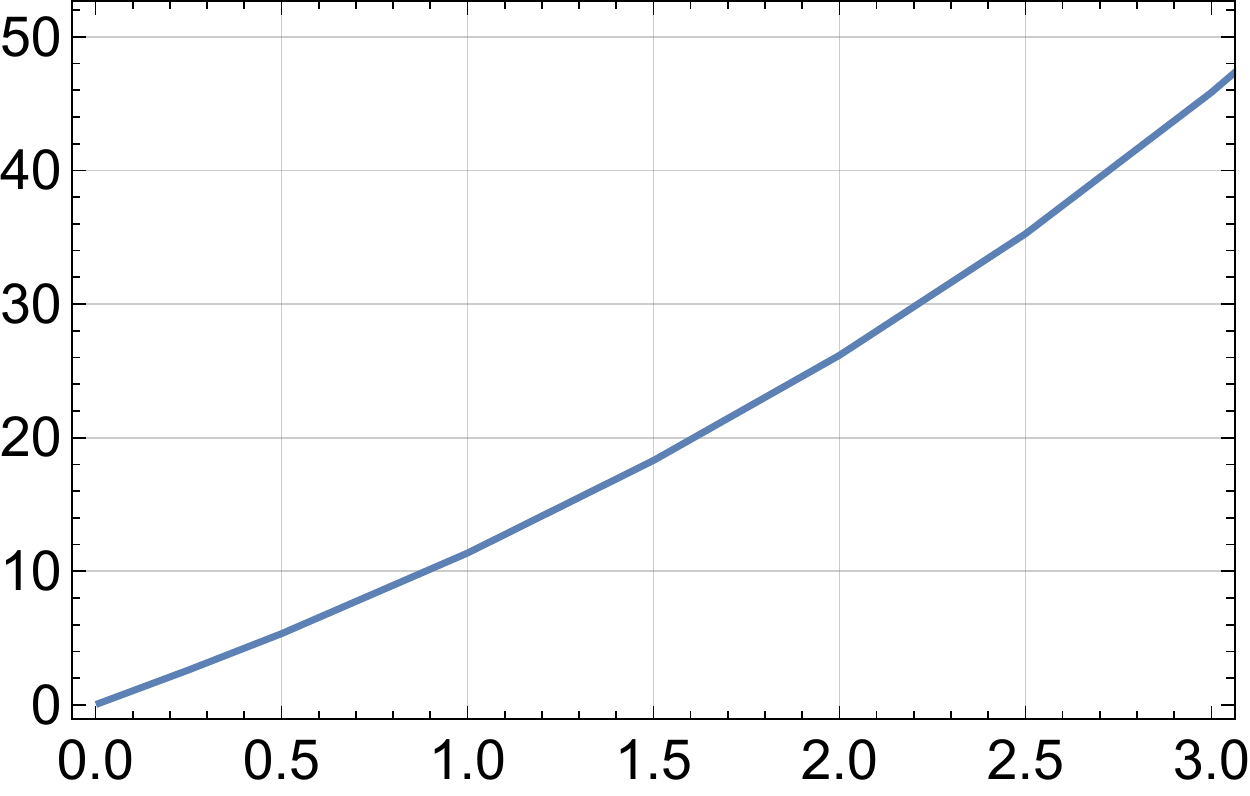}}
\put(-195,135){\large $\overline f$}
\put(-110,-15){\large $\sqrt{2}\, \xi$}
\hspace{0.05\textwidth}
{\includegraphics[width=0.45\textwidth]{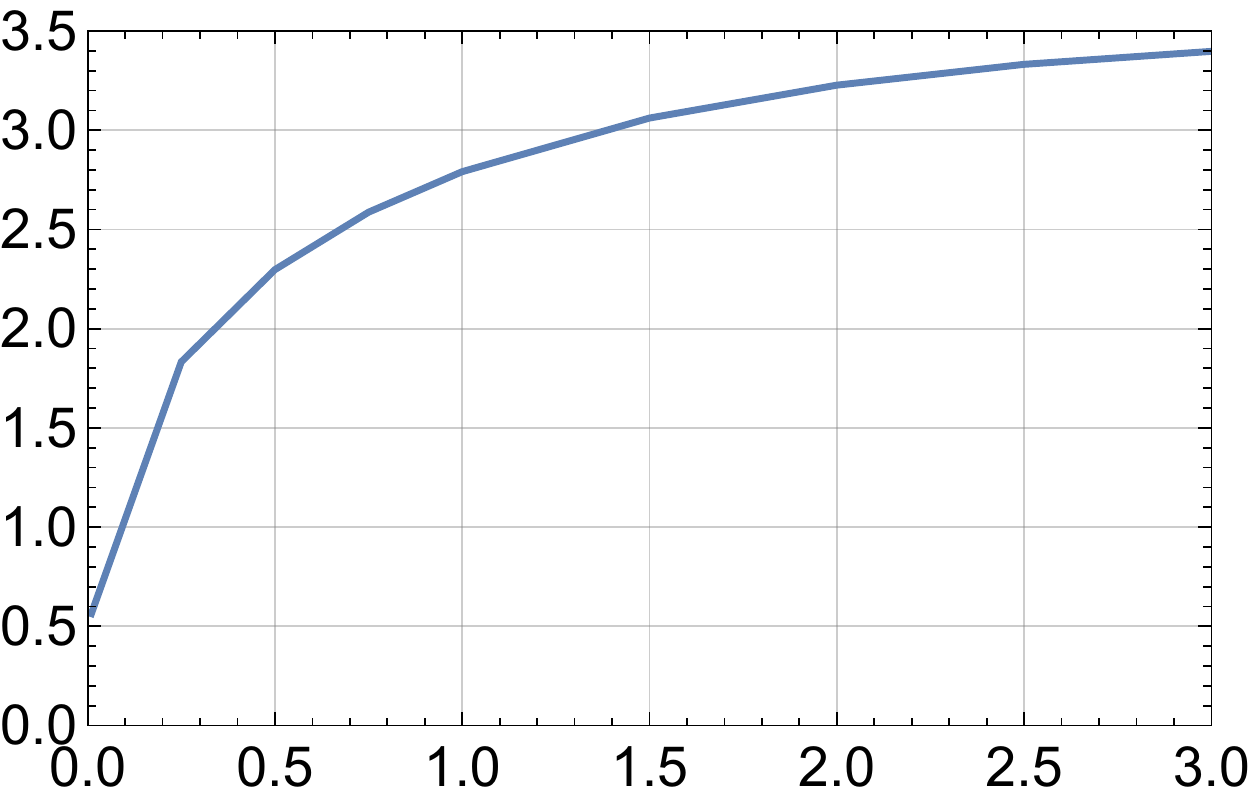}}
\put(-195,135){\large $\overline g'$}
\put(-110,-15){\large $\sqrt{2} \, \xi$}
\caption{\small Functions $f$ and $g'$ defined through Eqs.~\eqq{fff} and \eqq{ggg}, respectively.}\label{fig:fandgp}
\end{figure}

At asymptotically low temperature we can use the fact that the far IR is controlled by a Lifshitz geometry to verify the negativity of  $c_s^2$. The Lifshitz asymptotics imply that, at sufficiently low $T$, the entropy density must scale as $T^{3/z}$ with a possibly  charge-dependent coefficient, namely
\be
\label{introf}
s= f(\Qst) \, T^{3/7} \,.
\ee
The function $f(\Qst)$ can be easily extracted from the numerics and is shown in \fig{fig:fandgp}(left). Since
\be
\label{fff}
s= f(\Qst) \, T^{3/7}=-\frac{\partial F}{\partial T}\bigg\rvert_{\Qst}\,,
\ee
a simple integration of $T$ leads to 
 \be
 \label{FF}
F=-f(\Qst)\, \frac{7}{10} \, T^{10/7}+g(\Qst)\,,
\ee
where $g(\Qst)$ is a function of the charge only, related to the chemical potential through 
\be
\mu=\frac{\partial F}{\partial \Qst}\Bigg|_T=-f'(\Qst) \, \frac{7}{10}\, T^{10/7}+g'(\Qst)\,.
\label{ggg}
\ee
Using this formula we extract the function $g'(\Qst)$ from the numerically computed chemical potential. The result is shown \fig{fig:fandgp}(right). 
We can now compute all the quantities of interest. In particular 
\be
C_Q=\frac{3}{7} \, f(\Qst) \, T^{3/7} \sac \chi^{-1}=g''(\Qst) \,. 
\ee
Using \eqq{FF} and the standard thermodynamic relations for the energy and the pressure
\be\bal
E&=F+Ts \,, 
\,\\[2mm]
P&=-F+\mu \Qst \,, 
\eal\ee
we can compute \eqq{cscs} at $T=0$ with the result
\be
c_s^2(T=0)= \Qst \frac{  g''}{g'}  = \frac{\Qst}{\mu} \chi^{-1} \ .
\ee
Since $\mu$ is positive at low temperature (see \fig{EandP}) this result shows that at $T=0$  the speed of sound squared and the charge susceptibility change sign at exactly the same value of the charge density. 

\section{Discussion}
\label{disc}
We begin by addressing the fact that we used factors of $\ell_s$ and $\nc$ in our definitions of dimensionless quantities such as the temperature \eqq{TTrr}, the entropy density \eqq{TTT}, the charge density \eqq{eq.ratiodependence}, etc. The first observation is that the $\ell_s$ factors cancel out in dimensionless ratios of  physical quantities. For example, ignoring purely numerical factors we have that 
\be
\frac{s}{T^3} \sim \frac{\overline s}{\overline T^3} \, \nc^2 \sac 
\frac{\nq}{s} \sim \frac{\xi}{\overline s} \, \left( \frac{\nf}{\nc^3} \right)^{1/2} \sac \mbox{etc.}
\label{second}
\ee
Assuming that $\overline s / \overline T^3$ is of order 1, we see in the first ratio that not only $\ell_s$ cancels out  but also that we seemingly get the scaling of the entropy density with $\nc$ expected from the number of color degrees of freedom in the theory.  However, this scaling is actually ambiguous. The reason is that in the region where $\overline s / \overline T^3$ is of order 1, namely around the transition point between the LP and the Lifshitz geometries, so is the combination 
\be
\nf \, e^{\phi} \sim \lambda \frac{\nf}{\nc} \sim 1 \,,
\ee
where $\lambda$ is the 't Hooft coupling \eqq{coup}. This means that, parametrically,  at that scale one can freely replace factors of $\nf$ and/or $\nc$ with powers of  $\lambda$. The same ambiguity affects  other physical quantities such as the second ratio in \eqq{second}. Ultimately, this ambiguity arises from the lack of a privileged scale or a fixed point at which to anchor the value of the coupling.

\begin{figure}[t!!!]
\centering
{\includegraphics[width=0.55\textwidth]{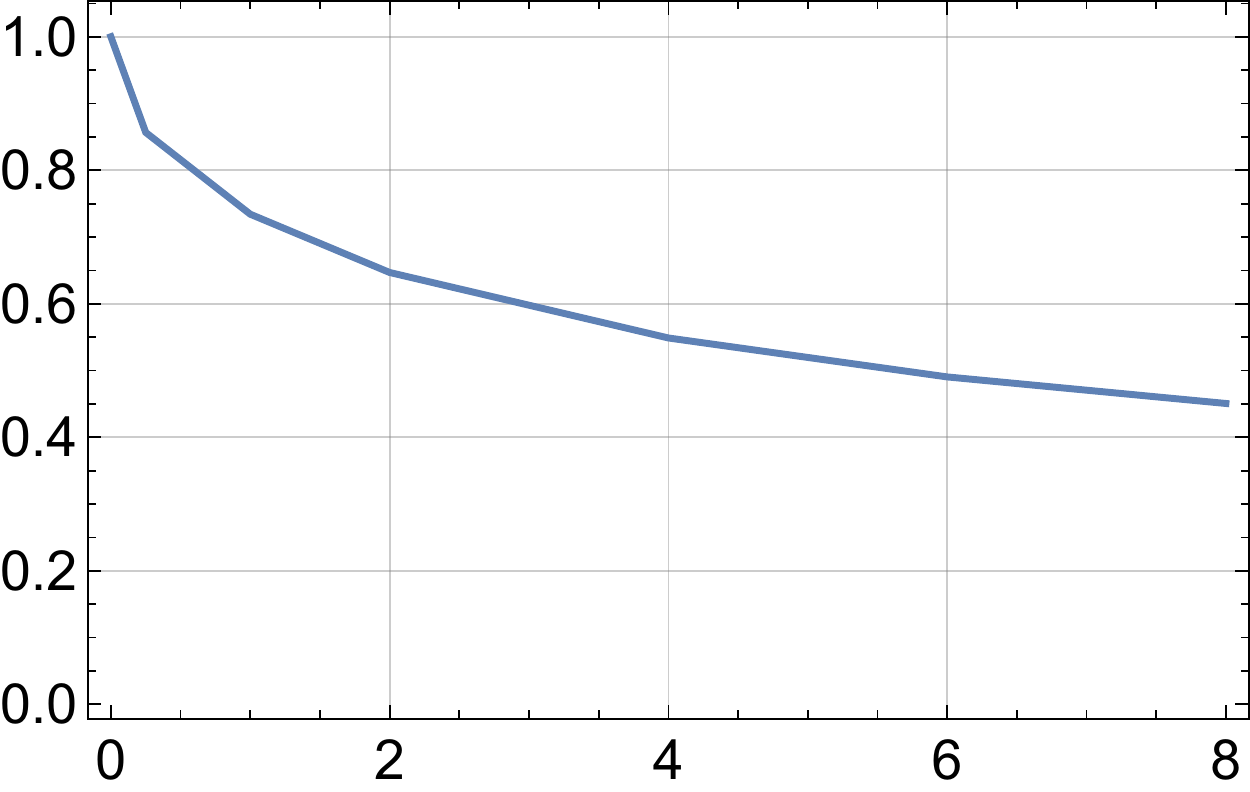}}
\put(-270,80){\Large $\frac{M}{M_0}$}
\put(-140,-15){\large $\sqrt{2} \, \xi$}
\caption{\small  Mass of a string stretching from the origin of a zero-temperature solution to the position of the LP as a function of the charge density, normalized to the value at $\xi=0$.
}
\label{LP}
\end{figure}

The second observation concerns the relation between dimensionful gauge theory quantities and the Landau pole scale. While the conventions we have chosen are convenient on the gravity side, from the gauge theory viewpoint it may be more natural to measure dimensionful quantities in units of the intrinsic scale of the theory, namely in units of $\lp$. In particular, it may be natural to construct the phase diagram by changing $\nq$ and $T$ while keeping $\lp$ fixed instead of $\ls$. In order to do this one must define the energy scale associated to the Landau pole. A convenient and well-motivated choice  is to define $\lp$ as the mass $M$ of a string stretching from the IR bottom of a zero-temperature geometry all the way up  to the LP \cite{Faedo:2016cih}. \fig{LP} shows this mass, normalized to the mass $M_0$ in the neutral case, as a function of $\xi$. 
The important message of this plot is that not only $M/M_0$ does not change parametrically as $\xi$ is varied but also that it changes rather smoothly. This means that, in order to refer all our dimensionful quantities to the LP scale instead of to $\ell_s$, we would need to do a rather mild $\xi$-dependent rescaling of our results. This would slightly distort our phase diagram but it would leave unchanged all the qualitative conclusions, which we now proceed to discuss. 

One of the most interesting outcomes of our analysis is the presence of instabilities in our system, as summarised in \fig{phases}. The most relevant ones are those that are present  at low temperature, since they are the least sensitive to the UV completion of the theory. 
One of these instabilities is associated to the  negative speed of sound squared,  $c_s^2< 0$, that is present in some regions of the phase diagram, most interestingly in Region IV. This indicates a dynamical instability for the hydrodynamic sound mode, whose dispersion relation takes the form
\be
\omega_s (k) \simeq \pm \, c_s \,  k  +\mathcal{O}(k^2) \,.
\ee
Negativity of $c_s^2$ implies that $c_s$ is purely imaginary. This indicates that, if perturbed by a small-amplitude fluctuation with sufficiently long wavelength, the system will  develop an inhomogeneous profile that will initially grow exponentially in time as
 $e^{|c_s| k t}$. Presumably, when the 
non-linearities become important the system will settle down to an equilibrium, inhomogeneous configuration in which translation invariance is broken spontaneously. It would be interesting to identify the endpoint of this instability under dynamical evolution along the lines of \cite{Attems:2017ezz,Janik:2017ykj}. Note that this is related, but not identical, to identifying all the possible equilibrium, inhomogeneous states of the system. For example, the latter may include states with domains of finite characteristic size, namely crystalline phases,  as well as phase-separated configurations in which two semi-infinite, homogenous phases  separated by an interface coexist. From a thermodynamic viewpoint one of these  inhomogeneous states will be absolutely preferred, rendering the rest metastable. However, the lifetime of some of these metastable states may be very long in general, and in particular this will be the case in our  large-$\nc$ limit. Moreover, to which final state the initial unstable configuration will evolve  is a dynamical question that depends on the ``landscape'' of configurations above and beyond the purely equilibrium ones. 

Similar considerations apply to the instability associated to the negative charge susceptibility, $\chi <0$. As explained in Appendix \ref{difu}, this is related to the charge diffusion constant through
\be
\label{DDD}
D = \frac{\sigma}{\chi} \, \frac{E+P}{ c_s^2 \, T \, C_Q}  \,.
\ee
In this expression $\sigma$ is the electrical conductivity, which must be positive in order for the divergence of the entropy current of first-order hydrodynamics to be non-negative, i.e.~in order for the second law of thermodynamics to hold. A negative value of $D$ indicates a dynamical instability towards charge clustering (anti-diffusion), thus also generating inhomogeneities. Our numerical results show that this is  the situation in several regions of the phase diagram \ref{phases}, most interestingly in Regions III and V.

To conclude, we reiterate that some of the instabilities that we have identified suggest the possible existence of quark matter crystalline phases in our model, but establishing this definitively requires further analysis.  We hope to report on these issues in the near future.

\section*{Acknowledgements}

We thank  Jorge Casalderrey-Solana, Roberto Emparan, Pavel Kovtun, Prem Kumar, Alfonso Ramallo and Javier G.~Subils for discussions, and the ``Centro de Ciencias de Benasque Pedro Pascual" (Benasque, Spain) for hospitality during the last stages of this work.  We are supported by grants 2014-SGR-1474, MEC FPA2013-46570-C2-1-P, MEC FPA2013- 46570-C2-2-P, CPAN CSD2007-00042 Consolider-Ingenio 2010, ERC Starting Grant HoloLHC-306605 and Maria de Maeztu Unit of Research Excellence distinction. JT is supported by the  Advanced ARC project ``Holography, Gauge Theories and Quantum Gravity'' and by the Belgian Fonds National de la Recherche Scientifique FNRS (convention IISN 4.4503.15). 

\appendix

\section{Supergravity with sources}\label{app.action}

In this Appendix we present the action and equations of motion we solve for. The ideas presented here follow closely the ones given in \cite{Benini:2007kg}, and the full equations of motion for a generic setup were presented in \cite{Cotrone:2012um}. The presentation of the Appendix is  the IIB counterpart to the IIA setup given in a corresponding Appendix in \cite{Faedo:2015urf}.

The starting point  is type IIB supergravity in the democratic formulation \cite{Townsend:1995gp}
\be
\bal
S& =\frac{1}{2\kappa^2}\int e^{-2\phi}(R*1 + 4\d\phi\wedge*\d\phi - \frac{1}{2} H \wedge * H) \\
& \qquad \qquad - \frac{1}{4} \left( \cG_1 \wedge *\cG_1+ \cG_3 \wedge *\cG_3 + \cG_5 \wedge *\cG_5 + \cG_7 \wedge *\cG_7 + \cG_9  \wedge *\cG_9 \right)  \ ,
\eal
\ee
where the Ramond-Ramond (RR) field strengths are defined as
\be
\cG_n = \d \cC_{n-1} - H \wedge \cC_{n-3}
\ee
with $n=1, 3, 5, 7, 9$ and $\cC_{-2}=0$. The equations of motion following from this action are for the RR field strengths
\be\label{eq.Geoms}
\bal
\d *\cG_1  & = - H \wedge * \cG_3 \ , \\
\d *\cG_3 & = - H \wedge * \cG_5 \ , \\
\d *\cG_5 & = - H \wedge * \cG_7 \ , \\
\d *\cG_7 & = - H \wedge * \cG_9 \ , \\
\d *\cG_9 & = 0 \ ,
\eal
\ee
and the  for the Neveu-Schwarz (NS) field 
\be\label{eq.NSequationunsourced}
\bal
\d \left( e^{-2\phi} *H \right) & = \frac{1}{2} \d \left( \cC_0 \wedge * \cG_3 + \cC_2 \wedge * \cG_5 + \cC_4 \wedge * \cG_7 + \cC_6 \wedge * \cG_9\right) \\
& = \frac{1}{2} \left( \cG_1 \wedge * \cG_3 + \cG_3 \wedge * \cG_5 + \cG_5 \wedge * \cG_7 + \cG_7 \wedge * \cG_9 \right) \ . 
\eal
\ee
In the last line we have used the equations of motion \eqref{eq.Geoms}. The duality relations to be imposed after the equations of motion have been obtained are
\be\label{eq.Gdualities}
\cG_9 = * \cG_1 \ , \qquad \cG_7 = - * \cG_3 \ , \qquad \cG_5 = * \cG_5 \ ,
\ee
with $**=1$ for odd forms. Then the equations of motion for $\cG_7$ and $\cG_9$ become the Bianchi identities for $\cG_1$ and $\cG_3$ (as usual $\cG_5$ is self-dual and its equation of motion is its Bianchi identity).

Add now the sources piece coming from the D7-branes. The WZ term gives a linear coupling of the worldvolume fields to the RR forms
\be\label{eq.WZaction}
S_\mt{WZ} = \frac{1}{2\kappa^2} \int \frac{1}{2} \left( \cC_8 - \cC_6 \wedge \cF + \frac{1}{2} \cC_4 \wedge \cF^2 - \frac{1}{6} \cC_2 \wedge \cF^3 + \frac{1}{24} \cC_0 \, \cF^4 \right) \wedge \Gamma \ .
\ee
Here $\Gamma$ is a two-form that describes the distribution of the D7-branes and 
\be
\cF= B + 2\pi\ls^2 \cA \ ,
\ee
with $\cA$ the BI field living in the worldvolume of the D7-branes. Note that
\be
\d \cF=H \ .
\ee
The linear source in the WZ term modifies the equations of motion for the RR forms which now read (we denote by $\ccF_n$ the RR field strengths in the presence of sources, as opposed to the unsourced $\cG_n$ ones)
\be\label{eq.Feoms}
\bal
\d *\ccF_1 & = - H \wedge * \ccF_3 + \frac{1}{24} \cF^4 \wedge \Gamma \ , \\
\d *\ccF_3 & = - H \wedge * \ccF_5 - \frac{1}{6} \cF^3 \wedge \Gamma \ , \\
\d *\ccF_5 & = - H \wedge * \ccF_7 + \frac{1}{2} \cF^2 \wedge \Gamma \ , \\
\d *\ccF_7 & = - H \wedge * \ccF_9 - \cF \wedge \Gamma \ , \\
\d *\ccF_9 & = \Gamma  \ .
\eal
\ee
In order to impose the equivalent duality relations to \eqref{eq.Gdualities}, i.e.
\be\label{eq.Fdualities}
\ccF_9 = * \ccF_1 \ , \qquad \ccF_7 = - * \ccF_3 \ , \qquad \ccF_5 = * \ccF_5 \ ,
\ee
we must modify the definitions of the RR field strengths to
\be\label{eq.Fdefinitions}
\bal
\ccF_1 & = \cG_1 + \gamma \ , \\
\ccF_3 & = \cG_3 + \cF \wedge \gamma \ , \\
\ccF_5 & = \cG_5 + \frac{1}{2} \cF^2 \wedge \gamma \ , \\
\ccF_7 & = \cG_7 + \frac{1}{6} \cF^3 \wedge \gamma \ , \\
\ccF_9 & = \cG_9 + \frac{1}{24} \cF^4 \wedge \gamma \ , 
\eal
\ee
where $\gamma$ is a one-form satisfying
\be
\d \gamma = - \Gamma \ .
\ee
In section \ref{sec.setup} we have written
\be
\gamma = \Qf \, \etakah \ , \qquad \Gamma = 2\,\Qf \,\Jkah  \ .
\ee

With this definition of the field strengths the equation of motion for the NS forms gets two extra contributions with respect to the one derived in Eqn.~\eqref{eq.NSequationunsourced}, first from the explicit $\cF$ terms in the WZ action \eqref{eq.WZaction} and second from the implicit $\cF$ terms in the definitions of the $\ccF_n$ in \eqref{eq.Fdefinitions}:
\be
\bal
\d\left( e^{-2\phi}*H \right) & = \frac{1}{2} \d \left( \cC_0 \wedge * \cG_3 + \cC_2 \wedge * \cG_5 + \cC_4 \wedge * \cG_7 + \cC_6 \wedge * \cG_9\right) \\
& \quad + \frac{1}{2} \left( * \ccF_3 + \cF \wedge *\ccF_5 + \frac{1}{2} \cF^2 \wedge * \ccF_7 + \frac{1}{6} \cF^3 \wedge * \ccF_9 \right) \wedge \gamma  \\
& \quad + \frac{1}{2} \left( \cC_6 - \cF \wedge \cC_4 + \frac{1}{2} \cF^2 \wedge \cC_2 - \frac{1}{6} \cF^3 \wedge \cC_0  \right) \wedge \Gamma  \\
& \quad + \text{DBI-terms}  \\
& = \frac{1}{2} \left( \ccF_1 \wedge * \ccF_3 + \ccF_3 \wedge * \ccF_5 + \ccF_5 \wedge * \ccF_7 + \ccF_7 \wedge * \ccF_9 \right)  \\
& \quad + \text{DBI-terms} \ , 
\eal
\ee
where in the last equality we have used the equations of motion \eqref{eq.Feoms}. The equation of motion for the dilaton and the metric can be obtained from the same democratic-formulation action, adding the DBI piece. Coming back to the non-democratic formulation of supergravity, the equations of motion for the different $\ccF_n$ and $H$ fields, as well as the ones for the dilaton and the metric, can be obtained from the  ten-dimensional action  found in Einstein frame in \cite{Cotrone:2012um}, where a five-dimensional   reduction over the compact manifold preserving the SU(2) structure of the Sasaki-Einstein manifold is given.

\section{Equations of motion for rescaled variables}\label{app.scaledeoms}

In this section we write the equations of motion for our ansatz explicitly, for string frame metric. We give the scaled quantities described in Sec.~\ref{sec.scalings}, and for simplicity we omit all bars in the different variables.

The scaled electric field satisfies
\be
 \At' = \frac{e^{\phi} \sqrt{-G_{tt} G_{rr}}\, (\ratio + \bfunc) }{ \sqrt{ G_{xx}^3 G_b^2 \, G_f + e^{2\phi} (\ratio + \bfunc)^2 }  } \ ,
\ee
and the $\bfunc$ equation of motion is
\be
\partial_y \left( \frac{\sqrt {-G_ {tt}}} {\sqrt {G_{rr}\,  G_f}} \frac{\bfunc'} {\sqrt {G_ {xx}^{3}}} \right) -  8\frac {\sqrt {-G_ {tt} G_{rr} G_f}} {\sqrt {G_ {xx}^3 G_b^4}}\bfunc - 
\frac {2e^\phi \sqrt {-G_ {tt} G_{rr}} (\ratio + \bfunc)} {\sqrt{G_ {xx}^3 G_b^2 G_f + e^{2\phi} (\ratio + \bfunc)^2}} = 0\,.
\ee

In order to write the remaining equations of motion let us define the combination
\be
\bal
{\cal F}_{A,B,C,D,E} =&  \frac{e^{2\phi}}{G_{xx}^3 G_b^4 G_f} \Bigg( 
- A\, G_b^2 \bfunc'^2 +\frac{B}{2} G_{xx}^3 G_{rr} G_b^4 + \frac{C}{2} G_{xx}^3 G_{rr} \\
& \qquad\qquad \quad + 8D\, \ratio^2 G_{rr} G_b^4 G_f- 8 E\, G_{rr} G_f \bfunc^2 \Bigg) \ . 
\eal
\ee
In terms of this, the metric and dilaton equations are
\be
\bal
0&=G_ {tt}'' + \log'\left[   e^{-2\phi} \frac{\sqrt{G_ {xx}^3 G_b^4 G_f}} {\sqrt {-G_ {tt} G_{rr}}} \right] G_ {tt}' + {\cal F}_{1,-1,-1,-1,1}\, G_ {tt} - \frac{2 e^\phi G_ {tt} G_{rr}}{\sqrt {G_ {xx}^3 G_b^4 G_f}} \frac{ G_ {xx}^3 G_b^2 G_f +   2 e^{2\phi} (\ratio + \bfunc)^2 } {\sqrt{G_ {xx}^3 G_b^2 G_f +   e^{2\phi} (\ratio + \bfunc)^2}} \ , \\
0&=G_ {xx}'' + \log'\left[   e^{-2\phi} \frac{\sqrt{-G_{tt}G_ {xx}^2 G_b^4 G_f}} {\sqrt {G_{rr}}} \right] G_ {xx}' + {\cal F}_{-1,-1,-1,1,-1}\, G_ {xx} 
+ \frac{2 e^\phi  G_ {xx}^{5/2} G_{rr} G_f } {\sqrt{G_ {xx}^3 G_b^2 G_f +   e^{2\phi} (\ratio + \bfunc)^2}} \ , \\
0&=G_ {b}'' + \log'\left[   e^{-2\phi} \frac{\sqrt{-G_{tt} G_ {xx}^3 G_b^2 G_f}} {\sqrt {G_{rr}}} \right] G_ {b}' + \left ( {\cal F}_{0,-1,1,-1,-1} + 4  \frac{G_f\, G_{rr}}{G_b^2}  -12 \frac{G_{rr}}{G_b} \right) G_ {b}\ , \\
0&=G_ {f}'' + \log'\left[   e^{-2\phi} \frac{\sqrt{-G_{tt} G_ {xx}^3 G_b^4 }} {\sqrt {G_ {f} G_{rr}}} \right] G_ {f}' + \left ({\cal F}_{-1,1,1,-1,1} - 8 \frac{G_f\, G_{rr}}{G_b^2} \right) G_ {f} - \frac{2 e^\phi G_ {xx}^{3/2} G_{rr} G_f^{3/2}}{\sqrt{G_ {xx}^3 G_b^2 G_f +   e^{2\phi} (\ratio + \bfunc)^2}} \ , \\
0&=\phi'' + \log'\left[   e^{-2\phi} \frac{\sqrt{-G_{tt} G_ {xx}^3 G_b^4 G_f}} {\sqrt {G_{rr}}} \right] \phi' + {\cal F}_{0,2,0,1,1}\,- \frac{2 e^\phi G_{rr}}{\sqrt {G_ {xx}^3 G_b^4 G_f}} \frac{ 2 G_ {xx}^3 G_b^2 G_f +   e^{2\phi} (\ratio + \bfunc)^2 } {\sqrt{G_ {xx}^3 G_b^2 G_f +   e^{2\phi} (\ratio + \bfunc)^2}} \ , 
\eal
\ee
together with a first-order constraint
\be
\bal
0=&\,\phi' \log'\left[ e^{-\phi} \sqrt{-G_{tt} G_{xx}^3 G_{rr} G_b^4 G_f} \right] - \frac{1}{2} \log' \left[ G_{xx}^{3/2} \right]  \log' \left[ \sqrt{-G_{tt} G_{xx} G_b^4 G_f} \right] \\
& - \frac{1}{2} \log' \left[ G_b^2 \right] \log' \left[ \sqrt{-G_{tt} G_b^{3/2} G_f} \right] - \frac{1}{2} \log' \left[ \sqrt{-G_{tt}} \right] \log' \left[ \sqrt{G_f} \right] + \frac{1}{4} {\cal F}_{-1,-1,-1,-1,1} \\
& - \frac{G_f\, G_{rr}}{G_b^2} + \frac{6G_{rr}}{G_b} - \frac{e^\phi G_{rr} \sqrt{G_{xx}^3 G_b^2 G_f+ e^{2\phi} (\ratio+\bfunc)^2}}{\sqrt{G_{xx}^3 G_b^4 G_f}} \ . 
\eal
\ee

\section{A probe in the asymptotic geometries}\label{app.probe}

In this Appendix we provide  more evidence supporting our claim that the zero-temperature IR geometry is dominated by the asymptotic solution \eqq{eq.IRlifshitz} and the UV by the LP geometry \eqref{eq.LandaupoleasHV}. To this end we note that the addition of a (charged) probe D7-brane in these  solutions does not have an important backreaction on the background configurations in the radial regimens of interest, i.e.~the origin for the IR geometry and the boundary for the UV one. An analogous discussion  for a lower-dimensional case was given in Ref.~\cite{Faedo:2015urf}, and we provide here a slightly different (but equivalent) argument.

Consider a set of $n_\mt{f}$ probe D7-branes with a worldvolume gauge field, $a$, turned on. The action describing this set of branes is (in string frame)
\be\label{eq.probeaction}
\begin{split}
S_\mt{D7} & = - \tens\, n_\mt{f} \int \d^8\zeta\,  e^{-\phi} \sqrt{-\det\left( {\cal P}[G]+2\pi\ls^2\, \d a + {\cal P}[B] \right)} \\
& \quad + \tens\, n_\mt{f} \int {\cal P}[\cC_8] - {\cal P}[\cC_6] \wedge \left( 2\pi\ls^2\, \d a + {\cal P}[B]  \right) \ ,
\end{split}
\ee
where  $\tens\,n_\mt{f}$ is the total tension, and we have assumed that there is a non-trivial potential $\cC_6$ in the background of the form given by Eqn.~\eqref{eqn.C6potential}:
\be
\cC_6 \supset \bfunc\, \d x^1 \wedge \d x^2 \wedge \d x^3 \wedge \Jkah \wedge \etakah \ .
\ee
 We also allow for a non trivial NS potential, which we will take in the $B_{rt}$ directions to be aligned with $\d a = a_t'(r) \d r \wedge \d t$. 

For a massless probe brane the embedding profile is  a constant in the transverse directions to the brane. Therefore, working in the static gauge we can write the probe action as
\be
S_\mt{D7} =  \int \d^3 x \, \d t \, \d r \left( -\sqrt{\cH_1}\, \sqrt{1- \cH_2 \left( a_t' + \frac{B_{rt}}{2\pi \ls^2} \right)^2} + \cH_3 - \cH_4 \left(a_t' + \frac{B_{rt}}{2\pi \ls^2} \right) \right) \ ,
\ee
where we have defined the following functions
\be
\bal
\cH_1 & =  - (\tens\, n_\mt{f} \, V_3)^2\,e^{-2\phi}\,G_{tt} \, G_{rr}\, G_{xx}^3 \, G_{b}^2\, G_f  \ ,\\
\cH_2 & =  \left(2\pi\ls^2\right)^2\, (- G_{tt} G_{rr})^{-1}  \ ,\\
\cH_3 & =  \tens\, n_\mt{f} \, V_3\,\frac{\Qf}{2} \frac{\sqrt{-G_{tt} \,G_{xx}^3 \,G_{rr} \,G_{b}^4}}{\sqrt{G_f}}  \ ,\\
\cH_4 & =  \tens\, n_\mt{f} \, V_3\, 2\pi\ls^2\, \bfunc \ ,
\eal
\ee
with $V_3$  the volume of the 3-cycle wrapped by the probe branes and the metric functions  those of the generic metric \eqref{eq.10dmetricgeneric}. The function $\cH_3$ comes from $\cC_8$ via Hodge-dualization of $F_1=\Qf \,\etakah$ and $\cH_4$ has its origin in  $\cC_6$.

The electric field on the worldvolume of the D7-branes has a first constant of motion given by
\be\label{eq.atprobe}
\frac{\delta S_\mt{D7}}{\delta a_t'} =  n_\mt{q} \quad \Rightarrow \quad   a_t' + \frac{B_{rt}}{2\pi \ls^2 } =  \frac{  n_\mt{q} +  \cH_4 }{ \sqrt{\cH_2} \, \sqrt{\cH_1 \cH_2 +  (n_\mt{q}+ \cH_4)}^2  } \ ,
\ee
whose backreacted version for the ansatz considered in this paper  can be found in Eq.~\eqref{eq.Atprimesolution}.
Working at fixed charge density $n_\mt{q}$  we can use the former expression to eliminate $a_t'$ in favor of $n_\mt{q}$ in the action by performing a Legendre transform. This results in
\be\label{eq.probeargument}
\tilde S_\mt{D7}  = S_\mt{D7} - \int \d t\, \d^3 x \, \d r \frac{\delta S_\mt{D7}}{\delta a_t'} a_t' =  \int \d t\, \d^3 x \, \d r \left( -\frac{\sqrt{\cH_1  \cH_2 + (n_\mt{q} + \cH_4)^2}}{ \sqrt{\cH_2}} + \cH_3 +n_\mt{q} \frac{B_{rt} }{2\pi\ls^2}  \right) \ .
\ee

\subsection{Charged flavorless setup and the IR geometry}

Consider now the behavior of \eqref{eq.probeargument} in the limit $n_\mt{f}\to0$, but keeping $n_\mt{q}$ finite. This is the limit described in Sec.~\ref{sec.IRisLifshitz}. The Legendre-transformed action becomes
\be\label{eq.NGapproximation}
\tilde S_\mt{D7} \to \int \d t\, \d^3 x \, \d r \left(- \frac{n_\mt{q}}{\sqrt{\cH_2}} + n_\mt{q} \frac{B_{rt} }{2\pi\ls^2} \right) = - \frac{1}{2\pi\ls^2}\int \d t\, \d r \sqrt{-G_{tt} G_{rr}} |\Xi|^{1/2} +\frac{1}{2\pi\ls^2} \int B \wedge \Xi \ ,
\ee
and we recognize this as a smeared Nambu-Goto action where $\Xi=n_\mt{q} \, \d x^1 \wedge \d x^2 \wedge \d x^3$ is a density of fundamental strings, extended in the radial direction and distributed on the spatial ones. The same limit can clearly be achieved asymptotically if one works in a radial regime of a geometry such that 
\be\label{eq.stringscondition}
\cH_1 \cH_2\to 0 \ , \qquad \sqrt{\cH_2}\, \cH_3\to 0 \ , \qquad \cH_4\to0 \ . 
\ee
The solution \eqref{eq.IRlifshitz} satisfies this criterium  when $r\to0$,  suggesting that near the origin this is a valid asymptotic solution. Actually,  one should consider the first correction to the function $\bfunc=0$ to ensure the requirement $\cH_4\to 0$ near the origin, i.e.~that the leading correction to this function is vanishing when $r\to0$ (otherwise the probe approximation would fail). In Eq.~\eqref{eq.IRbfunc} we show that  indeed $\bfunc\to0$ in the backreacted setup.

\subsection{Supersymmetric chargeless solution and UV geometry}

From the probe argument \eqref{eq.probeargument} we would expect that the supersymmetric solution is valid asymptotically if near the Landau pole $(n_\mt{q}+\cH_4)/(\cH_1 \cH_2)\to0$, provided the NS form vanishes.  Plugging all the values from the asymptotic expansion \eqref{eq.UVnumerics} in the $\cH_i$ functions we obtain
\be\label{eq.susyBGcondition}
\frac{n_\mt{q}+\cH_4}{\cH_1 \cH_2} =  (n_\mt{q}+\cH_4) \, r^{56/5} \ ,
\ee
which near the Landau pole does not go to zero unless there is a precise cancellation between $\cH_4$ and the charge density. In other words, it is not possible to obtain a parametrically small quotient of the charge versus D7-tension effects (such that the charge is subdominant in the UV) by just going to sufficiently large values of $r$. Gravitationally, this is an effect of having the end of the geometry at a finite proper distance from any point in the bulk. 

We argue now that the setup we have considered automatically tunes itself to avoid this  issue: the function $\bfunc$ goes to a constant value near the Landau pole that cancels exactly the contribution from the charge density, such that the backreacted version of \eqref{eq.susyBGcondition} indeed goes to zero as one approaches the Landau pole. To see this take the  full system of backreacted equations of motion, given in appendix~\ref{app.scaledeoms}, and expand around the supersymmetric solution. In particular one can consider a perturbative expansion in which $\Qst = \epsilon\, \Qst$ and $\bfunc = \epsilon\, \bfunc$, where $\epsilon$ is a book-keeping parameter to denote the inclusion of a small charge density in the supersymmetric setup. At first order in $\epsilon$ the equation of motion for $\bfunc$ decouples and reads
\be
\bfunc '' +2 \left( 3 - 2e^{2(\sff-\sfg)}  \right) \bfunc' - 2e^{2(\sff-\sfg)} \left( \Qf\, e^\phi + 4 e^{2(\sff-\sfg)} \right) \bfunc - \frac{\Qc\, \Qst}{2}e^{\phi+2(\sff-\sfg)} = 0 \ ,  
\ee
with all functions evaluated in the asymptotic solution \eqref{eq.UVnumerics}. Close to the Landau pole we can approximate the equation by
\be
\bfunc '' +6 \,\bfunc' - 2 \, e^{\phi + 2(\sff-\sfg)} \left( \Qf\, \bfunc + \frac{\Qc\, \Qst}{4}\right) =  {\cal O}(r^{-4})  \ ,  
\ee
and the particular solution is given at leading order by a constant
\be\label{eq.BpartUV}
\bfunc_{particular} = - \frac{\Qc\, \Qst}{4\, \Qf} + {\cal O}(r^{-4}) \ .
\ee
By plugging this value in \eqref{eq.Atprimesolution}, which is the backreacted version of \eqref{eq.probeargument}, we observe that at the Landau pole $\At'\to0$, which allows us to construct the asymptotic expansion \eqref{eq.UVnumerics} around the supersymmetric solution.

\section{Scaling solution and IR geometry}\label{app.Lifshitzmodes}

In this Appendix we show that  the configuration in Eq.~\eqref{eq.IRlifshitz} is an asymptotic solution in the presence of a finite number of flavors. One can perform an expansion of the full equations for small $\Qf$, and solve order by order. At order ${\cal O}(\Qf^0)$ we have  Eq.~\eqref{eq.IRlifshitz} with $\bfunc=0$ as a solution. To find the first order correction we expand the functions as
\be
\bal
G_Z(u) & = G_Z^{(0)}(r) \left( 1 + \Qf \, \gamma_Z(r) + {\cal O}(\Qf^2) \right) \ , \\
e^{\phi(u)} & = e^{\phi^{(0)}(r)} \left(  1 + \Qf \, \varphi(r) + {\cal O}(\Qf^2) \right)\ ,  \\
\bfunc (u) & = \Qf \, \beta(r) + {\cal O}(\Qf^2)  \ ,
\eal
\ee
where the metric components are $Z=\{tt,xx,rr,b,f\}$ and the superindex ${}^{(0)}$ refers to the solution \eqref{eq.IRlifshitz}. The equation of motion for $\beta(u)$ decouples and has the solution
\be\label{eq.betasolution}
\beta (u) = -  \frac{2^5\, 3^3}{34^{3/2}}   \frac{1}{\Qst^3} \frac{r^{12}}{L^{12}} + \beta_1 \, r^{10} + \beta_2 \, r^{-8} \ ,
\ee
with $\beta_{1,2}$ integration constants. For the $\Qf$-expansion to be well defined we need to set $\beta_2=0$ and we observe that as $r\to0$ the first-order correction $\beta$ vanishes, and in particular the explicit solution to the non-homogeneous part of the equation goes like $r^{12}$. The remaining first-order corrections satisfy a coupled system of equations, and the solutions are a combination of eight different modes which we group in pairs as
\be\label{eq.Lifshitzmodes}
\gamma_Z(r) = \sum_{i,\pm} \gamma_Z^{i,\pm} r^{\Delta_{i,\pm}} \ , \qquad \varphi(r) = \sum_{i,\pm} \varphi^{i,\pm} r^{\Delta_{i,\pm}} \ ,
\ee
with $i=1,2,3,4$. For fixed $i$ and sign $\pm$, the factors $\gamma_Z^{i,\pm}$ and $\varphi^{i,\pm}$ are not all independent, but given in terms of just two free parameters, with one of these parameters corresponding to a gauge fixing of the radial coordinate. 

We have grouped the expansion modes in pairs of fixed $i$. The two powers characterising each of the pairs add up to
$\Delta_{i,+} + \Delta_{i,-} =  -10$. The first of these pairs is given by the values
\be
\Delta_{1,\pm} = -5 \pm 5 \ ,
\ee
the $\Delta_{1,+}=0$ mode corresponding to a rescaling of time and the $\Delta_{1,-}=-10$ mode corresponding to turning on a temperature, with the corresponding non-vanishing coefficients given by
\be
\gamma_{tt}^{1,-}=-\gamma_{rr}^{1,-} \ , \qquad \varphi^{1,-} = 2 \gamma_{f}^{1,-} =2 \gamma_{b}^{1,-}= \frac{6}{5} \gamma_{xx}^{1,-}  \ ,
\ee 
where, once again, one of the two undetermined coefficients corresponds to a fixing of the radial coordinate. As seen from the negative value of the $\Delta_{1,-}$ coefficient this is a relevant mode that modifies the IR geometry; since in this appendix we are interested in the zero temperature solutions, with the Lifshitz geometry in the IR modified just by irrelevant deformations, we take the coefficients multiplying the $\Delta_{1,-}$ modes to be zero. The mode $\Delta_{1,+}=0$ is associated to a free $\gamma_{tt}^{1,+}$ coefficient, corresponding to the choice of $c_\mt{t}$ in Eq.~\eqref{eq.IRlifshitz}.

The remaining three pairs appearing in the solution are given by
\be
\bal
\Delta_{2,\pm} & = - 5 \pm \sqrt{\frac{5}{17} \left(  917 - 8 \sqrt{1279} \right)}  \ ,  \\
\Delta_{3,\pm} & =   - 5 \pm \sqrt{\frac{5}{17} \left(  917 + 8 \sqrt{1279} \right)} \ , \\
\Delta_{4,\pm} & =   - 5 \pm \sqrt{145}  \ .
\eal
\ee
 The corresponding coefficients $\gamma_Z^{i,\pm}$, $\varphi^{i,\pm}$ can be determined analytically. We give here only the relations between the two coefficients in the compact part of the manifold
\be
 \gamma_{f}^{2,\pm} = \gamma_{b}^{2,\pm}  \ , \qquad  \gamma_{f}^{3,\pm} = \gamma_{b}^{3,\pm}  \ , \qquad  \gamma_{f}^{4,\pm} =-4\gamma_{b}^{4,\pm} + \frac{5}{2} \varphi^{4,\pm} \ .
\ee
The modes $\Delta_{2,\pm}$ and $\Delta_{3,\pm}$ appeared already in the analysis made in \cite{Kumar:2012ui,Faedo:2014ana}. In particular $\gamma_{f}^{2,\pm}=\gamma_{b}^{2,\pm}$ and $\gamma_{f}^{3,\pm}=\gamma_{b}^{3,\pm}$, which guarantees that these deformations do not contribute to the squashing of the five-dimensional  compact manifold, preserving the full isometry group. However, ${\gamma_{f}^{4,\pm}=-4\gamma_{b}^{4,\pm}}+ \frac{5}{2} \varphi^{4,\pm}$ and these are the modes responsible for the breaking of the symmetry by means of the squashing of the compact manifold allowed in our ansatz \eqref{eq.10dmetricgeneric}, which is nothing but a consequence of the presence of flavor in the setup. Considering just the irrelevant deformations (those with $\Delta_{i,\pm}>0$) to have the Lifshitz solution \eqref{eq.IRlifshitz} in the IR implies choosing the free parameters so that they cancel the $\Delta_{i,-}$ modes, allowing only the $\Delta_{i,+}$ ones, with $i=2,3,4$. 

Thus, we have taken to zero all relevant deformations of the geometry, meaning that our perturbation holds always close to the origin, even if we relax the initial condition $\Qf\ll1$ that lead us to the perturbation equations we just solved. In other words, we have just proven from the  supergravity equations of motion with sources that there is a solution near the origin that asymptotes to the \eqref{eq.IRlifshitz} solution, with corrections given by the 5 irrelevant deformations $\Delta_{2,+}$, $\Delta_{3,+}$, $\Delta_{4,+}$, $c_\mt{t}$ and $\beta_1$.

\section{Charge diffusion constant}
\label{difu}
In this section we provide a short derivation of \eq{DDD}. We start with the expression derived on page 27 of \cite{Kovtun:2012rj}:
\be 
D=\frac{\sigma \left( \alpha_2 \beta_1 - \alpha_1 \beta_2 \right) }{c_s^2} \,,
\ee
where $\sigma$ is the electric conductivity and $\alpha_i, \beta_i$ are thermodynamic derivatives defined in  \cite{Kovtun:2012rj}. Using the expressions at the bottom of page 26 of that reference one can rewrite this as 
\be
D = \frac{\sigma}{c_s^2} \, \frac{(E+P)^2}{\det \chi_{ab}}  \,,
\ee
where  $\chi_{ab}$ is the susceptibility matrix 
\be\label{eq.suscepmat}
\chi_{ab}=
\begin{pmatrix}
T  \left( \frac{ \partial E}{\partial T}\right)_{\mu/T} & 0 & \left( \frac{ \partial E}{ \partial \mu }\right)_T \\[2mm]
0 & E + P & 0 \\[2mm]
T  \left( \frac{ \partial \Qst}{\partial T}\right)_{\mu/T} & 0 & \left( \frac{ \partial \Qst}{ \partial \mu }\right)_T 
\end{pmatrix} \,.
\ee
Note that $\chi_{33} = \chi$ as defined in the first equation in \eqq{stab1}. 
Using thermodynamic identities one can show that 
\begin{eqnarray}
\chi_{33} &=& H_{22} \,,\\
\chi_{13} &=& T H_{12} + \mu H_{22} \,, \\
\chi_{11} &=& T^2 H_{11} + \mu^2 H_{22} + 2 T \mu H_{12} \,,
\end{eqnarray}
where $H_{ij}$ are the components of the Hessian \eqref{eq.hessian}. It follows that the determinants are related through 
\be
\det \chi_{ab} = T^2 \, (E+P) \, \det H \,,
\ee
and therefore
\be
D = \frac{\sigma}{c_s^2} \, \frac{E+P}{T^2 \, \det H}  \ .
\ee
Using \eq{eq.detH} we finally arrive at

\be
D = \frac{\sigma}{\chi} \, \frac{E+P}{ c_s^2 \, T \, C_Q}  \ ,
\ee
which gives the Einstein relation in the presence of a finite chemical potential and charge density. Note that in the neutral case we have 
\be
\mu=\Qst=0 \sac E + P = T  s \sac c_s^2 = s /C_Q \,,
\ee
and therefore $D$ reduces to the familiar expression 
\be
D = \frac{\sigma}{\chi} \,.
\ee

\end{document}